\documentclass[11pt,a4paper]{article}
\pdfoutput=1
\usepackage{jheppub}
\usepackage[T1]{fontenc}
\usepackage{lmodern}
\usepackage{mathrsfs}
\usepackage{dsfont}
\usepackage{mathtools,bm}
\usepackage{microtype}
\usepackage{placeins}
\usepackage{xcolor}
\usepackage{tikz}
\usetikzlibrary{arrows.meta,calc}
\hypersetup{
 citecolor=blue!55!black,
 linkcolor=blue!55!black,
 urlcolor=blue!55!black
}

\newcommand{\one}{\mathbf 1}

\preprint{USTC-ICTS/PCFT-26-66}
\title{Renormalization Group Domain Walls from Phantom Symmetries}
\author[a,b]{Jin Chen}
\affiliation[a]{Department of Physics, Xiamen University, Xiamen 361005, China}
\affiliation[b]{Peng Huanwu Center for Fundamental Theory, Hefei, Anhui 230026, China}
\emailAdd{zenofox@gmail.com}
\abstract{
We develop a systematic construction of renormalization group (RG) domain walls between two-dimensional rational conformal field theories. Using phantom currents, we consider an auxiliary chiral extension linked to the folded ultraviolet--infrared theory by a topological interface. The construction of RG walls then reduces to identifying suitable automorphism-twisted Cardy states in the auxiliary theory. We find two such types of automorphisms: outer automorphisms exchanging identical current algebra factors, as in Gaiotto's construction, and continuous inner automorphisms generated by the zero modes of spin-one phantom currents. In the latter case, the RG domain walls generically admit a continuous defect conformal manifold. We further show that, when an RG flow can be orbifolded to another one, the orbifold may lift defect conformal moduli of the RG brane constructed using inner automorphisms. The two types of RG branes can then be mapped to each other via a topological interface defined by the orbifold. We demonstrate the two types of RG brane construction via various examples of flows between non-unitary minimal models, from parafermions to minimal models, and between generalized parafermions. Many physical observables of the RG walls are computed non-perturbatively, including boundary entropies and spectra, and UV--IR operator mixings. We also find interesting nontrivial boundary monodromy: a closed cycle in the defect conformal moduli space can carry an RG domain wall into other distinct non-trivial conformal interfaces.
}

\begin{document}
\maketitle
\raggedbottom

\section{Introduction}
\label{sec:introduction}

\begin{figure}[t]
 \centering
\begin{tikzpicture}[
 x=1.086189cm,y=1.243381cm,
 panel edge/.style={draw=black!55,line width=0.45pt},
 process arrow/.style={-{Stealth[length=2.4mm,width=1.6mm]},
   draw=black!75,line width=0.8pt},
 theory label/.style={font=\small,align=center},
 step label/.style={font=\footnotesize,align=center,inner sep=1pt},
 panel label/.style={font=\footnotesize,align=center},
]
 \definecolor{PCIRFill}{RGB}{247,226,226}
 \definecolor{PCUVFill}{RGB}{226,238,249}
 \definecolor{PCWall}{RGB}{181,43,114}

 \path[panel edge,fill=PCUVFill] (0,0) rectangle (3.2,2.05);
 \draw[black!45,densely dashed,line width=0.6pt]
   (1.6,0) -- (1.6,2.05);
 \node[theory label] at (0.8,1.03) {UV};
 \node[theory label] at (2.4,1.03) {UV};
 \node[step label,anchor=north] at (1.6,-0.13) {$x=0$};
 \node[panel label] at (1.6,-0.68) {(a) Uniform UV theory};

 \path[fill=PCIRFill] (5.3,0) rectangle (6.9,2.05);
 \path[fill=PCUVFill] (6.9,0) rectangle (8.5,2.05);
 \draw[panel edge] (5.3,0) rectangle (8.5,2.05);
 \draw[PCWall,dash pattern=on 4pt off 2pt,line width=1.8pt]
   (6.9,0) -- (6.9,2.05);
 \node[theory label,font=\footnotesize] at (6.1,1.03) {perturbed\\UV};
 \node[theory label] at (7.7,1.03) {UV};
 \node[step label,anchor=north] at (6.9,-0.13) {$x=0$};
 \node[panel label] at (6.9,-0.68) {(b) Along the flow};

 \path[fill=PCIRFill] (10.6,0) rectangle (12.2,2.05);
 \path[fill=PCUVFill] (12.2,0) rectangle (13.8,2.05);
 \draw[panel edge] (10.6,0) rectangle (13.8,2.05);
 \draw[PCWall,line width=1.8pt] (12.2,0) -- (12.2,2.05);
 \node[theory label] at (11.4,1.03) {IR};
 \node[theory label] at (13.0,1.03) {UV};
 \node[step label,anchor=north] at (12.2,-0.13) {$x=0$};
 \node[panel label] at (12.2,-0.68) {(c) Conformal RG wall};

 \draw[process arrow] (3.42,1.03) -- (5.08,1.03);
 \node[step label,anchor=south] at (4.25,1.24) {turn on $\lambda$};
 \node[step label,anchor=north] at (4.25,0.82) {for $x<0$};
 \draw[process arrow] (8.72,1.03) -- (10.38,1.03);
 \node[step label,anchor=south] at (9.55,1.24) {bulk and};
 \node[step label,anchor=north] at (9.55,0.82) {interface RG};
\end{tikzpicture}
 \caption{Half-space construction of an RG domain wall. A perturbation given by a relevant operator is switched on for $x<0$. The dashed magenta line denotes the interface during the flow, which need not be conformal. The solid magenta line denotes the RG wall after both the perturbed bulk and the interface interactions have reached their fixed points.}
 \label{fig:half-space-rg}
\end{figure}
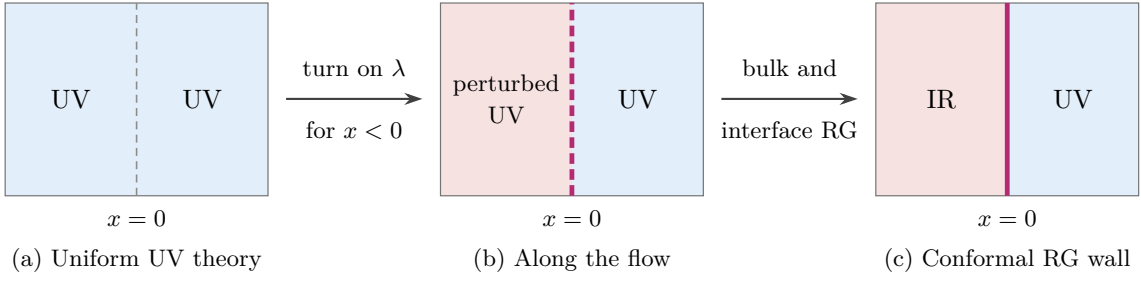

Conformal interfaces are codimension-one extended objects joining two conformal field theories (CFTs) while preserving the conformal symmetry along the defects. In two dimensions, they describe how two critical systems can be conformally coupled with each other along a line without introducing an energy scale. The energy transmission and reflection rates measure how strongly the two sides communicate, while correlation functions in the presence of the interface relate their operator content~\cite{Quella:2006de,Gaiotto:2012np}. Furthermore, conformal interfaces deserve to be studied in their own right, as their spectra, entanglement entropies, etc. encode much interesting non-perturbative information about the coupled systems. A particularly interesting class of conformal interfaces consists of renormalization group (RG) domain walls, which connect ultraviolet (UV) theories to their infrared (IR) fixed points along an RG flow~\cite{Zamolodchikov:1987ti,Fateev:1990bf,Fateev:1991bv,Nakayama:2024msv,Ambrosino:2025yug,Gaberdiel:2026sfg,Ambrosino:2026umb,Benedetti:2026drn}. One can construct such an RG domain wall in the following way. We start with a UV CFT containing a scalar relevant operator $\mathcal O_{\rm rel}$ of dimension $\Delta_{\mathcal O_{\rm rel}}<2$. We turn on a homogeneous perturbation by $\mathcal O_{\rm rel}$ only on half of spacetime, for example the left half, 
\begin{align}
 S=S_{\rm UV}
 +\lambda\int_{-\infty}^{0}dx\int_{-\infty}^{\infty}d\tau\,
 \mathcal O_{\rm rel}(x,\tau)\,,
 \label{eq:half-space-deformation}
\end{align}
where $S_{\rm UV}$ is the action of the undeformed theory, and the coupling constant $\lambda$ has mass dimension $2-\Delta_{\mathcal O_{\rm rel}}$. At long distances, the bulk on the left approaches the IR fixed point, while the bulk on the right remains at the UV fixed point. The junction at $x=0$ can then define an interface, extended along the temporal direction, between the two CFTs, as illustrated in Figure~\ref{fig:half-space-rg}~\cite{Gaiotto:2012np}. Correlators across the interface then encode how UV operators reorganize into IR ones. They thus provide a valuable non-perturbative description of the operator mixings obtained by diagonalizing the anomalous dimension matrix.

It is worth noting that reaching the IR fixed point in the perturbed bulk does not, by itself, guarantee the conformality of the domain wall. Renormalization group flow may generate further interaction terms localized along the wall, and these interface couplings have their own RG flow~\cite{Gaiotto:2012np,Fredenhagen:2006dn}. A conformal RG wall requires these interactions to reach a fixed point as well, where the beta functions of the interface couplings vanish. Equivalently, applying the folding trick, one can fold the UV and IR theories into a tensor product theory, where their RG domain wall then becomes a conformal boundary condition, which, in the closed string channel, we call an RG brane. It is not hard to see that an RG brane cannot be made of tensor products of the Cardy branes in the UV and IR theories, because such factorized branes, or interfaces after unfolding, allow no energy transmission between the left and right theories~\cite{Quella:2006de}. Therefore, the construction of an RG wall remains a non-trivial task, even when both of the UV and IR theories are rational CFTs (RCFTs) with exactly known conformal data. An RG domain wall, or equivalently the RG brane, is generically a non-factorizable and irrational conformal interface with respect to the chiral algebra of the product of the UV--IR theories in the folded picture~\cite{Gaiotto:2012np}.

\medskip
\noindent\textbf{Transmitted symmetries and phantom currents.}
Generalized symmetry provides important additional constraints on this problem, just as it does for bulk RG flows. In two dimensions, generalized symmetries are represented by topological defect lines (TDLs), which can be deformed without changing correlation functions. We here refer readers to references~\cite{Fuchs:2002cm,Fuchs:2001qc,Frohlich:2004ef,Fuchs:2012dt,Gaiotto:2014kfa,Tachikawa:2019TASI,Bhardwaj:2017xup,Chang:2018iay,Choi:2021kmx,Komargodski:2020mxz,Thorngren:2019iar,Thorngren:2021yso,Kaidi:2022cpf,Bhardwaj:2022yxj,Chang:2022hud,Kaidi:2023maf,Bhardwaj:2023wzd,Bhardwaj:2023ayw,Bhardwaj:2023kri,Chen:2023qnv,Shao:2023gho,Luo:2023ive,Cordova:2024iti,Chen:2025qub,Antinucci:2025fjp,Chen:2026wyn,Chen:2026cqz, Fukusumi:2026pkr} for thorough reviews and advanced topics on this subject. In the language of TDLs, invertible lines implement ordinary group symmetries, while non-invertible ones can fuse into a sum of lines. The finite collections of these lines, which are closed under fusion and equipped with necessary junction data, form fusion categories~\cite{Frohlich:2006ch,Chang:2018iay}. When a relevant perturbation preserves such a fusion category, its lines survive along the flow and must be realized in the IR theory. Matching their fusion and junction data provides many non-trivial constraints and consistency conditions on the flow. This approach has led to various families of proposed massless flows among (non-unitary) minimal models and coset CFTs~\cite{Nakayama:2024msv,Ambrosino:2025yug,Gaberdiel:2026sfg,Ambrosino:2026umb,Benedetti:2026drn, Fukusumi:2026pkr}. For the purpose of constructing RG domain walls, these preserved TDLs also provide important clues on how to construct such walls entirely algebraically, as first noted in the seminal paper of Gaiotto~\cite{Gaiotto:2012np}.

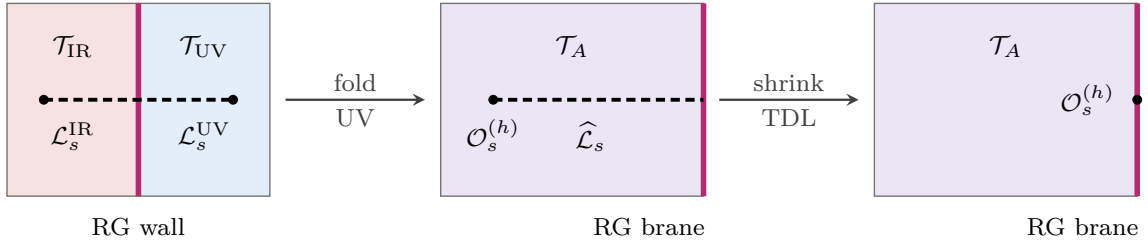
\begin{figure}[t]
\centering
\definecolor{PCIRFill}{RGB}{247,226,226}
\definecolor{PCUVFill}{RGB}{226,238,249}
\definecolor{PCFoldFill}{RGB}{235,229,244}
\definecolor{PCWall}{RGB}{181,43,114}
\resizebox{\linewidth}{!}{%
\begin{tikzpicture}[
  x=1.5cm,
  y=1.10cm,
  >=stealth,
  panel edge/.style={draw=black!55,line width=0.45pt},
  rg wall/.style={PCWall,line width=1.8pt},
  tdl/.style={black,line width=1.25pt,dash pattern=on 3pt off 1.8pt},
  endpoint/.style={black},
  process arrow/.style={->,black!75,line width=0.75pt},
  step label/.style={font=\scriptsize,fill=white,inner sep=1pt},
  theory label/.style={font=\scriptsize},
  defect label/.style={font=\scriptsize},
  wall label/.style={font=\scriptsize,align=center}
]
  \path[panel edge,fill=PCIRFill]
    (-1.00,1.00) rectangle (0,-1.00);
  \path[panel edge,fill=PCUVFill]
    (0,1.00) rectangle (1.00,-1.00);
  \draw[rg wall] (0,1.00) -- (0,-1.00);
  \node[theory label] at (-0.50,0.55) {$\mathcal T_{\rm IR}$};
  \node[theory label] at (0.50,0.55) {$\mathcal T_{\rm UV}$};
  \draw[tdl] (-0.72,0) -- (0.72,0);
  \fill[endpoint] (-0.72,0) circle[radius=1.60pt];
  \fill[endpoint] (0.72,0) circle[radius=1.60pt];
  \node[defect label] at (-0.50,-0.38) {$\mathcal L_s^{\rm IR}$};
  \node[defect label] at (0.50,-0.38) {$\mathcal L_s^{\rm UV}$};
  \node[wall label,anchor=north] at (0,-1.10) {RG wall};

  \draw[process arrow] (1.12,0) --
    node[step label,above=1.5pt]{fold}
    node[step label,below=1.5pt]{UV}
    (2.18,0);
  \path[panel edge,fill=PCFoldFill]
    (2.30,1.00) rectangle (4.30,-1.00);
  \draw[rg wall] (4.30,1.00) -- (4.30,-1.00);
  \node[theory label] at (3.30,0.55) {$\mathcal T_A$};
  \draw[tdl] (2.70,0) -- (4.30,0);
  \fill[endpoint] (2.70,0) circle[radius=1.60pt];
  \node[defect label,anchor=north] at (2.70,-0.08)
    {$\mathcal O_s^{(h)}$};
  \node[defect label] at (3.44,-0.38) {$\widehat{\mathcal L}_s$};
  \node[wall label,anchor=north east] at (4.40,-1.10) {RG brane};

  \draw[process arrow] (4.42,0) --
    node[step label,above=1.5pt]{shrink}
    node[step label,below=1.5pt]{TDL}
    (5.48,0);
  \path[panel edge,fill=PCFoldFill]
    (5.60,1.00) rectangle (7.60,-1.00);
  \draw[rg wall] (7.60,1.00) -- (7.60,-1.00);
  \node[theory label] at (6.60,0.55) {$\mathcal T_A$};
  \fill[endpoint] (7.60,0) circle[radius=1.60pt];
  \node[defect label,anchor=east] at (7.52,0)
    {$\mathcal O_s^{(h)}$};
  \node[wall label,anchor=north east] at (7.70,-1.10) {RG brane};
\end{tikzpicture}%
}
\caption{Folding and shrinking a transmitted TDL. Black dots denote chiral IR and anti-chiral UV endpoint fields before folding, and their combined chiral field $\mathcal O_s^{(h)}$ afterwards. The dashed line is the transmitted TDL, which becomes $\widehat{\mathcal L}_s$ in the folded theory $\mathcal T_A$. Shrinking its segment brings the endpoint field to the RG brane (magenta) as a boundary operator of the same dimension $h$.}
\label{fig:folded-phantom-current}
\end{figure}

To formulate this problem, let $s$ label a simple TDL preserved under an RG flow, with realizations $\mathcal L_s^{\rm UV}$ and $\mathcal L_s^{\rm IR}$ in the UV and IR bulk CFTs, denoted by $\mathcal T_{\rm UV}$ and $\mathcal T_{\rm IR}$, respectively. Upon folding the UV side, the matched pair becomes the diagonal line $\widehat{\mathcal L}_s=\mathcal L_s^{\rm IR}\boxtimes \overline{\mathcal L_s^{\rm UV}}$ in the product theory $\mathcal T_A=\mathcal T_{\rm IR}\boxtimes\overline{\mathcal T}_{\rm UV}$, where the overline denotes the orientation reversal. The intersection of the original line with the RG domain wall now becomes a topological junction at which $\widehat{\mathcal L}_s$ ends on the RG brane. Now consider a chiral IR endpoint field and an anti-chiral UV endpoint field on an open line crossing the interface. In the folded picture, these two endpoints combine into a single chiral endpoint field $\mathcal O_s^{(h)}$ of $\widehat{\mathcal L}_s$, with spin $h=h_{\rm IR}+h_{\rm UV}$, where $h_{\rm IR}$ and $h_{\rm UV}$ are the IR chiral and UV anti-chiral weights before the folding. This field is not an ordinary local bulk operator in $\mathcal T_A$. Instead, it carries an attached topological line, $\widehat{\mathcal L}_s$. Since the line can be shrunk topologically and the endpoint field is chiral, the non-local operator can then be brought to the RG brane without additional IR regularization. Shrinking the line segment to its boundary junction leaves a boundary operator of dimension $h$, as illustrated in Figure~\ref{fig:folded-phantom-current}. We will make extensive use of these operators with (half)-integer spin, dubbed ``phantom currents'', in the construction of the RG domain walls that we will review below~\cite{Antinucci:2025uvj,Furuta:2025ahl,Delmastro:2025ksn}.

Now let $\mathfrak A$ denote the chiral algebra of the theory $\mathcal T_A$. These endpoint fields can provide extra sectors to enlarge $\mathcal T_A$ to an auxiliary theory $\mathcal T_B$ with a chiral algebra $\mathfrak B\supset\mathfrak A$, so that the theory $\mathcal T_A$ is conformally embedded into $\mathcal T_B$. The chiral extension is realized by condensing a mutually local set of endpoint fields, which are simply the integer spin phantom currents, into the vacuum module of $\mathcal T_A$. Equivalently, the physical and auxiliary CFTs $\mathcal T_A$ and $\mathcal T_B$ are related by generalized orbifoldings of a Frobenius algebra $\mathfrak a$ defined by the set of TDLs associated with these phantom operators~\cite{Gaiotto:2020iye}. The extension preserves the total stress tensor of the UV and IR theories and incorporates the phantom sectors absent from the vacuum module of $\mathcal T_A$. These additional phantom operators can then mix with the UV and IR stress tensors and thereby enable the mixing at the RG domain wall. Therefore it is reasonable to seek a rational boundary state, in the auxiliary $\mathcal T_B$, to describe an RG brane for the physical $\mathcal T_A$. The resulting boundary state in $\mathcal T_B$ is then transported to the original folded theory through the topological interface associated with the condensation between $\mathcal T_A$ and $\mathcal T_B$. Indeed, this idea was pioneered in the seminal paper of Gaiotto, where he gave an algebraic description of the RG domain walls between consecutive Virasoro minimal models~\cite{Gaiotto:2012np}. The folded product algebra $\mathfrak A$ embeds conformally into an auxiliary coset algebra $\mathfrak B$ whose numerator contains two identical affine current algebra factors. Exchanging these factors defines an outer automorphism that mixes the embedded UV and IR stress tensors with a spin-two phantom operator. An automorphism-twisted Cardy boundary in $\mathcal T_B$ is transported back to $\mathcal T_A$, producing an RG wall candidate. It can reproduce UV--IR mixing coefficients non-perturbatively, in perfect agreement with conformal perturbation theory at large rank. This construction has also been generalized later to $N=1$ superconformal and $W$-algebra minimal models~\cite{Poghosyan:2014jia,Poghosyan:2022mfw,Poghosyan:2023brb}.

From a modern perspective, the phantom currents have many interesting applications to conformal interfaces independently of their uses in constructing RG walls. Spin-one phantom currents can generate exactly marginal boundary deformations, giving rise to a defect conformal manifold: a continuous family of interfaces between fixed bulk theories~\cite{Antinucci:2025uvj}. Spin-two phantom currents can mix with the UV and IR stress tensors as we have mentioned before. Consistency required by the generalized cluster decomposition principle provides a bootstrap approach to compute the energy transmission and reflection coefficients without knowing an explicit interface between two CFTs~\cite{Furuta:2025ahl}. These applications probe interfaces directly through low-spin operator data. Very recently, these two mechanisms combined together have been applied to calculate the energy transmission rates for interfaces between parafermion theories and minimal models in~\cite{Zhang:2026gqp}, where the transmission rate varies along a one-parameter family.

\medskip
\noindent\textbf{RG walls from automorphisms.}
In the construction of rational RG branes, these low-spin mechanisms are in fact part of a larger algebraic structure. Starting from the transmitted TDLs, we choose the condensable Frobenius algebra $\mathfrak a$ as before and define the conformal embedding from the physical theory $\mathcal T_A$ to its auxiliary theory $\mathcal T_B$, and correspondingly $\mathfrak A\subset \mathfrak B$. The central ansatz is as follows: Assume that the RG brane in $\mathcal T_A$ admits a rational realization in the auxiliary theory $\mathcal T_B$. The UV and IR operators then must mix with phantom currents in a consistent way, which is implemented by a single conformal automorphism $\Omega$ of $\mathfrak B$. That is to say, the mixing of phantom operators with quasi-primaries in the UV and IR vacuum modules must therefore be compatible with the action of the same automorphism at every conformal weight. This action must include not only the spin-one and spin-two fields, but also higher-spin phantom currents and vacuum quasi-primary fields, which transform covariantly under the global conformal group.

At each conformal weight, the quasi-primary fields in the extended vacuum form a representation of the conformal automorphism group of $\mathfrak B$, with $\Omega$ represented by a finite-dimensional mixing matrix. We determine these matrices by requiring that transforming fields before or after taking an operator product expansion (OPE) gives the same result. The resulting covariance equations will couple different weights. OPEs of lower-weight operators determine the action on the operators they generate, while additional generators introduce new matrix elements and constraints. This gives a bootstrap of the gluing automorphism from the local operator algebra. Once a suitable automorphism is identified, we construct the associated twisted Cardy states $|b;\,\Omega\rangle_B$ in the auxiliary theory. We call these states \emph{seed branes}, with $b$ labeling the elementary Cardy boundaries for the gluing $\Omega$. The topological interface $\check{\mathcal I}_{B\to A}$ determined by $\mathfrak a$, from $\mathcal T_B$ to $\mathcal T_A$, gives the boundary state $|R_{b,\Omega}\rangle_A$ through
\begin{align}
 |R_{b,\Omega}\rangle_A
 =\check{\mathcal I}_{B\to A}|b;\,\Omega\rangle_B \,.
 \label{eq:intro-boundary-transport}
\end{align}
Such a procedure will generically bring forth many conformal interfaces in $\mathcal T_A$ determined by the pairs of the specific brane labels and automorphisms $(b,\,\Omega)$. To further identify a boundary state $|R_{b,\Omega}\rangle_A$ with the right RG brane candidate, one may need to know its boundary entropy, partial UV--IR operator mixing results, or open-channel spectrum as extra input data.

Our examples exhibit two gluing mechanisms within this construction: $\mathsf a.$ Discrete outer automorphisms exchange identical current algebra factors and give Gaiotto-type RG branes $|R_{\rm out}\rangle$ in the folded product theory $\mathcal T_A$; $\mathsf b.$ Continuous inner automorphisms are generated by the zero modes of spin-one phantom currents: exponentiating these zero modes defines continuous families of seed branes twisted by inner automorphisms in $\mathcal T_B$, and, after topological transport back to $\mathcal T_A$, produces continuous families of conformal boundaries, $|R_{\rm in}(\vec\varphi)\rangle$. Here $\vec\varphi=(\varphi_1,\ldots,\varphi_d)$ denotes the $d$ independent real parameters of the defect conformal manifold. The undeformed RG brane selected by the half-space bulk flow lies at a distinguished point $\vec\varphi_*$, which we dub the \emph{RG angle}. Moving along the moduli space corresponds to exactly marginal deformations of this brane, with the two bulk CFTs fixed. The families will interpolate between the RG brane and factorized interfaces, where the two CFTs obey independent boundary conditions and no energy is transmitted.

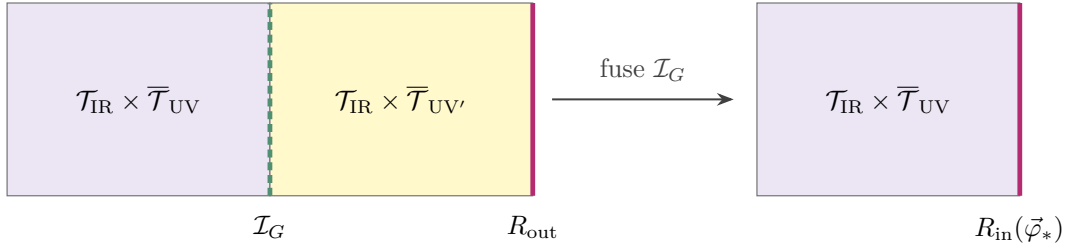
\begin{figure}[tbp]
\centering
\definecolor{PCOrbifoldFill}{RGB}{255,249,204}
\definecolor{PCFoldFill}{RGB}{235,229,244}
\definecolor{PCWall}{RGB}{181,43,114}
\definecolor{PCTopWall}{RGB}{81,143,114}
\begin{tikzpicture}[
 x=1.022296cm,y=1.213775cm,>=Stealth,
 panel edge/.style={draw=black!55,line width=0.45pt},
 rg wall/.style={PCWall,line width=1.8pt},
 top wall/.style={PCTopWall,line width=1.8pt,
   dash pattern=on 3pt off 1.8pt},
 process arrow/.style={->,black!75,line width=0.75pt},
 theory label/.style={font=\small},
 wall label/.style={font=\small,anchor=north,inner sep=2pt}
]
 \path[panel edge,fill=PCFoldFill] (0,0) rectangle (3.4,2.1);
 \path[panel edge,fill=PCOrbifoldFill] (3.4,0) rectangle (6.8,2.1);
 \draw[top wall] (3.4,0) -- (3.4,2.1);
 \draw[rg wall] (6.8,0) -- (6.8,2.1);
 \node[theory label] at (1.7,1.05)
  {$\mathcal T_{\rm IR}\times\overline{\mathcal T}_{\rm UV}$};
 \node[theory label] at (5.1,1.05)
  {$\mathcal T_{\rm IR}\times\overline{\mathcal T}_{\rm UV'}$};
 \node[wall label] at (3.4,-0.16) {$\mathcal I_G$};
 \node[wall label] at (6.8,-0.16) {$R_{\rm out}$};
 \draw[process arrow] (7.05,1.05) --
  node[above=3pt,font=\small] {fuse $\mathcal I_G$} (9.4,1.05);
 \path[panel edge,fill=PCFoldFill] (9.7,0) rectangle (13.1,2.1);
 \draw[rg wall] (13.1,0) -- (13.1,2.1);
 \node[theory label] at (11.4,1.05)
  {$\mathcal T_{\rm IR}\times\overline{\mathcal T}_{\rm UV}$};
 \node[wall label,anchor=north east] at (13.75,-0.16)
  {$R_{\rm in}(\vec\varphi_*)$};
\end{tikzpicture}
\caption{Transport between RG branes for orbifold related RG flows. The green dashed line is the topological interface $\mathcal I_G$ between the original and orbifold UV theories, with the common IR factor unchanged. Fusing it into the outer RG brane (magenta) removes the intervening orbifold strip and produces the inner RG brane at the distinguished moduli point $\vec\varphi_*$.}
\label{fig:G_I_G}
\end{figure}

\medskip
\noindent\textbf{Orbifold relations between RG walls.}
These two constructions are related when the underlying RG flows are connected by a finite orbifold. Let $G$ be a finite (invertible) symmetry group of the UV theory that can be gauged and preserves the relevant perturbations. Gauging $G$ gives an orbifold UV theory $\mathcal T_{\rm UV'}=\mathcal T_{\rm UV}/G$ and the corresponding perturbed theory. We consider pairs of flows with the same IR fixed point $\mathcal T_{\rm IR}$, whose RG branes admit inner and outer constructions, respectively. In the corresponding auxiliary theories $\mathcal T_B$ and $\mathcal T_{B'}$, the orbifold can project out the spin-one currents responsible for the continuous rotations in $\mathcal T_B$, while selected finite rotations survive as outer automorphisms of $\mathcal T_{B'}$ with a smaller chiral algebra $\mathfrak B'\subset\mathfrak B$. Let $\mathcal I_G$ denote the associated topological interface, defined from the orbifold UV theory to the original UV theory. Fusing it with the outer RG brane gives the undeformed inner RG brane at its RG angle $\vec\varphi_*$,
\begin{align}
 \mathcal I_G|R_{\rm out}\rangle
 =|R_{\rm in}(\vec\varphi_*)\rangle \,.
 \label{eq:RGorb_map}
\end{align}
Thus such an orbifold construction can help identify the RG point and its Cardy label within the continuous family once the RG brane on the other side is known. The brane transport is depicted in Figure~\ref{fig:G_I_G}, which relates the two RG branes at the inner brane's RG angle.

\medskip
\noindent\textbf{Examples and main results.}
We first revisit the outer automorphism construction of Gaiotto-type RG walls, originally proposed for successive unitary Virasoro minimal models~\cite{Gaiotto:2012np}. This construction later has been generalized for $\mathcal N=1$ superconformal minimal models~\cite{Poghosyan:2014jia}, $W_3$ minimal models~\cite{Poghosyan:2022mfw,Poghosyan:2023brb}, and diagonal $\widehat{\mathfrak{su}}(2)$ cosets at general levels~\cite{Stanishkov:2016pvi}. We add to these examples an infinite family of non-unitary Virasoro minimal model flows~\cite{Nakayama:2024msv}. For coprime positive integers $q$ and $I$ with $1<I\leq q-2$, the family is
\begin{align}
 \mathcal M_{q,\,q+I}^{\rm UV}\  
 \longrightarrow\  \mathcal M_{q-I,\,q}^{\rm IR}\,.
 \label{eq:intro-nonunitary-flow}
\end{align}
Their admissible fractional-level coset realizations combine into an auxiliary theory containing two identical level-one $\widehat{\mathfrak{su}}(2)$ factors. Exchanging these factors gives the outer automorphism and extends the Gaiotto construction to this non-unitary family.

Next, for RG branes with defect conformal manifolds that are defined by continuous inner automorphisms, our main family consists of massless flows from the $\mathbb Z_k$ parafermion CFT $\mathcal P_k$~\cite{Fateev:1985mm,Fateev:1987zh} to the unitary Virasoro minimal model $\mathcal M_{k+1,\,k+2}$ for $k\geq3$:
\begin{align}
 \mathcal P_k=
 \frac{\widehat{\mathfrak{su}}(2)_k}
 {\widehat{\mathfrak u}(1)_k}\ 
 \longrightarrow\ \mathcal M_{k+1,\,k+2}\,,\qquad k\geq3\,,
 \label{eq:intro-parafermion-flow}
\end{align}
These flows were inferred from integrability~\cite{Fateev:1990bf,Fateev:1991bv}. They remain
non-perturbative at large $k$, where the UV and IR central charges approach two and one, respectively. Condensing the matched phantom sectors gives the auxiliary algebra proposed in~\cite{Gaiotto:2012np, Zhang:2026gqp},
\begin{align}
 \mathfrak B_k=\frac{\widehat{\mathfrak{su}}(2)_{k-1}
 \times\widehat{\mathfrak{su}}(2)_1}
 {\widehat{\mathfrak u}(1)_k}\,,
 \label{eq:intro-parafermion-extension}
\end{align}
whose spin-one phantom current $Q$ generates the inner automorphisms $\Omega_\varphi=\operatorname{Ad}(e^{i\varphi Q_0})$. Here $Q_0$ is the current zero mode, normalized so that $\Omega_\varphi$ has period $2\pi$ on $\mathfrak B_k$, $\varphi$ is the angular modulus of the boundary state, and $\operatorname{Ad}$ denotes conjugation on fields. We explicitly construct the RG branes with defect modulus $\varphi$, determine their boundary entropies, and specify the distinguished point $\varphi_*=\pi$ for the RG angle, when $k$ is odd. For $k=3$, we further compute their UV--IR operator mixings as a function of $\varphi$ and extract the low-lying open-channel spectrum numerically from the self-annulus amplitude. The latter gives the scaling dimensions of boundary primaries living on the interface. At the RG angle $\varphi_*=\pi$, the mixing coefficients agree with the calculation based on a duality available only for $k=3$~\cite{Zhang:2026gqp}. In contrast, our present construction also applies at an arbitrary point $\varphi$ in the defect moduli space and for all odd $k$.

These parafermion flows also illustrate how the orbifold transport relates RG branes constructed from inner and outer automorphisms. For odd $k$, a charge conjugation symmetry preserves the parafermionic perturbation. Gauging this symmetry on the UV side leaves the IR theory unchanged and projects the spin-one phantom current out of the auxiliary algebra. The finite inner rotation $\Omega_\pi$ survives as an outer automorphism of the orbifold auxiliary algebra. At $k=3$, the orbifold flow is $\mathcal M_{5,\,6}\to\mathcal M_{4,\,5}$, whose outer RG brane is already included in Gaiotto's construction. Fusing the orbifold topological interface with this brane gives the inner RG brane at $\varphi_*=\pi$ and determines its Cardy label. More generally, when the RG brane of an orbifold related flow is known, this transport determines both the gluing automorphism and the Cardy label of its image, providing a way to locate the RG brane within a continuous family. We further illustrate this procedure with the proposed massless flow from $\widehat{\mathfrak{su}}(3)_2/\widehat{\mathfrak u}(1)^2$, known as a generalized parafermion theory, to the three-state Potts model, 
\begin{align}
    \frac{\widehat{\mathfrak{su}}(3)_2}
 {\widehat{\mathfrak u}(1)^2}\ 
 \longrightarrow\ \frac{\widehat{\mathfrak{su}}(2)_3}{\widehat{\mathfrak{u}}(1)_3}
\end{align}
studied through massless scattering and the thermodynamic Bethe ansatz~\cite{Ahn:2025xvb, Ahn:2022pia}. Its auxiliary extension
\begin{align}
    \mathfrak B=\frac{\widehat{\mathfrak{su}}(2)_1^3}{\widehat{\mathfrak{u}}(1)_3}
\end{align}
has two spin-one phantom currents and thus a two-parameter family of inner rotations. A $\mathbb Z_3$-orbifold of the UV theory gives the $W_3$ minimal model $W_3(5,6)$, relating the flow to $W_3(5,6)\to W_3(4,5)$, where $W_3(4,5)$ is an equivalent description of the previous Potts model. The auxiliary algebra for this orbifold flow is 
\begin{align}
\mathfrak B'=
 \frac{\widehat{\mathfrak{su}}(3)_1
 \times\widehat{\mathfrak{su}}(3)_1
 \times\widehat{\mathfrak{su}}(3)_1}
 {\widehat{\mathfrak{su}}(3)_3}\,.
 \label{eq:intro-generalized-extension}
\end{align}
Exchanging two numerator factors defines the seed used in the $W_3$ RG-wall construction~\cite{Poghosyan:2022mfw,Poghosyan:2023brb}. Transporting this seed back identifies a boundary in the original generalized parafermion and Potts product. Its gluing combines a particular inner rotation with a reflection of the two spin-one currents. This example shows how a flow with a larger defect conformal manifold can be studied through a related discrete construction.


The remainder of this paper is organized as follows. Section~\ref{sec:construction} presents the general construction, from phantom current condensation and the automorphism bootstrap to twisted Cardy states and their transport to the folded UV--IR theory. Section~\ref{sec:outer-walls} studies RG walls defined by outer automorphisms, including Gaiotto-type constructions for unitary and non-unitary minimal models. Section~\ref{sec:inner-walls} develops the continuous inner-automorphism families for RG branes between parafermion and minimal models. Section~\ref{sec:orbifold-transport} studies transport between orbifold-related flows, showing how it relates inner and outer constructions and identifies RG branes within continuous families. We apply this procedure to parafermion flows at odd $k$ and to a generalized parafermion example. Section~\ref{sec:discussion} summarizes our results and discusses further applications and extensions.

\section{RG domain walls from phantom current condensation}
\label{sec:construction}


\subsection{Transmitted phantom currents and condensation}
\label{sec:phantom-condensation}

Starting with a UV RCFT $\mathcal T_{\rm UV}$, we deform the theory by a relevant primary operator $\phi_i$,
\begin{align}
 S=S_{\rm UV}+\lambda\int_{x<0}\!\!\!\!{\rm d}^2x\,\,\phi_{i}(x,t)\,.
 \label{eq:construction-halfspace}
\end{align}
A topological defect line $\mathcal L_s^{\rm UV}$ survives the
perturbation if it can be moved through insertions of $\phi_{i}$ without changing correlation functions, where $s$ labels all such simple lines. The preserved TDLs will furnish a fusion category, denoted by $\mathscr C_{\rm pre.}$. For Verlinde lines in $\mathcal T_{\rm UV}$, which we assume to be a diagonal RCFT, the preservation condition can be cast in terms of its modular matrix $S^{\rm UV}$~\cite{Chang:2018iay}:
\begin{align}
 \frac{S^{\rm UV}_{si}}{S^{\rm UV}_{\textbf{v}i}}
 =\frac{S^{\rm UV}_{s\textbf{v}}}{S^{\rm UV}_{\textbf{v}\textbf{v}}}
 \equiv D_s\,.
 \label{eq:preserved-line-test}
\end{align}
Here $i$ labels the deformation primary operator $\phi_i$, $\textbf{v}$ is the vacuum label, and $D_s$ is the quantum dimension of the line $\mathcal L_s^{\rm UV}$. We match the surviving lines to their IR realizations, $\mathcal L_s^{\rm UV}\leftrightarrow\mathcal L_s^{\rm IR}$. Thus the matched lines have the same fusion rules and multiplicities, given by
\begin{align}
 \mathcal L_s^{\rm UV}\otimes\mathcal L_t^{\rm UV}
 \simeq\bigoplus_uN_{st}^{u}\,\mathcal L_u^{\rm UV}\,,
 \qquad
 \mathcal L_s^{\rm IR}\otimes\mathcal L_t^{\rm IR}
 \simeq\bigoplus_uN_{st}^{u}\,\mathcal L_u^{\rm IR}\,.
 \label{eq:transmitted-fusion}
\end{align}
The indices run over matched simple lines, and $N_{st}^{u}$ counts independent junctions and multiplicities. Their associativity data, e.g. $F$-symbols, are also part of the matching.

An endpoint field is defined together with a topological line
terminating on it. We consider a UV anti-chiral endpoint operator
${\psi_s^{\rm UV}}$ with weights $(0, h_s^{\rm UV})$ and an IR
chiral endpoint ${\psi_s^{\rm IR}}$ with weights $(h_s^{\rm IR},0)$, connected by the transmitted line. We now fold the UV side and obtain the UV--IR product theory $\mathcal T_A$ with its chiral algebra $\mathfrak A$ given by
\begin{align}
 \mathcal T_A=\mathcal T_{\rm IR}\boxtimes
 \overline{\mathcal T}_{\rm UV}\,,
 \qquad \mathfrak A=\mathfrak A_{\rm IR}\boxtimes\mathfrak A_{\rm UV}\,.
 \label{eq:folded-product-algebra}
\end{align}
Here $\mathfrak A_{\rm IR,UV}$ are the two chiral algebras of the UV and IR theories, and the bar on the UV factor denotes orientation reversal. The UV anti-chiral endpoint consequently becomes chiral after folding. As shown in Figure~\ref{fig:folded-phantom-current}, we write
\begin{align}
 \begin{split}
 X_s=\psi_s^{\rm IR}\otimes\overline{\psi_s^{\rm UV}}\,,
 \qquad h_s=h_s^{\rm IR}+h_s^{\rm UV}\,,\qquad
 \widehat{\mathcal L}_s
 =\mathcal L_s^{\rm IR}\boxtimes
   \overline{\mathcal L_s^{\rm UV}}\,.
 \end{split}
 \label{eq:folded-phantom-sector}
\end{align}
Here the operator $X_s$ becomes the endpoint chiral operator of the defect line $\widehat{\mathcal L}_s$, with a folded weight $h_s$. Obviously, in $\mathcal T_A$, it is a quasi-local operator and does not belong to the vacuum algebra $\mathfrak A$. When $X_s$ has (half-)integer spin, these endpoint fields are the phantom currents of Refs.~\cite{Antinucci:2025uvj,Furuta:2025ahl,Delmastro:2025ksn}.

For the endpoint sectors, let $\mathcal V_s^{\rm IR}$ and $\overline{\mathcal V_s^{\rm UV}}$ be the irreducible modules associated with the two chosen endpoint fields. Their tensor product is then the $\mathfrak A$-module containing the folded phantom current $X_s$, i.e.
\begin{align}
 \mathcal X_s=\mathcal V_s^{\rm IR}\otimes\overline{\mathcal V_s^{\rm UV}}\,,
 \qquad X_s\in\mathcal X_s\,.
 \label{eq:phantom-endpoint-module}
\end{align}
It is worth noting that the endpoints $X_s\in \mathcal X_s$ may have multiplicities, as will be seen in Section~\ref{sec:orbifold-generalized}. We therefore use $\alpha$ below to label the distinct irreducible endpoint modules. When there is one such module for each matched line, $\alpha$ is simply $s$. We now select a compatible collection of these integer-weight modules, including the vacuum module $\mathcal X_0=\mathfrak A$ itself. Their sum, with $n_\alpha$ copies of each selected module, is
\begin{align}
 \mathfrak a=\bigoplus_\alpha n_\alpha\mathcal X_\alpha\,,
 \qquad n_\alpha\in\mathbb Z_{\geq0}\,,\qquad n_0=1\,.
 \label{eq:condensable-phantom-algebra}
\end{align}
The collection $\mathfrak a$ has to be chosen so that the OPEs of its fields are associative and mutually local. Condensing the fields in the other modules of $\mathfrak a$ into the vacuum sector $\mathcal X_0=\mathfrak A$ will define an enlarged chiral algebra denoted by $\mathfrak B$, corresponding to our auxiliary theory $\mathcal T_B$ mentioned in Section~\ref{sec:introduction}. The extension is conformal and satisfies
\begin{align}
 \mathfrak A\subset \mathfrak B\,,\qquad \mathfrak B\big|_\mathfrak A\simeq\mathfrak a\,,
 \qquad T_B=T_{\rm IR}+T_{\rm UV}\,,\qquad
 c_B=c_A=c_{\rm IR}+c_{\rm UV}\,.
 \label{eq:auxiliary-extension}
\end{align}
Here $\mathfrak B\big|_\mathfrak A$ is the vacuum module of $\mathcal T_B$ viewed as an $\mathfrak A$-module. Equivalently, these data define a condensable Frobenius algebra $\mathcal L_\mathfrak a$ consisting of TDLs with endpoints $X_\alpha$. The condensation of $\mathfrak a$ corresponds to the generalized orbifolding or gauging of these non-invertible symmetry lines\,\footnote{Throughout the paper, we will not distinguish notationally between the condensation algebra $\mathfrak a$ and its associated Frobenius algebra $\mathcal L_\mathfrak a$. The intended meaning should not be confused in contexts.}~\cite{Frohlich:2006ch,Huang:2014ixa,Diatlyk:2023fwf}.

The auxiliary theory $\mathcal T_B$ is still rational and has a single identity vacuum module. The extension enlarges the vacuum sector but preserves the total stress tensor and central charge as it should. Equivalently, we obtain $\mathfrak B$ by taking the full local OPE closure of $\mathfrak A$ and the selected phantom sectors $\mathcal X_{\alpha\neq 0}$. Higher weight operators produced by these OPEs must therefore enter the extended vacuum as well, where the allowed junctions determine which fusion channels occur. This operator description makes it possible to study the boundary mixing of all the condensed fields within a single chiral algebra.

\subsection{Automorphisms and OPE covariance equations}
\label{sec:all-weight-mixing}
\label{sec:graded-bootstrap}

As we have discussed, the UV and IR operators will in principle have non-trivial mixings at the RG domain wall. On the other hand, when the RG wall admits transmitted TDLs, their endpoint operators can be also absorbed into the boundary. Therefore, our central ansatz is that a single conformal automorphism\,\footnote{In the paper, we only consider the unitary automorphism group.} $\Omega$ of $\mathfrak B$ implements the boundary mixing of fields in the original vacuum and condensed phantom sectors. In the folded picture, we schematically express this mixing for a UV--IR product operator $\mathcal O^{\rm IR} \overline{\mathcal O^{\rm UV}}\in \mathfrak A$ as
\begin{align}
    \mathcal O^{\rm IR} \overline{\mathcal O^{\rm UV}}\quad\xrightarrow[]{\text{after mixing on the wall}}\quad \Omega\left(\mathcal O^{\rm IR} \overline{\mathcal O^{\rm UV}}\right)\,,
\end{align}
where both factors are understood as chiral fields after folding the UV side. A straightforward example is the mixing of the UV and IR stress tensors in the presence of spin-two phantom currents~\cite{Furuta:2025ahl}. In $\mathcal T_B$, we rewrite the two chiral stress tensors as
\begin{align}
    t_{\rm IR}=T_{\rm IR}\otimes\overline{\mathbf 1_{\rm UV}}\,,\quad 
    t_{\rm UV}=\mathbf 1_{\rm IR}\otimes\overline{T_{\rm UV}}
\end{align}
with OPE
\begin{align}
 t_i(z)t_j(0)\sim\delta_{ij}
 \left(\frac{c_i}{2z^4}+\frac{2\,t_i(0)}{z^2}
 +\frac{\partial t_i(0)}{z}\right)\,,
 \qquad i,j\in\{\mathrm{IR},\mathrm{UV}\}\,,
 \label{eq:factor-stress-normalization}
\end{align}
where $c_i$ is the central charge of factor $i$. Although $\Omega$ fixes
their sum, it can mix the individual stress tensors with each other and
with the phantom fields. Given $\Omega$, we can compute energy reflection and transmission, directly in the auxiliary theory, in terms of the correlator $K_{ij}$ 
\begin{align}
 \big\langle t_i(z)\,\Omega(t_j)(0)\big\rangle_{\mathfrak B}
 =\frac{K_{ij}(\Omega)}{z^4}\,,\qquad{\rm or}\quad
 K_{ij}(\Omega)
 =\langle0|L_2^{(i)}\,\Omega\!\left(L_{-2}^{(j)}\right)|0\rangle\,,
 \label{eq:stress-overlap-definition}
\end{align}
where $L_n^{(i)}$ are the modes of stress tensor $t_i$. The diagonal entries in $K_{ij}(\Omega)$ describe reflection and the off diagonal entries describe transmission, with the conventional normalization~\cite{Quella:2006de},
\begin{align}
 \mathcal R=\frac{2(K_{\rm IR,IR}+K_{\rm UV,UV})}
 {c_{\rm IR}+c_{\rm UV}}\,,\qquad
 \mathcal T=\frac{2(K_{\rm IR,UV}+K_{\rm UV,IR})}
 {c_{\rm IR}+c_{\rm UV}}\,.
 \label{eq:reflection-transmission}
\end{align}
Since the total stress tensor is invariant under the action of $\Omega$, these coefficients also satisfy
\begin{align}
 \sum_jK_{ij}=\frac{c_i}{2}\,,\qquad
 \sum_iK_{ij}=\frac{c_j}{2}\,,\qquad
 \mathcal R+\mathcal T=1\,.
\end{align}
In particular, if both of the two factor stress tensors are fixed, the interface is then completely reflective. In contrast, an automorphism that mixes the two stress tensors nontrivially can allow energy transmission.

Now consider the quasi-primary space of each conformal weight $h$,
\begin{align}
 \mathcal Q_h(\mathfrak B)=\{v\in \mathfrak B:L_0v=hv,\ L_1v=0\}\,,
 \qquad d_h=\dim\mathcal Q_h(\mathfrak B)\,.
 \label{eq:quasi-primary-space}
\end{align}
The automorphism $\Omega\in\operatorname{Aut}(\mathfrak B)$ acting on $\mathcal Q_h(\mathfrak B)$ will furnish a matrix representation of $\Omega$ itself. In a basis
$\{\mathcal O_i^{(h)}\}_{i=1}^{d_h}$, we can write it as
\begin{align}
 \Omega\bigl(\mathcal O_i^{(h)}\bigr)
 =\sum_{j=1}^{d_h}(M_h)_{ij}\mathcal O_j^{(h)}\,.
 \label{eq:automorphism-matrix-convention}
\end{align}

We can determine the allowed matrices by requiring that the action of $\Omega$ be covariant with respect to the OPEs of these quasi-primaries, leading to the \emph{OPE covariance equations}:
\begin{align}
 \Omega(u_{(n)}v)=(\Omega u)_{(n)}(\Omega v)\,,
 \qquad u(z)v(0)=\sum_{n\in\mathbb Z}z^{-n-1}(u_{(n)}v)(0)\,,
 \quad {\rm for}\quad u,\,v\in\mathcal Q(\mathfrak B)\,.    
 \label{eq:full-mode-covariance}
\end{align}
Here $\mathcal Q(\mathfrak B)$ denotes the sum of the quasi-primary spaces over all conformal weights, and $u_{(n)}v$ is the coefficient of $z^{-n-1}$ in the OPE of $u$ and $v$. Let $C_{ij}^{\ \ k}(h_1,h_2;h_3)$ be the coefficients of $\mathcal O_k^{(h_3)}$ in the OPE of $\mathcal O_i^{(h_1)}$ and $\mathcal O_j^{(h_2)}$. After suppressing the weight arguments, the OPE covariance equations become
\begin{align}
 \sum_k C_{ij}^{\ \ k}(M_{h_3})_{k\ell}
 =\sum_{a,b}(M_{h_1})_{ia}(M_{h_2})_{jb}C_{ab}^{\ \ \ell}\,.
 \label{eq:graded-OPE-covariance}
\end{align}
Here $i,a=1,\ldots,d_{h_1}$, $j,b=1,\ldots,d_{h_2}$ and $k,\ell=1,\ldots,d_{h_3}$. Each triple of weights gives a finite-dimensional polynomial system. In particular, the most singular identity channel preserves the two-point matrix
$G_h$:
\begin{align}
 \langle\mathcal O_i^{(h)}(z)\mathcal O_j^{(h)}(0)\rangle
 =\frac{(G_h)_{ij}}{z^{2h}}\,,
 \qquad M_hG_hM_h^{\mathsf T}=G_h\,.
 \label{eq:gram-covariance}
\end{align}

One can solve Eq.~\eqref{eq:graded-OPE-covariance} for $M_h$ in increasing weight, once the OPE coefficients in $\mathfrak B$ are known. It can thus determine all allowable automorphisms acting on the entire vacuum algebra $\mathfrak B$. For a continuous family of automorphisms, we may linearize the equations at the identity, $M_h(t)=\mathbf1_{d_h}+tD_h+O(t^2)$, with real parameter $t$ and infinitesimal matrices $D_h$. The equations reduce to
\begin{align}
 \begin{split}
 \sum_k C_{ij}^{\ \ k}(D_{h_3})_{k\ell}
 &=\sum_a(D_{h_1})_{ia}C_{aj}^{\ \ \ell}
  +\sum_b(D_{h_2})_{jb}C_{ib}^{\ \ \ell}\,,\\
 D_hG_h+G_hD_h^{\mathsf T}&=0\,.
 \end{split}
 \label{eq:linearized-ope-covariance}
\end{align}
This will determine the infinitesimal free parameters, with respect to which the matrices $D_h$ are the generators of the continuous parts of the automorphism group $\operatorname{Aut}(\mathfrak B)$. Zero modes of spin-one phantom currents can give such infinitesimal actions and exponentiate to inner automorphisms. On the other hand, a set of finite discrete solutions to \eqref{eq:graded-OPE-covariance} captures other disconnected components of the whole $\operatorname{Aut}(\mathfrak B)$. They are outer automorphisms and may permute inequivalent modules in $\mathcal T_B$. When $\operatorname{Aut}(\mathfrak B)$ is not known, we will use these OPE covariance equations to determine its representation matrices in the UV--IR and phantom operator basis level by level.

\subsection{Twisted seed branes}
\label{sec:twisted-seeds}

For a given automorphism $\Omega$, we now use it to define $\Omega$-twisted boundary states in $\mathcal T_B$. Here it is worth noting that an $\Omega$-twisted boundary state should not be confused with boundary states defined in the $\Omega$-twisted defect Hilbert space. Rather, it is defined by $\Omega$-twisted Ishibashi states and still belongs to the ordinary Hilbert space of $\mathcal T_B$. Let $\mathscr P_B$ be the set of ordinary irreducible $\mathfrak B$-module labels, with $\mathcal V_\mathbf v=\mathfrak B$ the vacuum module. The bulk Hilbert space thus decomposes as
\begin{align}
 \mathcal H_B=\bigoplus_{\mu\in\mathscr P_B}
 \mathcal V_\mu\otimes\overline {\mathcal V}_\mu\,,
 \label{eq:auxiliary-bulk-spectrum}
\end{align}
for the diagonal RCFT $\mathcal T_B$. The character of the module $\mathcal V_\mu$ is 
\begin{align}
\chi_\mu(\tau)=\operatorname{Tr}_{\mathcal V_\mu}q^{L_0-c_B/24}\,,
\end{align}
where $\tau$ is the modular parameter and $q=e^{2\pi i\tau}$. The ordinary modular matrix is then defined by
\begin{align}
    \chi_\mu(-1/\tau)=\sum_\nu S^B_{\mu\nu}\chi_\nu(\tau)\,.
\end{align}

Now we review the construction of the $\Omega$-twisted Ishibashi states~\cite{Ishibashi:1988kg,Ishikawa:2005ea}. Consider a quasi-primary chiral current $W$ of conformal weight $h$ in $\mathfrak B$. Its mode expansion is given by $W(z)=\sum_nW_nz^{-n-h}$. An $\Omega$-twisted Ishibashi state $|\mu;\Omega\rangle\!\rangle$ is defined by
\begin{align}
 \left(\Omega(W)_n-(-1)^h\overline W_{-n}\right)
 |\mu;\Omega\rangle\!\rangle=0\,.
 \label{eq:twisted-ishibashi-gluing}
\end{align}
For an inner automorphism $\Omega_\varphi=\operatorname{Ad}(U_\varphi)$ generated by the zero mode of a spin-one current $Q$, $U_\varphi=e^{i\varphi Q_0}$, we can obtain the state directly from the ordinary Ishibashi state,
\begin{align}
 |\mu;\Omega_\varphi\rangle\!\rangle
 =(U_\varphi\otimes\overline{\mathbf1})
 |\mu\rangle\!\rangle\,.
 \label{eq:inner-ishibashi-rotation}
\end{align}
Indeed, commuting $U_\varphi$ through the left modes reduces Eq.~\eqref{eq:twisted-ishibashi-gluing} to the ordinary Ishibashi condition\footnote{One may instead place the automorphism twist on the right-moving fields. Our convention is equivalent to $\left(W_n-(-1)^h\overline{\Omega^{-1}(W)}_{-n}\right) |\mu;\Omega\rangle\!\rangle=0$, by inverting $\Omega\to \Omega^{-1}$} $(W_n-(-1)^h\overline{ W}_{-n})|\mu\rangle\!\rangle=0$.

The inner automorphism acts within ordinary modules themselves. So, for the pairs of chiral and antichiral modules in~\eqref{eq:auxiliary-bulk-spectrum}, the allowed Ishibashi labels precisely run over the full set $\mathscr P_B$, i.e.
\begin{align}
 \mathscr P_B^\Omega=
 \{\mu\in\mathscr P_B:\Omega(\mathcal V_\mu)\simeq \mathcal V_\mu\}=\mathscr P_B\,.
 \label{eq:inner-ishibashi-support}
\end{align}
Their overlap is, by definition, normalized to the ordinary character,
\begin{align}
 \langle\!\langle\mu;\Omega|
 \widetilde q^{\frac12(L_0+\bar L_0-c_B/12)}
 |\nu;\Omega\rangle\!\rangle
 =\delta_{\mu\nu}\chi_\mu(-1/\tau)\,,
 \qquad \widetilde q=e^{-2\pi i/\tau}\,.
 \label{eq:ishibashi-character-normalization}
\end{align}
A Cardy state twisted by an inner automorphism is then given by
\begin{align}
 |b;\Omega_\varphi\rangle_B
 =\sum_{\mu\in\mathscr P_B}
 \frac{S^B_{b\mu}}{\sqrt{S^B_{\textbf{v}\mu}}}
 |\mu;\Omega_\varphi\rangle\!\rangle\,,
 \qquad b,\,\mu\in\mathscr P_B\,.
 \label{eq:inner-twisted-cardy}
\end{align}

On the other hand, for a finite order outer automorphism, the Cardy state coefficients are instead supplied by twisted modular data \cite{Ishikawa:2005ea}. Since now an outer automorphism $\Omega$ may permute different modules of $\mathfrak B$, only those that are fixed points under $\Omega$ can support a twisted Ishibashi state. For each fixed ordinary module $\mathcal V_\mu$, we collect their labels in the set
\begin{align}
 \mathscr P_B^\Omega=
 \{\mu\in\mathscr P_B:\Omega(\mathcal V_\mu)\simeq \mathcal V_\mu\}\,.
 \label{eq:fixed-ishibashi-support}
\end{align}
Let $U_\mu$ implement the action of $\Omega$ on its states in $\mathcal V_\mu$. We can define the $U_\mu$-intertwined character,
\begin{align}
 U_\mu W_nU_\mu^{-1}=\Omega(W)_n\,,
 \qquad
 \chi_\mu^\Omega(\tau)=
 \operatorname{Tr}_{\mathcal V_\mu}
 \left(U_\mu q^{L_0-c_B/24}\right)\,.
 \label{eq:twining-character}
\end{align}
The trace is a character with the symmetry operator inserted along the time direction, or the corresponding TDL along the spatial direction. A modular transformation gives characters of $\Omega$-twisted modules, in which fields return transformed by $\Omega$ after going around the spatial circle. Let $a$ label these modules and $\chi_a^{\rm tw}$ be their characters. With compatible phase choices, their modular transformation is defined by the twisted modular matrix $\Psi^\Omega$ as follows:
\begin{align}
 \chi_a^{\rm tw}(\tau)=
 \sum_{\mu\in\mathscr P_B^\Omega}
 \Psi^\Omega_{a\mu}\,\chi_\mu^\Omega(-1/\tau)\,.
 \label{eq:twisted-modular-transformation}
\end{align}
Accordingly, the outer $\Omega$-twisted Cardy states are
\begin{align}
 |b;\Omega\rangle_B
 =\sum_{\mu\in\mathscr P_B^\Omega}
 \frac{\Psi^\Omega_{b\mu}}{\sqrt{S^B_{\textbf{v}\mu}}}
 |\mu;\Omega\rangle\!\rangle\,,
 \label{eq:outer-twisted-cardy}
\end{align}
where the closed-channel labels $\mu$ are ordinary modules fixed by $\Omega$, whereas the boundary labels $b$ come from the twisted theory. The twisted characters appear in the open-channel between an ordinary and an $\Omega$-twisted state.

For both inner and outer $\Omega$, we can use them to construct $\Omega$-twisted Cardy branes, $|b;\Omega\rangle_B$, in $\mathcal T_B$. The Cardy condition requires nonnegative integer multiplicities in the ordinary open-channel~\cite{Cardy:1989ir}:
\begin{align}
 Z_{ab;\Omega}^{B,\,\text{open}}(q)=
 {}_B\langle a;\Omega|
 \widetilde q^{\frac12(L_0+\bar L_0-c_B/12)}
 |b;\Omega\rangle_B
 =\sum_{\nu\in\mathscr P_B}n_{ab}^{\nu}\,\chi_\nu(\tau)\,,
 \qquad n_{ab}^{\nu}\in\mathbb Z_{\geq0}\,.
 \label{eq:seed-cardy-condition}
\end{align}
We call $|b;\Omega\rangle_B$ a \emph{seed brane}. Throughout the paper, we assume that the RG brane is constructed from a simple seed, with one boundary identity, $n_{bb}^{\textbf{v}}=1$. The brane label $b$ has to be also specified in addition to $\Omega$.

\subsection{Brane transport from \texorpdfstring{$\mathcal T_B$}{TB} to \texorpdfstring{$\mathcal T_A$}{TA}}
\label{sec:condensation-transport}
Once the seed brane is determined, we pull it back to the physical product theory $\mathcal T_A$ through the topological interface
\begin{align}
    \check{\mathcal I}_{B\to A}:\quad \mathcal T_B\ \xrightarrow[]{\qquad}\ \mathcal T_A\,,
\end{align}
associated with the conformal extension $\mathfrak A\subset \mathfrak B$. To see how this operator acts on the $\Omega$-twisted Cardy branes in $\mathcal T_B$, let $\mathcal V_i^A$ denote the irreducible modules of $\mathfrak A$ with labels $i\in\mathscr P_A$, and write $\mathcal V_\mu^B\equiv \mathcal V_\mu$ for the auxiliary modules of $\mathfrak B$ introduced in~\eqref{eq:auxiliary-bulk-spectrum}. For simplicity, we consider the embedding branchings of $\mathfrak A\subset \mathfrak B$ without multiplicities,
\begin{align}
 \mathcal V_\mu^B\Big|_\mathfrak A=\bigoplus_{i\in\operatorname{Br}(\mu)}\mathcal V_i^A\,,
 \label{eq:transport-branching}
\end{align}
where $\operatorname{Br}(\mu)\subset\mathscr P_A$ is the set of $\mathfrak A$-representations contained in module $\mathcal V_\mu^B$. For each branch, we choose an inclusion $e_{\mu i}$, together with its dual $\check e_{i\mu}$, to extract that branch,
\begin{align}
 &e_{\mu i}:\mathcal V_i^A\longrightarrow \mathcal V_\mu^B\,,\qquad
 \check e_{i\mu}:\mathcal V_\mu^B\longrightarrow \mathcal V_i^A\,,\notag\\[5pt]
\text{satisfying}\quad &\check e_{i\mu}e_{\mu j}=\delta_{ij}\mathbf1_{\mathcal V_i^A}\,,\qquad
 P_{\mu i}=e_{\mu i}\check e_{i\mu}\,,\qquad
 \sum_{i\in\operatorname{Br}(\mu)}P_{\mu i}=\mathbf1_{\mathcal V_\mu^B}\,,
 \label{eq:transport-chiral-maps}
\end{align}
where $P_{\mu i}$ is a chiral projector that projects chiral states onto the chosen $\mathfrak A$-submodule $\mathcal V_i^{A}$. The maps can be normalized using the two-point correlators, so that $\check e_{i\mu}=e_{\mu i}^{\dagger}$. When a branch has multiplicity, our inclusion maps in~\eqref{eq:transport-chiral-maps} must also keep its copy index to specify the embeddings.

On the other hand, the full bulk sector $\mathcal V_\mu^B\otimes\overline{\mathcal V _\mu^B}$ contains all different pairs of branches $\mathcal V_i^A\otimes\overline {\mathcal V_j^A}$. To return to the diagonal bulk theory $\mathcal T_A$, we must retain only the pairs with indices $i=j$. The corresponding projectors are thus defined as
\begin{align}
 \Pi_{\mu,ij}=P_{\mu i}\otimes\overline{P_{\mu j}}\,,\qquad
 i,j\in\operatorname{Br}(\mu)\,,
 \label{eq:transport-bulk-projectors}
\end{align}
where the bars denote the anti-chiral pieces in the folded theory. With the canonical interface normalization discussed in~\cite{Gaiotto:2012np}, the pullback, restricted to the module $\mathcal V_\mu^B\otimes\overline{\mathcal V _\mu^B}$, is given by
\begin{align}
 \left.\check{\mathcal I}_{B\to A}\right|_{\mathcal V_\mu^B\otimes\overline{\mathcal V _\mu^B}}
 =\sum_{i\in\operatorname{Br}(\mu)}\rho_{\mu i}
 \left(\check e_{i\mu}\otimes\overline{\check e_{i\mu}}\right)
 \Pi_{\mu,ii}\,,\qquad {\rm with}\quad
 \rho_{\mu i}&=\sqrt{\frac{S^B_{\mathbf v\mu}}{S^A_{\mathbf vi}}}\,,
 \label{eq:transport-interface-projector}
\end{align}
where $\mathbf v$ denotes the vacuum label as before in the indicated theory and $S^A$, $S^B$ are the corresponding modular matrices. In principle, the interface simply first projects a state in $\mathcal T_B$ onto an allowed diagonal pair and then uses the maps $\check e_{i\mu}\otimes\overline{\check e_{i\mu}}$ to identify it with a state of $\mathcal T_A$. For descendant states, for example, labeled by $\alpha$ and $\beta$, this gives
\begin{align}
 \check{\mathcal I}_{B\to A}
 \left(e_{\mu i}|i,\alpha\rangle\otimes
       \overline{e_{\mu j}|j,\beta\rangle}\right)
 =\delta_{ij}\rho_{\mu i}
 |i,\alpha;\overline{i,\beta}\rangle_A\,.
 \label{eq:transport-state-action}
\end{align}
It implies that the projection only matches the representation labels, but does not require a match of descendant states $\alpha=\beta$.

Reversing the interface $\check{\mathcal I}_{B\to A}$ gives its dual $\mathcal I_{A\to B}$:
\begin{align}
    {\mathcal I}_{A\to B}:\quad \mathcal T_A\ \xrightarrow[]{\qquad}\ \mathcal T_B\,,
\end{align}
which simply maps a physical product state in $\mathcal T_A$ to
\begin{align}
 \mathcal I_{A\to B}|i,\alpha;\overline{i,\beta}\rangle_A
 =\sum_{\mu:\,i\in\operatorname{Br}(\mu)}\rho_{\mu i}
 e_{\mu i}|i,\alpha\rangle\otimes
 \overline{e_{\mu i}|i,\beta\rangle}\,.
 \label{eq:transport-forward-interface}
\end{align}
The two topological interfaces themselves are not invertible. The fusion of them gives
\begin{align}
 \check{\mathcal I}_{B\to A}\circ\mathcal I_{A\to B}
 &\simeq\mathcal L_{\mathfrak a}
 \qquad\text{in }\mathcal T_A\,,\label{eq:interface-fusion-A}\\
 \mathcal I_{A\to B}\circ\check{\mathcal I}_{B\to A}
 &\simeq\mathcal L_{\mathfrak b}
 \qquad\text{in }\mathcal T_B\,,\label{eq:interface-fusion-B}
\end{align}
where $\mathcal L_\mathfrak a$ is the condensable Frobenius algebra defined in Section~\ref{sec:phantom-condensation}, i.e.
\begin{align}
    \mathcal L_\mathfrak a=\bigoplus_s\widehat{\mathcal L}_s\,.
\end{align}
On the other hand, $\mathcal L_\mathfrak b$ is the quantum Frobenius algebra in $\mathcal T_B$, whose condensation will recover $\mathcal T_A$ from $\mathcal T_B$. It consists of emergent quantum non-invertible TDLs obtained by gauging $\mathcal L_\mathfrak a$, which are neither Verlinde lines nor associated with $\mathfrak B$-modules. In particular, in the description by $\mathfrak A$-defects, the quantum defect $\mathcal L_{\mathfrak b}$ is represented by a bimodule over
$\mathcal L_{\mathfrak a}$ as follows~\cite{Frohlich:2006ch,Brunner:2014lua,Chen:2024ulc,Chen:2026cqz}:
\begin{align}
    \mathcal L_{\mathfrak b}
    \longleftrightarrow
    \mathcal L_{\mathfrak a}\otimes\mathcal L_{\mathfrak a}\,.
\end{align}

Equipped with the topological interface $\check{\mathcal I}_{B\to A}$, 
we can obtain the RG brane in $\mathcal T_A$ by fusing it to the seed brane from $\mathcal T_B$:
\begin{align}
 |R_{b,\Omega}\rangle_A
 =\check{\mathcal I}_{B\to A}|b;\Omega\rangle_B\,.
 \label{eq:seed-brane-transport}
\end{align}
Due to the topological nature of $\check{\mathcal I}_{B\to A}$, its image is still conformal. Unfolding it thus gives a conformal interface between the UV and IR theories. We can now apply the projector~\eqref{eq:transport-chiral-maps} directly to the Ishibashi states in the seed brane. Then
\begin{align}
 \check{\mathcal I}_{B\to A}|\mu;\Omega\rangle\!\rangle_B
 =\sum_{i\in\operatorname{Br}(\mu)}\rho_{\mu i}
 \left(\check e_{i\mu}U_\mu e_{\mu i}
       \otimes\overline{\mathbf1}\right)|i\rangle\!\rangle_A\,.
 \label{eq:transport-twisted-ishibashi}
\end{align}
The operator $\check e_{i\mu}U_\mu e_{\mu i}$ sends a physical chiral state into the auxiliary module, rotates it by $U_\mu$, and extracts its component back in the original branch. In general $U_\mu$ mixes different $\mathfrak A$ submodules, so it cannot be commuted with the branch projector. Writing also $\Psi^\Omega_{b\mu}=S^B_{b\mu}$ when $\Omega$ is an inner gluing, the transported boundary state is therefore
\begin{align}
 |R_{b,\Omega}\rangle_A
 =\sum_{\mu\in\mathscr P_B^\Omega}
  \sum_{i\in\operatorname{Br}(\mu)}
  \frac{\Psi^\Omega_{b\mu}}{\sqrt{S^A_{\mathbf vi}}}
  \left(\check e_{i\mu}U_\mu e_{\mu i}
        \otimes\overline{\mathbf1}\right)|i\rangle\!\rangle_A\,.
 \label{eq:transport-expanded-boundary}
\end{align}
For a fixed physical representation label $i$ and descendant labels $\alpha$ and $\beta$, the overlap with $|R_{b,\Omega}\rangle_A$ reads
\begin{align}
 {}_A\langle i,\alpha;\overline{i,\beta}|R_{b,\Omega}\rangle_A
 =\sum_{\substack{\mu\in\mathscr P_B^\Omega\\ i\in\operatorname{Br}(\mu)}}
 \frac{\Psi^\Omega_{b\mu}}{\sqrt{S^A_{\mathbf vi}}}
 \langle i,\alpha|\check e_{i\mu}U_\mu e_{\mu i}|i,\beta\rangle\,.
 \label{eq:transport-physical-overlap}
\end{align}
Therefore one can either evaluate the overlap between external states of $\mathcal T_A$ and the pullback RG brane, or, equivalently, first embed the states into $\mathcal T_B$ and evaluate their overlaps with the seed brane. One can see more concrete examples in~\cite{Gaiotto:2012np, Poghosyan:2014jia, Poghosyan:2023brb}, and also computations in Section~\ref{sec:inner-walls}.

Finally, we can also evaluate the self-annulus partition functions in both the open and closed channels,
\begin{align}
Z_{{\rm RG},{\rm RG}}^{A,\,\text{closed/open}}&=
 {}_A\langle R_{b,\Omega}|
 \widetilde q^{\frac12(L_0+\bar L_0-c_A/12)}
 |R_{b,\Omega}\rangle_A
 =
 {}_B\langle b;\Omega|
 \widetilde q^{\frac12(L_0+\bar L_0-c_B/12)}
 \mathcal L_{\mathfrak b}|b;\Omega\rangle_B\,\notag\\[5pt]
 &=\text{Tr}_{\mathcal H_{\rm RG,RG}}q^{L_0-c_A/24}
 \label{eq:transported-annulus}
\end{align}
where $\mathcal H_{\rm RG,RG}$ is the open channel Hilbert space with the RG brane at both ends, and in the second equality we have used Eq.~\eqref{eq:interface-fusion-B}. The self-annulus partition function encodes many interesting features of the conformal interface itself. In a companion paper, we will reveal the logarithmic BCFT structure of the spectra of conformal interfaces between non-unitary RCFTs~\cite{Chen:logarithmicPoint}.

\par\bigskip
\noindent
To determine a seed brane by its twist automorphism and brane label, extra input data are usually necessary. In Gaiotto's construction, the seed brane was chosen as its boundary state overlaps agree with the UV--IR operator mixing obtained from perturbative anomalous dimensions~\cite{Gaiotto:2012np}. For continuous families, such comparisons in our parafermion example help locate the RG brane at the correct RG angle in the defect moduli space. In Section~\ref{sec:orbifold-transport} we also employ an orbifold topological interface to transport a known RG brane and thereby determine an unknown one for an orbifold related flow.

Last but not least, it is worth mentioning that not every RG domain wall can be constructed in terms of identifying automorphisms in an auxiliary theory. In Section~\ref{sec:nonunitary-gaiotto}, we will discuss non-unitary examples where the auxiliary theory has no non-trivial automorphism or admits only fully reflective automorphisms. This hence excludes the construction of RG walls in our set-ups. Nevertheless, the walls could still exist and be intrinsically irrational or require a different extension mechanism.

\FloatBarrier
\section{Outer automorphisms and Gaiotto-type RG walls}
\label{sec:outer-walls}

In this section, we will consider the construction of RG domain walls from outer automorphism groups. We will first review Gaiotto's unitary RG walls between two consecutive minimal models, and then apply the same mechanism to study a family of non-unitary RG flows. The non-unitary cases also provide examples in which phantom currents exist but the auxiliary theories do not admit appropriate automorphisms that can be used for the constructions of RG walls.

\subsection{Consecutive unitary minimal models}
\label{sec:unitary-gaiotto}

\noindent\textbf{Phantom sectors and the auxiliary coset.}
We use the convention $\mathcal M_{m,n}$, with coprime integers $2\leq m<n$, to label a minimal model. It has central charge and primary weights
\begin{align}
 c_{m,n}=1-\frac{6(n-m)^2}{mn}\,,\qquad
 h_{r,s}^{(m,n)}=\frac{(nr-ms)^2-(n-m)^2}{4mn}\,.
 \label{eq:sec3-minimal-conventions}
\end{align}
The Kac labels obey $1\leq r<m$, $1\leq s<n$, with $(r,s)\sim(m-r,n-s)$, and $(1,1)$ labels the vacuum. A scalar bulk primary has diagonal left and right weights $(h_{r,s},h_{r,s})$. The unitary flow of interest is given by
\begin{align}
 \mathcal M_{k+2,\,k+3}^{\rm UV}
 \xrightarrow{\ \phi_{1,3}^{\rm UV}\ }
 \mathcal M_{k+1,\,k+2}^{\rm IR}\,,\qquad k\geq2\,.
 \label{eq:sec3-unitary-flow}
\end{align}
The two theories have the coset formulations
\begin{align}
 \mathcal T_{\rm UV}=
 \frac{\widehat{\mathfrak{su}}(2)_k\times
       \widehat{\mathfrak{su}}(2)_1}
      {\widehat{\mathfrak{su}}(2)_{k+1}}\,,\qquad
 \mathcal T_{\rm IR}=
 \frac{\widehat{\mathfrak{su}}(2)_{k-1}\times
       \widehat{\mathfrak{su}}(2)_1}
      {\widehat{\mathfrak{su}}(2)_k}\,.
 \label{eq:sec3-unitary-cosets}
\end{align}
The perturbation $\phi_{1,3}^{\rm UV}$ preserves the UV lines $\mathcal L_{r,1}^{\rm UV}$, which are matched to IR lines $\mathcal L_{1,r}^{\rm IR}$, for $r=1,\ldots,k+1$~\cite{Gaiotto:2012np}. Their folded endpoint modules and lowest weights are given by
\begin{align}
 \mathcal X_r=\mathcal V_{1,r}^{\rm IR}\otimes\overline {\mathcal V}_{r,1}^{\rm UV}\,,
 \qquad
 h(\mathcal X_r)=h_{1,r}^{\rm IR}+h_{r,1}^{\rm UV}
 =\frac{(r-1)^2}{2}\,.
 \label{eq:sec3-unitary-endpoints}
\end{align}
Thus odd $r$ gives integer-spin phantom currents, beginning at weights $2,8,18,\ldots$ for $r=3,5,7,\ldots$, while even $r$ gives half-integer spins. The associated TDLs with odd labels have the same fusion rules as the lines of integer isospins in $\widehat{\mathfrak{su}}(2)_{k}$, often denoted $\mathrm{SO}(3)_{k}$.

Let $A_k=\mathcal T_{\rm IR}\boxtimes \overline{\mathcal T}_{\rm UV}$ denote the folded product physical theory. The compatible Frobenius algebra on the odd phantom sectors is
\begin{align}
 \mathfrak a_k=\bigoplus_{\substack{1\leq r\leq k+1\\r\ {
\rm odd}}}\mathcal X_r\,.
\end{align}
Condensation of $\mathfrak a_k$ precisely gives the conformal embedding used in Gaiotto's construction, i.e. our auxiliary theory $B_k$
\begin{align}
 A_k\xhookrightarrow{\ \mathfrak a_k\ }
 B_k=
 \frac{\widehat{\mathfrak{su}}(2)_{k-1}^{(1)}\times
       \widehat{\mathfrak{su}}(2)_1^{(2)}\times
       \widehat{\mathfrak{su}}(2)_1^{(3)}}
      {\widehat{\mathfrak{su}}(2)_{k+1}}\,.
 \label{eq:sec3-gaiotto-extension}
\end{align}
Here $(1),\, (2),\, (3)$ label the numerator factors, and the denominator is the diagonal piece of the numerator.

\par\medskip
\noindent\textbf{The outer automorphism from the OPE.}
For the case at hand, it is manifest that the automorphism $\Omega=(23)$ exchanges the last two numerator $\widehat{\mathfrak{su}(2)}_1$ factors in Eq.~\eqref{eq:sec3-gaiotto-extension}. Nevertheless, it is instructive to recover this automorphism directly from the OPE covariance equations~\eqref{eq:full-mode-covariance}. Let $X$ be the spin-two phantom current of $\mathcal X_3$. We normalize it as
\begin{align}
    X=\frac{1}{\sqrt{N_k}} \phi_{1,3}^{\rm IR}\otimes\overline{\phi_{3,1}^{\rm UV}}\,,\qquad
    N_k=3k(k-1)(k+4)(k+5)\,,
    \label{eq:X_norm}
\end{align}
so that $\langle X(z)X(0)\rangle=z^{-4}$. We suppress the folding bar on $T_{\rm UV}$ when both stress tensors are regarded as chiral fields of $B_k$. In the ordered basis $(T_{\rm IR},T_{\rm UV},X)$, the two-point matrix is $G_2=\operatorname{diag}(c_{\rm IR}/2,c_{\rm UV}/2,1)$, where
\begin{align}
 c_{\rm IR}=1-\frac6{(k+1)(k+2)}\,,\qquad
 c_{\rm UV}=1-\frac6{(k+2)(k+3)}\,.
 \label{eq:sec3-unitary-central-charges}
\end{align}
For simplicity, write $u*v$ for the coefficient of $z^{-2}$ in the OPE of two spin-two operators $u$ and $v$. The complete set of $z^{-2}$ coefficients in the spin-two OPEs is given by
\begin{align}
 \begin{gathered}
 T_{\rm IR}*T_{\rm IR}=2T_{\rm IR}\,,\qquad
 T_{\rm UV}*T_{\rm UV}=2T_{\rm UV}\,,\qquad
 T_{\rm IR}*T_{\rm UV}=0\,,\\
 T_{\rm IR}*X=\frac{k}{k+2}X\,,\qquad
 T_{\rm UV}*X=\frac{k+4}{k+2}X\,,\\
 X*X=\frac{2k(k+1)}{(k-1)(k+4)}T_{\rm IR}
 +\frac{2(k+3)(k+4)}{k(k+5)}T_{\rm UV}
 -\frac{4(k-2)(k+6)}{\sqrt{ N_k}}X\,.
 \end{gathered}
 \label{eq:sec3-unitary-products}
\end{align}
Solving their covariance equations together with
$M_2G_2M_2^{\mathsf T}=G_2$, we find that, for $k>2$, there are only two solutions, the identity and
\begin{align}
 M_2((23))=
 \begin{pmatrix}
 \dfrac3{(k+2)(k+4)}&\dfrac{(k-1)(k+3)}{k(k+2)}&
 \dfrac{\sqrt{ N_k}}{k(k+2)(k+4)}\\[5pt]
 \dfrac{(k+1)(k+5)}{(k+2)(k+4)}&\dfrac3{k(k+2)}&
 -\dfrac{\sqrt{ N_k}}{k(k+2)(k+4)}\\[5pt]
 \dfrac{6(k+1)(k+5)}{(k+4)\sqrt{ N_k}}&
 -\dfrac{6(k-1)(k+3)}{k\sqrt{ N_k}}&
 \dfrac{k^2+4k-6}{k(k+4)}
 \end{pmatrix}\,.
 \label{eq:sec3-gaiotto-M2}
\end{align}
The matrix squares to one and agrees with the coset exchange action in Ref.~\cite{Gaiotto:2012np} after normalizing $X$ in Eq.~\eqref{eq:X_norm}. Thus, for $k>2$, we only have one non-trivial $\mathbb Z_2$-automorphism solution, say $\Omega=(23)$. On the other hand, for $k=2$, there are six solutions. Setting
\begin{align}
 M_2(\sigma,\epsilon)=
 \begin{pmatrix}
 1/8&5/8&\sigma\sqrt7/8\\[2pt]
 7/8&3/8&-\sigma\sqrt7/8\\[2pt]
 \epsilon\sigma\sqrt7/2&-5\epsilon\sigma/(2\sqrt7)&\epsilon/2
 \end{pmatrix}\,,\qquad \sigma,\epsilon=\pm1\,,
 \label{eq:sec3-ising-tim-six-matrices}
\end{align}
The six solutions are given by
\begin{align}
 &M_2(e)=\mathbf1_3\,,\qquad M_2((12))=\operatorname{diag}(1,1,-1)\,,\notag\\
 &M_2((23))=M_2(+,+)\,,\quad M_2{((13))}=M_2(-,+)\,,\notag\\
 &M_2{((123))}=M_2(-,-)\,,\quad M_2{((132))}=M_2(+,-)\,.
 \label{eq:sec3-ising-tim-S3}
\end{align}
Therefore, overall we have
\begin{align}
 \operatorname{Aut}(\mathfrak B_k)=
 \begin{cases} S_3\,,&k=2\,,\\
 \mathbb Z_2\,,&k>2\,,
 \end{cases}
 \label{eq:sec3-gaiotto-aut}
\end{align}
where $\mathfrak B_k$ is the chiral algebra of the auxiliary theory $B_k$.

\par\medskip
\noindent\textbf{Seed branes and their pullbacks.}
We label an integrable $\widehat{\mathfrak{su}}(2)_\ell$ module by the dimension $a=1,\ldots,\ell+1$ of its horizontal highest-weight representation. Its modular matrix is
\begin{align}
 S^{(\ell)}_{ab}=\sqrt{\frac2{\ell+2}}
 \sin\frac{\pi ab}{\ell+2}\,.
 \label{eq:sec3-affine-S}
\end{align}
Ordinary $\mathfrak B_k$ modules carry labels $[t,d,e;s]$, with
$t=1,\ldots,k$, $d,e=1,2$, $s=1,\ldots,k+2$, subject to
\begin{align}
 t+d+e+s\in2\mathbb Z\,,\qquad
 [t,d,e;s]\sim[k+1-t,3-d,3-e;k+3-s]\,.
 \label{eq:sec3-coset-labels}
\end{align}
The vacuum module is $\textbf{v}=[1,1,1;1]$, and the transposition $(23)$ exchanges the labels $d$ and $e$. So there are $\lfloor {(k+1)^2/}{2}\rfloor$ $(23)$-twisted Ishibashi states. One can check that the seed brane used in~\cite{Gaiotto:2012np},
\begin{align}
 \left|\textbf{v};\,{(23)}\right\rangle_{B_k}=
 \sum_{[t,d,d;s]}
 \sqrt{2S^{(k-1)}_{1t}S^{(k+1)}_{1s}}
 \,\left|[t,d,d;s];(23)\right\rangle\!\rangle\,,
 \label{eq:sec3-gaiotto-seed}
\end{align}
is the same, up to phases on twisted-Ishibashi states, as \eqref{eq:outer-twisted-cardy} with the identity-label, where the sum contains one representative per identification orbit, with $t-s$ even\,\footnote{For odd $k$ there is one additional $(23)$-fixed module $[(k+1)/2,1,2;(k+3)/2]$, fixed through field identification. Its coefficient in this particular seed vanishes.}.

The Frobenius algebra $\mathfrak a_k$ helps define the topological interfaces $\mathcal I_{A_k\to B_k}$ and $\check{\mathcal I}_{B_k\to A_k}$. The latter gives the physical boundary
\begin{align}
 |R_{(23)}\rangle_{A_k}=
 \check{\mathcal I}_{B_k\to A_k}
 \left|\textbf{v};\,{(23)}\right\rangle_{B_k}\,.
 \label{eq:sec3-gaiotto-pullback}
\end{align}
This is the RG brane in Gaiotto's original construction~\cite{Gaiotto:2012np}. In particular its $g$-factor and energy transmission are~\cite{Furuta:2025ahl}
\begin{align}
 \begin{aligned}
 g_{(23)}&={}_{A_k}\langle0|R_{(23)}\rangle_{\!A_k}
 =\frac{\sqrt{S^{(k-1)}_{11}S^{(k+1)}_{11}}}{S^{(k)}_{11}}\,,\\
 \mathcal T_{(23)}&=
 \frac{(k-1)(k+1)(k+3)(k+5)}
 {(k+2)^2(k^2+4k-3)}\,,\qquad
 \mathcal R_{(23)}=1-\mathcal T_{(23)}\,,
 \end{aligned}
 \label{eq:sec3-unitary-g-and-T}
\end{align}
where we've used the fact that the vacuum interface coefficient is $\sqrt{S^B_{\textbf{v}\textbf{v}}/S^A_{\textbf{v}\textbf{v}}}=1/(\sqrt2 S^{(k)}_{11})$ to obtain the $g$-function of the RG brane.

\par\medskip
\noindent\textbf{The exceptional Ising--TIM case.}
For $k=2$, the automorphism group is enhanced to $S_3$, and thus there are more non-trivial ways to define twisted-Cardy states in the auxiliary theory ${B_2}$. Accordingly, after pulling back, there would be more non-factorizable conformal interfaces between Ising and TIM CFTs.
So far, for simplicity, we have focused only on branes with the vacuum label but different automorphism elements. We will present a broader classification of rational Ising--TIM interfaces in a separate work~\cite{AntunesChenRong:interfaces}.

Recall that we have found all six $S_3$-automorphism actions on the weight-two quasi-primaries in the vacuum module of $B_2$ (see~\eqref{eq:sec3-ising-tim-six-matrices} and~\eqref{eq:sec3-ising-tim-S3}). From them, it's easy to compute the energy transmission rate,
\begin{align}
    \mathcal T_e=\mathcal T_{(12)}=0\,,\quad \mathcal T_{\rm etc.}=\frac{35}{48}\,.
\end{align}
So both the identity $e$- and $(12)$-twisted Cardy states, after pullback to $A_2$, give totally reflective interfaces that we will not consider here. On the other hand, the rest will give two non-trivial branes in $\mathcal T_{A_2}$. To see this, it's worth noting that, from $A_2$ to $B_2$, the condensation is performed by gauging the diagonal piece of the $\mathbb Z_2\times\mathbb Z_2$ symmetry in Ising and TIM. In $B_2$, one can check that the quantum symmetry is just the transposition $(12)\in S_3$. Therefore we have
\begin{align}
    \mathcal L_\mathfrak b=\mathbf1\oplus\mathcal L_{(12)}\,, \quad
    \text{and } \quad A_2=B_2^{\langle(12)\rangle}\,,
\end{align}
where $\mathcal L_{(12)}$ denotes the quantum symmetry defect implementing that permutation. Further notice that the four remaining seed branes are connected to each other as
\begin{align}
    \mathcal L_{(12)}\left|\textbf{v};\,{(23)}\right\rangle_{B_2}=\left|\textbf{v};\,{(13)}\right\rangle_{B_2}\,,\qquad
    \mathcal L_{(12)}\left|\textbf{v};\,{(123)}\right\rangle_{B_2}=\left|\textbf{v};\,{(132)}\right\rangle_{B_2}\,.
\end{align}
Consequently, when brought back to $A_2$, these four seeds only give two different non-factorizable branes
\begin{align}
 |R_{(23)}\rangle_{A_2}=\check{\mathcal I}_{B_2\to A_2}\left|\textbf{v};\,{(23)}\right\rangle_{B_2}\,,\qquad
 |R_{(123)}\rangle_{A_2}=\check{\mathcal I}_{B_2\to A_2}\left|\textbf{v};\,{(123)}\right\rangle_{B_2}\,.
 \label{eq:sec3-transposition-orbit}
\end{align}
Both of them have the same transmission rate $\mathcal T=35/48$, but their boundary entropies differ from each other. We explicitly compute them as
\begin{align}
 g_{(13)}=g_{(23)}=
 \left(2-\frac2{\sqrt5}\right)^{1/4}\,,\qquad
 g_{(123)}=\sqrt2\,g_{(23)}\,.
 \label{eq:sec3-cycle-entropy}
\end{align}

\par\medskip

\subsection{Non-unitary minimal-model flows}
\label{sec:nonunitary-gaiotto}

\noindent\textbf{Phantom sectors and the auxiliary coset.}
Now we turn to constructing RG domain walls between non-unitary minimal models, whose RG flows were studied in~\cite{Ahn:1992qi,Lassig:1991an,Martins:1992ht,Dorey:2000zb,Nakayama:2024msv}. These UV--IR pairs can be parametrized by a triple $(q,\,k_{\rm flow},\,I)$, satisfying
\begin{align}
 p_\pm\equiv k_{\rm flow}q\pm I\,,
 \qquad q\geq 2\,,
 \quad k_{\rm flow}\in\tfrac12\mathbb Z_{>0}\,,
 \quad I>0\,.
 \label{eq:sec3-nonunitary-parameters}
\end{align}
The integers $p_\pm\geq2$ must both be coprime to $q$, and the plus and minus signs specify the UV and IR models, respectively. We retain our previous notation for $\mathcal M_{m,n}$, and define primaries $\phi_a^\pm$, $a=1,\ldots,q-1$, as
\begin{align}
 \phi_a^\pm=
 \begin{cases}
 \phi_{a,1}\text{ in }\mathcal M_{q,\,p_\pm}\,,&p_\pm>q\,,\\
 \phi_{1,a}\text{ in }\mathcal M_{p_\pm,\,q}\,,&p_\pm<q\,.
 \end{cases}
 \label{eq:sec3-nonunitary-label-dictionary}
\end{align}
The corresponding Verlinde lines are denoted $\mathcal L_a^\pm$. 

To trigger an RG flow, the deformation operator used here is $\phi_{1,2k_{\rm flow}+1}$. For integer $k_{\rm flow}$, the preserved UV--IR TDL family contains all $\mathcal L_a^{\pm}$ for $a=1,\ldots,q-1$, while, for half-integer $k_{\rm flow}$, it contains only the lines with odd $a$. The endpoint field $X_a=\phi_a^-\otimes\overline{\phi_a^+}$ has factor weights and folded weight given below,
\begin{align}
 h_a^\pm=\frac{a-1}{4}
 \left(k_{\rm flow}(a+1)-2\pm\frac Iq(a+1)\right)\,,
 \quad
 H_a=h_a^-+h_a^+=
 \frac{(a-1)(k_{\rm flow}(a+1)-2)}2\,.
 \label{eq:sec3-nonunitary-weights}
\end{align}
To construct interesting RG branes in our setup, we need to require that there is at least one spin-two phantom current, so that, after condensation, this operator would be able to mix with the stress tensors of UV and IR theories. Therefore, setting $H_a=2$ reduces \eqref{eq:sec3-nonunitary-weights} to
\begin{align}    (a-1)(k_{\rm flow}(a+1)-2)=2H_a=4\quad \Longrightarrow\quad
    (k_{\rm flow},a)=\left(\frac{1}{2},5\right),\ (1,3),\ (2,2)\,.
\end{align}
We find three classes of RG flows for which we may consider constructing RG domain walls. Overall, Table~\ref{tab:nonunitary-phantoms} summarizes the three values of $k_{\rm flow}$ admitting a weight-two matched phantom operator, where every listed sequence is truncated by $a<q$. In particular, the case of $k_{\rm flow}=1/2$ is special in that it also contains a spin-zero phantom current. We will discuss more details at the end of this subsection. 
\begin{table}[t]
\centering
\small
\setlength{\tabcolsep}{4pt}
\renewcommand{\arraystretch}{1.5}
\begin{tabular}{c|c|c|c|c}
$k_{\rm flow}$ & UV $\longrightarrow$ IR & perturbation
& integer-weight $X_a$ & fusion rules\\\hline
$\tfrac12$ &
$\mathcal M_{q/2+I,\,q}\longrightarrow\mathcal M_{q/2-I,\,q}$ &
$\phi_{2,1}$ & odd $a$: $0,0, 2,6,\ldots$ & $\mathrm{SO}(3)_{q-2}$\\
$1$ &
$\mathcal M_{q,\,q+I}\longrightarrow\mathcal M_{q-I,\,q}$ &
$\phi_{1,3}$ & odd $a$: $0, 2,8,18,\ldots$ & $\mathrm{SO}(3)_{q-2}$\\
$2$ &
$\mathcal M_{q,\,2q+I}\longrightarrow\mathcal M_{q,\,2q-I}$ &
$\phi_{1,5}$ & all $a$: $0,2,6,12,\ldots$ & $\mathrm{SU}(2)_{q-2}$
\end{tabular}
\caption{Matched bosonic phantom sectors in the non-unitary flow family. The first ``0'' in the fourth column is for the identity sector. The first row has $0<I<q/2$, and the other rows use $0<I<q$, with admissible endpoints and a perturbation present in the Kac table. At $k_{\rm flow}=1$ the even-labelled TDLs are also preserved, but their matched endpoint weights are half-integer.}
\label{tab:nonunitary-phantoms}
\end{table}

\par\medskip
\noindent\textbf{Non-unitary Gaiotto walls.}
For $k_{\rm flow}=1$, we set $\nu=q/I-2$. The two minimal models then have the consecutive admissible-level coset constructions~\cite{Arakawa:2018iyk}
\begin{align}
 \begin{aligned}
 \mathcal M_{q-I,\,q}^{\rm IR}
 =\frac{\widehat{\mathfrak{su}}(2)_{\nu-1}
          \times\widehat{\mathfrak{su}}(2)_1}
         {\widehat{\mathfrak{su}}(2)_\nu}\,,\qquad
 \mathcal M_{q,\,q+I}^{\rm UV}
 =\frac{\widehat{\mathfrak{su}}(2)_\nu
          \times\widehat{\mathfrak{su}}(2)_1}
         {\widehat{\mathfrak{su}}(2)_{\nu+1}}\,.
 \end{aligned}
 \label{eq:sec3-fractional-cosets}
\end{align}
Condensing the odd-$a$ phantom sectors again removes the intermediate denominator and numerator, and gives an auxiliary theory,
\begin{align}
 B_{q,I}=
 \frac{\widehat{\mathfrak{su}}(2)_{\nu-1}
       \times\widehat{\mathfrak{su}}(2)_1^{(2)}
       \times\widehat{\mathfrak{su}}(2)_1^{(3)}}
      {\widehat{\mathfrak{su}}(2)_{\nu+1}}\,.
 \label{eq:sec3-fractional-extension}
\end{align}
The weight-two OPEs are the rational continuation of Eq.~\eqref{eq:sec3-unitary-products}. The solution to their covariance equation is also simply Eq.~\eqref{eq:sec3-gaiotto-M2} with $k=\nu$. Since now the levels $\nu$ are fractional, there are no exceptional enhanced automorphisms as the TIM-to-Ising flow. But the exchange $\Omega=(23)$ is unchanged, leading to a non-trivial $\mathbb Z_2$ automorphism group on the auxiliary chiral algebra $\mathfrak B_{q,I}$. 

As in the unitary case, the auxiliary theory still admits additional phantom currents of half-integer spin, which can be further condensed via fermionization. One then will obtain a fermionic extension as the tensor product of an $N=1$ supersymmetric non-unitary minimal model and a free fermion,
\begin{align}
    B_{q,I}\quad \xrightarrow[\text{half-integer spin condensation}]{\mathcal X_{a\in 2\mathbb Z}\,,\quad h_a=(a-1)^2/2} \quad \widetilde{B}_{q,I}=\text{SM}(q-I,q+I)\,\boxtimes\,\text{Maj.}\,,
\end{align}
where ``Maj.'' denotes a free Majorana fermion. Many computations can be facilitated in this fermionic basis as already noted in~\cite{Gaiotto:2012np}. Here we'll omit more technical details on this construction for brevity.

\par\medskip
\noindent\textbf{An explicit example: $\mathcal M_{5,\,7}\to
\mathcal M_{3,\,5}$.}
Take $q=5$, $I=2$, so that $\nu=1/2$. Both endpoint theories
are non-unitary. In this paragraph and below, we denote the physical product theory by $A=\mathcal M_{3,\,5}\boxtimes\overline{\mathcal M}_{5,\,7}$ and the auxiliary theory by
\begin{align}
 B=
 \frac{\widehat{\mathfrak{su}}(2)_{-1/2}
       \times\widehat{\mathfrak{su}}(2)_1^{(2)}
       \times\widehat{\mathfrak{su}}(2)_1^{(3)}}
      {\widehat{\mathfrak{su}}(2)_{3/2}}\,.
 \label{eq:sec3-q5-example}
\end{align}
The preserved TDLs are Fibonacci-like lines satisfying
\begin{align}
    \mathcal L_{3}^{\pm}\times\mathcal L_{3}^{\pm}=\mathbf 1+\mathcal L_{3}^{\pm}\,.
\end{align}
Gauging their diagonal combination corresponds to condensing the module $\mathcal X_{3}\equiv \mathcal V_{(1,3;3,1)}^A$, with weights $(h_{\rm IR},h_{\rm UV})=(1/5,9/5)$, into the vacuum of $A$, where the semicolon separates the IR and UV Kac labels. It thus defines the Frobenius algebra $\mathfrak a=\mathbf 1\oplus\mathcal X_3$ and its associated topological interface $\mathcal I_{A\to B}$.

The modules in $B$ can be labelled by a pair of integers $i=(\alpha, s)$ with $\alpha=0,1$ and $s=1,\dots,6$. Restricting to $A$, they give the following branching decomposition:
\begin{align}
 \begin{aligned}
 \mathcal V^B_{0,s}\big|_A
 &=\bigl(\mathcal V^{\rm IR}_{1,1}\otimes\overline {\mathcal V}^{\rm UV}_{1,s}\bigr)
   \oplus
   \bigl(\mathcal V^{\rm IR}_{1,3}\otimes\overline {\mathcal V}^{\rm UV}_{3,s}\bigr)\,,\\
 \mathcal V^B_{1,s}\big|_A
 &=\bigl(\mathcal V^{\rm IR}_{1,4}\otimes\overline {\mathcal V}^{\rm UV}_{1,s}\bigr)
   \oplus
   \bigl(\mathcal V^{\rm IR}_{1,2}\otimes\overline {\mathcal V}^{\rm UV}_{3,s}\bigr)\,.
 \end{aligned}
 \label{eq:sec3-q5-branching}
\end{align}
The modular matrix is
\begin{align}
 S^B_{(\alpha,s),(\beta,t)}=
 \frac{(-1)^{\alpha\beta+\alpha+\beta+s+t}}{\sqrt7}
 \sin\frac{5\pi st}{7}\,.
 \label{eq:sec3-q5-S}
\end{align}
In the weight-two subspace of the vacuum module of $B$, we once again have three quasi-primary fields $(T_{\rm IR},\, T_{\rm UV},\,X)$ with $X\in \mathcal X_3$ the spin-two phantom current. In the non-unitary CFT, the two-point correlator $\langle X(z)X(0)\rangle=-\frac{297}{16z^4}$ implies a negative norm. With this normalization, we can solve the OPE covariance equation to obtain 
\begin{align}
    M_2(\Omega)=
 \begin{pmatrix}
 4/15&-7/5&8/45\\
 11/15&12/5&-8/45\\
 11&21&-5/3
 \end{pmatrix}\,,\qquad M_2(\Omega)^2=\mathbf1_3\,.
 \label{eq:sec3-q5-M2}
\end{align}
It preserves the indefinite Gram matrix $G_2=\operatorname{diag}(-3/10,11/70,-297/16)$ and all OPE products. 

Using $\Omega$, we can now follow the standard procedure to construct the seed brane in $B$ and then pull it back to $A$:
\begin{align}
&|\textbf{v};\Omega\rangle_B=
 \sum_{i\in\mathscr P}\sqrt{2S^B_{\textbf{v}i}}\,
 |i;\Omega\rangle\!\rangle\,,\qquad
 |R_\Omega\rangle_A=\check{\mathcal I}_{B\to A}|\textbf{v};\Omega\rangle_B\,,\notag\\
 &\text{with }\quad \mathscr P=\{(0,1),(0,3),(0,5),(1,2),(1,4),(1,6)\}
 \label{eq:sec3-q5-seed}
\end{align}
Here we collect some physical observables of the RG brane $ |R_\Omega\rangle_A$: Its energy transmission and reflection rates and boundary entropy are given by
\begin{align}
    \mathcal T(R_\Omega)=\frac{77}{25}\,,\qquad
 \mathcal R(R_\Omega)=-\frac{52}{25}\,,\qquad
 g_{R_\Omega}={}_A\langle0|R_\Omega\rangle_A
 =\sqrt{\frac{(5-\sqrt5)\sin(2\pi/7)}{\sqrt7}}\,.
\end{align}
More interestingly, we compute the open-channel partition function of the RG brane, 
\begin{align}
    Z_{{R_\Omega}{R_\Omega}}^{A,{\rm open}}(q)= {}_A\langle R_\Omega|\widetilde q^{(L_0+\bar L_0-c_B/12)/2}
                      |R_\Omega\rangle_A\,,
 \label{eq:sec3-q5-physical-annulus}
\end{align}
from which we numerically extract a few low-lying spectral levels. The computation methods are summarized in Appendix~\ref{app:open-spectrum}. We find both a negative conformal weight and a pair of complex conformal weights in the spectrum:
\begin{align}
 h_0\simeq-0.08457\,,\qquad
 h_\pm\simeq0.3359\pm0.2533\,i\,.
 \label{eq:sec3-q5-open-candidates}
\end{align}
Both the negative reflection ratio and the complex conformal weights require further physical explanation in the context of non-unitary CFTs. In a companion paper, we will discuss an interesting phenomenon: when a conformal interface between two non-unitary CFTs admits a continuous defect moduli space, a pair of operators with complex conjugate conformal weights may coalesce into a logarithmic pair at a certain exceptional point in the defect conformal manifold~\cite{Chen:logarithmicPoint}.

\par\medskip
\noindent\textbf{Obstructions at $k_{\rm flow}=2$ and $1/2$.}
The existence of integer-spin phantom currents does not always produce an RG domain wall constructed via the automorphism mechanism. A simple example is
\begin{align}
 \mathcal M_{3,7}^{\rm UV}
 \xrightarrow{\ \phi_{1,5}^{\rm UV}\ }
 \mathcal M_{3,5}^{\rm IR}\,,\qquad
 (q,\,k_{\rm flow},\,I)=(3,\,2,\,1)\,.
 \label{eq:sec3-kflow2-example}
\end{align}
The nontrivial preserved TDL is an invertible $\mathbb Z_2$-line. Its endpoint $X_2=\phi_{2,1}^{\rm IR}\otimes\overline{\phi_{2,1}^{\rm UV}}$
has total weight $h_2=h_{\rm IR}+h_{\rm UV}=3/4+5/4=2$. We normalize $X_2$ so that $\langle X_2(z)X_2(0)\rangle=1/z^4$. In the basis $(T_{\rm IR},\,T_{\rm UV},\, X_2)$, using their OPE, one can solve the covariance equation~\eqref{eq:graded-OPE-covariance} and find that there are two solutions
\begin{align}
    M_2=\mathbf 1_3\,,\quad {\rm or} \quad \operatorname{diag}(1,1,-1)\,,
\end{align}
implying that the auxiliary theory has a $\mathbb Z_2$-automorphism group. However, the non-trivial element fixes both the stress tensors $T_{\rm IR}$ and $T_{\rm UV}$ and hence has zero transmission. More generally, for admissible $k_{\rm flow}=2$ parameters $0<I/q<1$, we check that the same covariance equation gives these two weight-two actions: $X_a\mapsto(-1)^{a-1}X_a$, while keeping the UV/IR stress tensors intact. Therefore the RG domain wall, if it exists, cannot be realized in terms of an automorphism-twisted seed brane in the corresponding auxiliary theory.

It is worth mentioning that this example allows a direct comparison with the solution discussed in~\cite{Furuta:2025ahl}, in which the spin-two phantom current can mix with the UV/IR stress tensors to obtain a non-zero transmission rate: 
\begin{align}
  T_{\rm IR}'=\frac{156}{281}T_{\rm IR}+\frac{21}{281}T_{\rm UV}+\gamma X_2\,,
 \quad
 T_{\rm UV}'=\frac{125}{281}T_{\rm IR}+\frac{260}{281}T_{\rm UV}-\gamma X_2\,,
 \quad \gamma^2=-\frac{15600}{281^2}\,.\notag
\end{align}
However, it is not hard to check that such mixing does not respect the OPE covariance equations, which involve the full OPEs, including the phantom operator self-product, rather than only the OPEs of the stress tensors. It would be interesting to investigate a different mechanism to construct such an RG brane beyond the group automorphism approach. 

\par\bigskip

For $k_{\rm flow}=1/2$, the smallest example is $\mathcal M_{3,\,5}\to\mathcal M_{2,\,5}$, with $q=5$, $I=1/2$ and perturbation $\phi_{2,1}^{\rm UV}$. The matched endpoint fields, apart from the identity operators, have factor weights $(-1/5,1/5)$ and thus total weight zero. The auxiliary theory obtained by condensation has the Urod conformal vertex algebra $\mathcal U$~\cite{Bershtein:2013oka}. The underlying vertex algebra is $\widehat{\mathfrak{su}}(2)_1$, but its stress tensor is defined as
\begin{align}
 T_{\mathcal U}=T_{\rm Sug}+\frac12\partial H
                 +\epsilon\partial^2e\,,\qquad
 \epsilon\ne0\,,\qquad c_{\mathcal U}=-5\,.
 \label{eq:sec3-urod-conformal-vector}
\end{align}
Here $e,H,f$ are the $\widehat{\mathfrak su}(2)_1$ raising, Cartan and lowering currents, normalized by $H_{(0)}e=2e$ and $e_{(1)}f=\mathbf1$, and $T_{\rm Sug}$ is their Sugawara stress tensor. Applying the OPE covariance equation up to level two, one can find that the conformal automorphism group of $\mathcal U$ is trivial.

The two examples fail for different reasons: the first retains a fully reflective automorphism, whereas Urod has no nontrivial automorphism at all. Therefore neither of them provides a transmitting conformal interface in our construction. This, however, does not exclude their proposed RG interfaces.

\FloatBarrier

\section{Inner automorphisms and RG walls with defect conformal moduli}
\label{sec:inner-walls}

So far, we have only considered the UV--IR theories with phantom operators starting from conformal weight $h=2$. When there are spin-one phantom currents, these currents will lead to a different realization of the construction. Once condensed into the auxiliary vacuum, they become ordinary chiral currents in the auxiliary $\mathcal T_B$. Let $Q^a$, $a=1,\ldots,d$, denote these currents. Their zero modes commute with the total Virasoro algebra and generate inner rotations
\begin{align}
 U(\vec\varphi)=\exp\!\left(i\sum_{a=1}^d\varphi_a Q^a_0\right)\,,
 \qquad \Omega_{\vec\varphi}=\operatorname{Ad}
 U(\vec\varphi)\,,
 \label{eq:sec4-inner-orbit}
\end{align}
where the real parameters $\varphi_a$ are boundary couplings~\cite{Antinucci:2025uvj}. For a given seed brane, defined by a fixed $\Omega_{\vec\varphi_0}$-twist, $U(\vec\varphi\,)$ can act on its chiral part while preserving the gluing~\eqref{eq:twisted-ishibashi-gluing} along a family of boundary conditions with the $\Omega_{\vec\varphi+\vec\varphi_0}$-twist. The brane is therefore marginally deformed along the direction of $\vec\varphi$ in the moduli space.

On the other hand, these continuous inner automorphisms can also be seen directly from Eq.~\eqref{eq:graded-OPE-covariance}. In $\mathcal T_B$, using OPEs, one can compute the commutators between phantom current zero modes and other quasi-primary operators, 
\begin{align}
    i\left[Q^a_0,\,\mathcal O_i^{(h)}\right]=\sum_j(D_h^a)_{ij}\mathcal O_j^{(h)}\,,\qquad{\rm for}\quad \mathcal O^{(h)}_i\in\mathcal Q_{h}(B)\,.
\end{align} 
Meanwhile, the linearized covariance equations~\eqref{eq:linearized-ope-covariance} form an underdetermined system of linear equations with a $d$-dimensional null space. The general solution to Eq.~\eqref{eq:linearized-ope-covariance} is precisely $\sum_a\varphi_a D^{a}_h$, parametrized by $d$ free parameters $\varphi_a$, $a=1,2,\dots,d$. Thus the free parameters $\vec\varphi$ of the local OPE solutions acquire a direct physical meaning as marginal deformation couplings of the boundary.

\subsection{RG walls from \texorpdfstring{$\mathbb Z_N$}{ZN}-parafermions to minimal models}
\label{sec:parafermion-walls}

\paragraph{Phantom sectors and the auxiliary coset.}
Our main example is the massless flow
\begin{align}
 \mathcal P_k=
 \frac{\widehat{\mathfrak{su}}(2)_k}{\widehat{\mathfrak u}(1)_k}
 \longrightarrow
 \mathcal M_{k+1,\,k+2}=
 \frac{\widehat{\mathfrak{su}}(2)_{k-1}\times
       \widehat{\mathfrak{su}}(2)_1}
      {\widehat{\mathfrak{su}}(2)_k}\,,\qquad k\geq3\,,
 \label{eq:sec4-pf-flow}
\end{align}
generated by a pair of parafermion bilinear operators of weights
$(1-1/k,1-1/k)$~\cite{Fateev:1990bf,Fateev:1991bv},
\begin{align}
 S=S_{\mathcal P_k}-\lambda\int d^2x\,
 \left(e^{i\pi/k}\,\phi_{1,2}^{\rm UV}+e^{-i\pi/k}\phi_{1,-2}^{\rm UV}\right)\,,
 \qquad \lambda>0\,.
 \label{eq:sec4-massless-phase}
\end{align}
The two primaries $\phi_{1,\pm2}^{\rm UV}$ are $\mathbb Z_k$-charge conjugates of each other, and their coupling phases $e^{\pm i\pi/k}$ will be important in Section~\ref{sec:odd-k-branes}. For a primary in $\mathcal P_k$, we use its coset construction label $[r,m]$, with $r=1,\ldots,k+1$, $m\in\mathbb Z_{2k}$, $r+m$ odd, and $[r,m]\sim[k+2-r,m+k]$. Its conformal weight is 
\begin{align}
    h_{[r,m]}=\frac{r^2-1}{4(k+2)}-\frac{m^2}{4k}\pmod 1\,.
\end{align}
For minimal models, their modular data has been reviewed in Sec.~\ref{sec:outer-walls}. 

The deformation~\eqref{eq:sec4-massless-phase} preserves the following Verlinde lines:
\begin{align}
    \mathcal L^{\rm UV}_{[2a+1,0]}\,,\quad{\rm and}\quad 
    \mathcal L^{\rm IR}_{1,2a+1}\,,\quad {\rm with} \quad
    a=0,\ldots,\lfloor k/2\rfloor
\end{align}
Their common fusion rules are those of $\mathrm{SO}(3)_k$. In the folded picture, denote the physical product theory by $A_k=\mathcal M_{k+1,\,k+2}\boxtimes\overline{\mathcal P}_k$. The matched phantom sectors are given below:
\begin{align}
 \mathcal X_a=\mathcal V^{\rm IR}_{1,2a+1}\boxtimes
       \overline{\mathcal V}^{\rm UV}_{[2a+1,0]}\,\qquad
       \Longrightarrow\qquad \mathfrak a_k=\bigoplus_{a=0}^{\lfloor k/2\rfloor}\mathcal X_a
 \label{eq:sec4-matched-sector}
\end{align}
with conformal weights
\begin{align}
 h^{\rm UV}_a=\frac{a(a+1)}{k+2}\,,\qquad
 h^{\rm IR}_a=a^2-\frac{a(a+1)}{k+2}\,,\qquad
 h_a=a^2\,.
 \label{eq:sec4-phantom-weights}
\end{align}
The sectors admit the diagonal condensable algebra $\mathfrak a_k=\bigoplus_a\mathcal X_a$. The condensation of $\mathfrak a_k$ gives
\begin{align}
 A_k\subset B_k\,,\qquad
 B_k=
 \frac{\widehat{\mathfrak{su}}(2)_{k-1}\times
       \widehat{\mathfrak{su}}(2)_1}
      {\widehat{\mathfrak u}(1)_k}\,,
 \qquad \mathcal V_{\textbf{v}}^B=\bigoplus_{a=0}^{\lfloor k/2\rfloor}
 \mathcal X_a\,.
 \label{eq:sec4-pf-extension}
\end{align}
This is the coset conformal extension proposed in~\cite{Gaiotto:2012np,Zhang:2026gqp}. The central charges are
\begin{align}
 c_{\rm UV}=\frac{2(k-1)}{k+2}\,,\qquad
 c_{\rm IR}=\frac{(k-1)(k+4)}{(k+1)(k+2)}\,,\qquad
 c_{B_k}=\frac{3(k-1)}{k+1}\,.
 \label{eq:sec4-central-charges}
\end{align}
In particular, the $a=1$ phantom field becomes a spin-one current in $B_k$. For a generic module in $B_k$, the branching formula is given by
\begin{align}
 \mathcal V^{B_k}_{[t,d,m]}\bigg|_{A_k}
 =\bigoplus_{\substack{r=1,\ldots,k+1\\t+d+r\ {\rm odd}}}
 \mathcal V^{\mathcal M_{k+1,k+2}}_{t,r}\otimes
 \overline{\mathcal V}^{\mathcal P_k}_{[r,m]}\,.
 \label{eq:sec4-annulus-branching}
\end{align}
Here we refer readers to~\cite{Zhang:2026gqp} for a more detailed analysis of these phantom sectors and their fusion rules.

\paragraph{The automorphism from low-weight OPEs.}
Let $J^a$ and $j^a$ be the numerator affine currents  of $\widehat{\mathfrak {su}}(2)_{k-1}$ and $\widehat{\mathfrak {su}}(2)_{1}$ in $B_k$, respectively. We normalize their Cartan OPEs as
$J^3(z)J^3(0)\sim(k-1)/(2z^2)$ and
$j^3(z)j^3(0)\sim1/(2z^2)$, and write
\begin{align}
 H=J^3+j^3\,,\qquad
 Q=\frac{J^3-(k-1)j^3}{k}\,,\qquad
 Q(z)Q(0)\sim\frac{\kappa_k}{z^2}\,,\qquad
 \kappa_k=\frac{k-1}{2k}\,.
 \label{eq:sec4-relative-current}
\end{align}
The current $H$ generates the denominator $\widehat{\mathfrak u}(1)_k$, whereas $Q$ survives in the coset and spans $\mathcal Q_1(B_k)$. The OPE covariance equation~\eqref{eq:graded-OPE-covariance} on its identity-channel OPE gives 
\begin{align}
    M_1(Q)^2=1\qquad\Longrightarrow\qquad M_1(Q)=\pm 1\,.
\end{align}
So an automorphism sends $Q$ to $Q$ or $-Q$. The simultaneous Weyl reflection, denoted by $\mathsf w$, of the numerator factors realizes the second solution $M_1=-1$:
\begin{align}
 \mathsf w:\quad
 (J^3,j^3)\mapsto(-J^3,-j^3)\,,\qquad
 (J^\pm,j^\pm)\mapsto(J^\mp,j^\mp)\,,\qquad Q\mapsto-Q\,.
 \label{eq:sec4-weyl}
\end{align}
We call $\mathsf w$ the charge conjugation automorphism in $B_k$. 

Now let's further solve the OPE covariance equations in the component connected to the identity, where $M_1=1$. At weight two, one can collect four quasi-primaries in the following basis
\begin{align}
 \mathcal Q_2=(T_{\rm UV},T_{\rm IR},W,X)\,,\qquad
 G_2=\operatorname{diag}\left(\frac{c_{\rm UV}}2,
                 \frac{c_{\rm IR}}2,1,1\right)\,,
 \label{eq:sec4-weight-two-basis}
\end{align}
where the field $W$ is the relative Virasoro descendant of $Q$, orthogonal to $\partial Q$,
\begin{align}
 W=-\frac{i}{\sqrt{2(k-1)}}
       (kL_{-1}^{\rm UV}-2L_{-1}^{\rm IR})\,Q\,.
 \label{eq:sec4-relative-descendant}
\end{align}
The phase chosen here normalizes its two-point correlator
$\langle W(z)W(0)\rangle=z^{-4}$. The field $X$ is the other normalized weight-two quasi-primary in this endpoint sector. It is obtained by acting with the UV spin-three $W^{(3)}$-current mode $W^{(3),{\rm UV}}_{-1}$ on the endpoint of $Q$,
\begin{align}
    X\propto W^{(3),{\rm UV}}_{-1}Q\,.
\end{align}
The charge conjugation $\mathsf w$ fixes both stress tensors and $X$, and reverses $W$.

Put $h_U=2/(k+2)$ and $h_I=k/(k+2)$, the two factor weights of $Q$, and define $T_U=T_{\rm UV}$ and $T_I=T_{\rm IR}$. We now spell out the OPEs of the operators in $\mathcal Q_2$ at their second order poles,
\begin{align}
 \begin{aligned}
 & T_U*T_U=2T_U\,,\quad T_I*T_I=2T_I\,,\quad T_U*T_I=0\notag\\
 &T_U*W =W\,, \quad T_I*W =W\,, \quad T_U*X=(h_U+1)X\,,\quad T_I*X=h_I X\notag\\
 &W*W=\frac2{c_{\rm UV}}T_U+\frac2{c_{\rm IR}}T_I+\rho_kX\,,\quad
 W*X=\rho_k W\,,\quad\notag\\ &X*X=\frac{2(h_U+1)}{c_{\rm UV}}T_U
        +\frac{2h_I}{c_{\rm IR}}T_I+\sigma_k X\,,
 \end{aligned}
\end{align}
with
\begin{align}
 \rho_k=\sqrt{\frac{2(k-2)}{(k-1)(k+4)}}\,,\qquad
 \sigma_k=\frac{3\sqrt2(k-4)}{\sqrt{(k-1)(k-2)(k+4)}}\,.
 \label{eq:sec4-ope-constants}
\end{align}
The computational details have been given in~\cite{Zhang:2026gqp}. We here only list the results. Substitution in the linearized OPE equations~\eqref{eq:linearized-ope-covariance} gives a system of rank
15 for the 16 entries of $\delta M_2$. The solution can be written down in terms of one real free parameter $\varphi$, 
\begin{align}
    M_2(\varphi)=\mathbf1_{4}+\varphi D_2+O(\varphi^2)\,,
\end{align}
with
\begin{align}
 D_2=
 \begin{pmatrix}
 0&0&p_k&0\\
 0&0&-p_k&0\\
 -q_k&r_k&0&-\lambda_k\\
 0&0&\lambda_k&0
 \end{pmatrix}\,,
 \quad
 \begin{aligned}
 p_k&=\frac{\sqrt{2(k-1)}}{k+2}\,,&
 q_k&=\sqrt{\frac2{k-1}}\,,\\
 r_k&=\frac{2\sqrt2(k+1)}{(k+4)\sqrt{k-1}}\,,&
 \lambda_k&=\sqrt{\frac{k-2}{k+4}}\,.
 \end{aligned}
 \label{eq:sec4-D2}
\end{align}
One can check that the solution~\eqref{eq:sec4-D2} precisely coincides with the action of the zero mode of $Q$ on the spin-two quasi-primaries, as it should:
\begin{align}
 \begin{aligned}
 iQ_0T_U&=p_kW\,,&iQ_0T_I&=-p_kW\,,\\
 iQ_0W&=-q_kT_U+r_kT_I-\lambda_kX\,,&iQ_0X&=\lambda_kW\,.
 \end{aligned}
 \label{eq:sec4-zero-mode-action}
\end{align}
In particular, notice that $D_2^3=-D_2$, which gives
\begin{align}
 M_2(\Omega_\varphi)=e^{\varphi D_2}=\mathbf1_4+\sin\varphi D_2
                    +(1-\cos\varphi)D_2^2\,,\quad{\rm with}\quad\varphi\sim\varphi+2\pi\,.
 \label{eq:sec4-M2}
\end{align}
Therefore, promoting the solution to a finite transformation by exponentiating the zero mode $Q_0$,
\begin{align}
    \Omega_\varphi\equiv {\operatorname {Ad}}\left(e^{i \varphi Q_0}\right)=\operatorname{Ad}\left(e^{i\varphi J^3_0}\right)\bigg|_{B_k}
\end{align}
we have a continuous $\rm U(1)$ inner automorphism parameterized by the free parameter $\varphi$.

There are no further automorphism parameters hidden at higher weights. Indeed, the chiral algebra of $B_k$, $\mathfrak B_k$, is generated under all OPE modes by the two neutral numerator fields $E_\pm=:J^\pm j^\mp:$ of weight two. They realize the algebra $D_{k-1,2}$ discussed in~\cite{adamovic2001regularitycertainvertexoperator}, while the coset realization and its branching are given in~\cite{yamada2019mathbbz2kcodevertexoperatoralgebras}. It can be shown that every higher-weight field is obtained from the weight-one and weight-two data, including fields arising in singular OPE coefficients. Consequently, the automorphism group of $\mathfrak B_k$ is given by
\begin{align}
 \operatorname{Aut}(\mathfrak B_k)=\mathrm U(1)^{Q_0}\rtimes
     \mathbb Z_2^{\mathsf w}\simeq\mathrm O(2)\,,\quad {\rm and}\quad
 \mathsf w\,\Omega_\varphi\mathsf w=\Omega_{-\varphi}\,.
 \label{eq:sec4-O2}
\end{align}

\paragraph{TIM--Potts RG angle and the seed brane.}
Now we will determine the seed brane for the special case $k=3$, i.e. the flow from the 3-state Potts to the TIM CFT. Recall that the irreducible $\mathfrak B_k$ modules are labelled by a triple
\begin{align}
 [t,d,m]\,,\qquad
 \begin{gathered}
 1\leq t\leq k\,,\quad d=1,2\,,\quad m\in\mathbb Z_{2k}\,,
 \quad t+d+m\equiv0\pmod2\,,\\
 [t,d,m]\sim[k+1-t,3-d,m+k]\,.
 \end{gathered}
 \label{eq:sec4-B-labels}
\end{align}
Here $t$ and $d$ label the modules of the two affine numerators $\widehat{\mathfrak {su}}(2)_{k-1}$ and $\widehat{\mathfrak {su}}(2)_{1}$, and $m$ is twice the Cartan charge of the denominator $\widehat{\mathfrak u}(1)_k$, so that $m=2H_0$ modulo $2k$. So there are $k^2$ primaries in $B_k$, with vacuum module $\textbf{v}=[1,1,0]$. For $a=[t,d,m]$ and $\mu=[u,e,n]$, the modular matrix is
\begin{align}
 S^{B_k}_{a\mu}=\sqrt{\frac{2}{k(k+1)}}
 \sin\frac{\pi tu}{k+1}\,
 (-1)^{(d-1)(e-1)}e^{i\pi mn/k}\,.
 \label{eq:sec4-B-Smatrix}
\end{align}
We can implement the inner automorphism on every module by $\Omega_\varphi=e^{i\varphi Q_0}$, and thus prepare $k^2$ $\Omega_\varphi$-twisted Ishibashi states,
\begin{align}
    |\mu;\,\Omega_{\varphi}\rangle\!\rangle\equiv \left(U_{\mu,\varphi}\otimes\overline{\one}\right)|\mu\rangle\!\rangle\,,
\end{align}
from the ordinary ones $|\mu\rangle\!\rangle$, where $U_{\mu,\varphi}$ is the unitary representation of $\Omega_\varphi$ on each module $\mu$. We thus have $k^2$ branes in $B_k$ and their pullbacks in $A_k$,
\begin{align}
 \begin{split}
 |b_a(\varphi)\rangle_{B_k}
 =\sum_\mu\frac{S^{B_k}_{a\mu}}{\sqrt{S^{B_k}_{\textbf{v}\mu}}}
       |\mu;\Omega_\varphi\rangle\!\rangle\,,\qquad
 |R_a(\varphi)\rangle_{A_k}
 =\check{\mathcal I}_{B_k\to A_k}|b_a(\varphi)\rangle_{B_k}\,.
 \end{split}
 \label{eq:sec4-inner-seeds}
\end{align}

At $k=3$, the flow joins 3-state Potts $\mathcal P_3$ to tri-critical Ising $\mathcal M_{4,\,5}$. The UV charge conjugation orbifold of $\mathcal P_3$ is the diagonal tetra-critical Ising model $\mathcal M_{5,\,6}$, so we can use Gaiotto's $\mathcal M_{5,\,6}\to\mathcal M_{4,\,5}$ wall, discussed in Section~\ref{sec:unitary-gaiotto}, to partially fix the choice of the seed brane here.  First, for both the inner $\Omega_\varphi$ and outer $\Omega_\varphi\mathsf w$ automorphisms in $B_k$, one can compute the energy transmission rate from~\eqref{eq:stress-overlap-definition} and \eqref{eq:reflection-transmission}:
\begin{align}
    \mathcal T_k(\Omega_\varphi)=\mathcal T_k(\Omega_\varphi\mathsf w)=\frac{8(k+1)}{3(k+2)^2}(1-\cos\varphi)\,.
\end{align}
On the other hand, it is known that the transmission rate through the RG wall between $\mathcal M_{5,\,6}$ and $\mathcal M_{4,\,5}$ is given in~\eqref{eq:sec3-unitary-g-and-T}, namely $\mathcal T(\mathcal M_{5,6}\to \mathcal M_{4,5})=\frac{64}{75}$. Since the orbifolding from $\mathcal M_{5,\,6}$ to $\mathcal P_3$ is implemented by a completely transparent topological interface, we can fix the undeformed RG brane at a specific point in moduli space, namely the RG angle
\begin{align}
    \varphi_*=\pi\,.
\end{align}

Next, let's partially fix the brane label from the $g$-factors. For clarity, we give all the seed data needed for this comparison. An ordered set of the nine ordinary labels is
\begin{align}
 \mathscr P_3^{\Omega_\varphi}=\{110,\,112,\,114,\,121,\,123,\,125,\,211,\,213,\,215\}\,,
 \label{eq:sec4-pf3-label-order}
\end{align}
where ``$tdm$'' abbreviates $[t,d,m]$. There are nine elementary seeds in the inner branch, see also~\eqref{eq:sec4-inner-seeds}. On the other hand, the outer charge conjugation $\mathsf w$ maps the labels:
\begin{align}
    \mathsf w:\quad [t,\,d,\,m]\ \longmapsto\ [t,\,d,\,-m]\,.
\end{align}
Up to the equivalence relation, there are three invariant $\Omega_\varphi\mathsf w$-twisted modules 
\begin{align}
    \mathscr P_3^{\Omega_\varphi\mathsf w}=\{110_\mathsf w,\,123_\mathsf w,\,213_\mathsf w\}\,,
    \label{eq:w-twist-cardy}
\end{align}
supporting Ishibashi states. For example, the $\Omega_\varphi\mathsf w$-twisted Ishibashi state associated with $110$ is defined by $|110_\mathsf w\rangle\!\rangle\equiv \Omega_\varphi\mathsf w\,\otimes\overline{\one}\,|110\rangle\!\rangle$. So the three elementary outer seeds are given by
\begin{align}
 |b^\mathsf w_a(\varphi)\rangle
 =\sum_{\mu\in\mathscr P_3^{\Omega_\varphi\mathsf w}}
  \frac{\Psi^\Omega_{a\mu}}{\sqrt{S^{B_3}_{\mathbf v\mu}}}
  |\mu;\Omega_\varphi\mathsf w\rangle\!\rangle\,,\qquad
 \Psi^\Omega=\frac12
 \begin{pmatrix}1&-1&\sqrt2\\1&-1&-\sqrt2\\\sqrt2&\sqrt2&0\end{pmatrix}\,,
 \label{eq:sec4-outer-three-seeds}
\end{align}
where $\Psi^\Omega$ is the $\mathsf w$-twisted modular matrix, obtained from the charge conjugation twisted coset Cardy construction
\cite{Ishikawa:2001zu}. We here for brevity omit the detailed derivation. However, one can explicitly check that their self-annuli are $\chi_{110}^{B_3}+\chi_{112}^{B_3}+\chi_{114}^{B_3}$ for $b_1^\mathsf w,b_2^\mathsf w$, and the sum of all six $t=1$ characters for $b_3^\mathsf w$. Each of them satisfies the Cardy condition, and has one open vacuum channel.

To compute the $g$-factors of the 12 candidate branes in $\mathcal T_{A_3}$, one notices that the topological interface from $B_3$ to $A_3$ has the vacuum amplitude
\begin{align}
 (\check{\mathcal I}_{B_3\to A_3})_{00}
 =\sqrt{\frac{S^{B_3}_{\mathbf v\mathbf v}}{S^{A_3}_{\mathbf v\mathbf v}}}\,,\qquad
 S^{B_3}_{\mathbf v\mathbf v}=\frac1{\sqrt{12}}\,,\qquad
 S^{A_3}_{\mathbf v\mathbf v}=\frac{5-\sqrt5}{20\sqrt3}\,.
 \label{eq:sec4-pf3-vacuum-interface}
\end{align}
Reading the vacuum coefficients from
Eqs.~\eqref{eq:sec4-inner-seeds} and \eqref{eq:sec4-outer-three-seeds}
therefore gives the physical $g$-functions
\begin{align}
    g_a=\frac{S^{B_3}_{a\mathbf v}}{\sqrt{S^{A_3}_{\mathbf v\mathbf v}}}\,,\qquad {\rm and}\quad 
    g_a^\mathsf w=\frac{\Psi_{a1}}{\sqrt{S^{A_3}_{\mathbf v\mathbf v}}}\,.
\end{align}
Define
\begin{align}
 g_0(\mathcal M_{5,6}\to \mathcal M_{4,5})=\sqrt\frac{5+\sqrt5}{4\sqrt3}\,,
 \label{eq:sec4-pf3-g0}
\end{align}
as a benchmark, which is the $g$-factor of the Gaiotto wall between $\mathcal M_{5,6}$ and $\mathcal M_{4,5}$ computed in~\eqref{eq:sec3-unitary-g-and-T} for $k=3$. Then all twelve candidates are summarized by
\begin{align}
 \begin{array}{c|c|c}
 \text{component}&\text{seed labels}&g\\\hline
 \text{inner}&110,112,114,121,123,125&g_0\\
 \text{inner}&211,213,215&\sqrt2\,g_0\\
 \text{outer}&110_\mathsf w,123_\mathsf w&\sqrt3\,g_0\\
 \text{outer}&213_\mathsf w&\sqrt6\,g_0
 \end{array}\,.
 \label{eq:sec4-pf3-g-table}
\end{align}
Further notice that the interface from $\mathcal M_{5,\,6}$ to Potts is defined by the Verlinde/charge conjugation $\mathbb Z_2$-orbifolding, so the seed brane should have $g$-factor,
\begin{align}
    g(R_{\rm seed})=g({\mathcal I_{\mathcal M_{5,6}\to \mathcal P_3}})\cdot g_0(\mathcal M_{5,6}\to \mathcal M_{4,5})=\sqrt 2g_0
    \label{eq:g-factor_orb}
\end{align}
Every outer seed has a larger value. Meanwhile, the 6 inner ones, including the vacuum-labeled seed, whose $g$-factors are $g_0$,  are also excluded. 

All  three remaining candidates $\{211,\,213,\,215\}$ can be the seed brane. The choice among them depends on the deformation in \eqref{eq:sec4-massless-phase} that we will explain at length in Sections~\ref{sec:odd-k-branes} and~\ref{sec:boundary-monodromy}. We will now impose an additional constraint on the seed brane: it is \emph{$\mathsf w$-invariant}. Let $\mathcal L_\mathsf w$ denote the unitary TDL action implementing $\mathsf w$ on  physical states in $B_3$. We require the seed brane to satisfy
\begin{align}
   \mathcal L_\mathsf w|R_{\rm seed}\rangle=|R_{\rm seed}\rangle\,.
    \label{eq:w-inv_seed}
\end{align}
Since $\mathsf w\,Q_0\mathsf w=-Q_0$, its action on all the inner Cardy candidates is
\begin{align}
 \mathcal L_\mathsf w|b_{[t,d,m]}(\varphi)\rangle
 =|b_{[t,d,-m]}(-\varphi)\rangle\,.
 \label{eq:sec4-pf3-full-reflection}
\end{align}
Further notice that, on each module $\mu$,
\begin{align}
 \left.e^{-2\pi iQ_0}\right|_\mu
 =(-1)^{e-1}e^{-i\pi n/3}\,.
 \label{eq:sec4-pf3-reflection-phase}
\end{align}
Multiplying the matrix elements of the S-matrix~\eqref{eq:sec4-B-Smatrix} by this phase gives
\begin{align}
 (-1)^{e-1}e^{-i\pi n/3}
 S^{B_3}_{[t,d,-m],\mu}
 =S^{B_3}_{[t,3-d,-m-1],\mu}\,.
 \label{eq:sec4-pf3-reflected-row}
\end{align}
Therefore we have $\mathcal L_\mathsf w$ acting on the 9 inner twisted Cardy states as
\begin{align}
    \mathcal L_\mathsf w\left|b_{[t,d,m]}(\pi)\right\rangle=\left|b_{[t,3-d,-m-1]}(\pi)\right\rangle
\end{align}
Using the equivalence relation, one finds
\begin{align}
\mathcal L_\mathsf w\left|b_{211}(\pi)\rangle=|b_{211}(\pi)\right\rangle\,,\quad
 \mathcal L_\mathsf w\left|b_{213}(\pi)\rangle=|b_{215}(\pi)\right\rangle\,,\,\quad
 \mathcal L_\mathsf w\left|b_{215}(\pi)\rangle=|b_{213}(\pi)\right\rangle\,.
 \label{eq:sec4-pf3-fixed-seed}
\end{align}
Overall, the selected seed brane for an undeformed RG brane in $A_3$ is isolated by the brane label $[2,\,1,\,1]$ with an inner automorphism $\Omega_\pi$ at the moduli point $\varphi_*=\pi$. We spell out this seed as well as the pullback RG brane as follows:
\begin{align}
 &\left|b_{211}(\varphi)\right\rangle_{B_3}
 =3^{-1/4}\sum_{\mu\in\mathscr P'}
       e^{i\pi m_\mu/3}|\mu;\Omega_\varphi\rangle\!\rangle\,,
      \qquad {\rm }\quad 
 \left|R(\varphi)\right\rangle_{A_3}=\check{\mathcal I}_{B_3\to A_3}\left|b_{211}(\varphi)\right\rangle_{B_3}\,,\notag\\
 &{\rm with}\quad \mathscr P'=\{110,\,112,\,114,\,121,\,123,\,125\}\,,
 \label{eq:sec4-pf3-selected-family}
\end{align}
where the coefficients of all three $t=2$ modules vanish. This remains true along the entire deformation, not only at the RG angle $\varphi_*$.

\FloatBarrier
\paragraph{RG brane spectrum.}
Now with the RG brane at hand, we aim to extract its spectrum encoded in its open-channel partition function. Recall from \eqref{eq:transported-annulus}
that \begin{align}
Z_{{\rm RG_\varphi},{\rm RG_\varphi}}^{A,\,\text{close}}&=
 {}_{A_3}\langle R(\varphi)|
 \widetilde q^{\frac12(L_0+\bar L_0-c_B/12)}
 |R(\varphi)\rangle_{A_3}\notag\\[5pt]
 &=
 {}_{B_3}\langle b_{211}(\varphi)|
 \widetilde q^{\frac12(L_0+\bar L_0-c_B/12)}
 (1+\mathcal L_\mathsf q)|b_{211}(\varphi)\rangle_{B_3}\,,
\end{align}
where we have used 
\begin{align}
    \mathcal I_{A_3\to B_3}\circ\check{\mathcal I}_{B_3\to A_3}=\mathcal L_\mathfrak b=\mathbf 1\oplus\mathcal L_{\mathsf q}
\end{align} 
Here $\mathcal L_{\mathsf q}$ is the quantum non-invertible symmetry line that emerges in $B_3$ after we gauge the diagonal Verlinde lines in $A_3$,
\begin{align}
    \mathcal L_\mathfrak a=\mathbf 1\oplus \operatorname{diag}\left(\mathcal L^{\rm TIM}_{1,3}\boxtimes\mathcal L^{\rm P}_{[3,0]}\right)\,,
\end{align}
where $\mathcal L^{\rm P}_{[3,0]}$ and $\mathcal L^{\rm TIM}_{1,3}$ are two Fibonacci lines transmitted through the RG wall. For later use, we summarize the eigenvalues of $\mathcal L^{\rm P}_{[3,0]}$ and $\mathcal L^{\rm TIM}_{1,3}$ on the primaries of the TIM and Potts in Table~\ref{tab:sec4-fibonacci-eigenvalues}.
\begin{table}[htbp]
 \centering
 \renewcommand{\arraystretch}{1.2}
 \begin{tabular}{c c c c|c c c c}
 \multicolumn{4}{c|}{TIM}&\multicolumn{4}{c}{three-state Potts}\\
 field&$(t,r)$&$h$&$\mathcal L^{\rm TIM}_{1,3}$
 &field&$[r,m]$&$h$&$\mathcal L^{\rm P}_{[3,0]}$\\\hline
 $1$&$(1,1)$&$0$&$\gamma$
 &$1$&$[1,0]$&$0$&$\gamma$\\
 $\epsilon$&$(1,2)$&$1/10$&$-\gamma^{-1}$
 &$\varepsilon$&$[3,0]$&$2/5$&$-\gamma^{-1}$\\
 $\epsilon'$&$(1,3)$&$3/5$&$-\gamma^{-1}$
 &$\psi$&$[1,2]$&$2/3$&$\gamma$\\
 $\epsilon''$&$(1,4)$&$3/2$&$\gamma$
 &$\psi^\dagger$&$[1,4]$&$2/3$&$\gamma$\\
 $\sigma_T$&$(2,2)$&$3/80$&$-\gamma^{-1}$
 &$\sigma_P$&$[2,1]$&$1/15$&$-\gamma^{-1}$\\
 $\sigma'_T$&$(2,1)$&$7/16$&$\gamma$
 &$\sigma_P^\dagger$&$[2,5]$&$1/15$&$-\gamma^{-1}$
 \end{tabular}
 \caption{Primary labels, conformal weights and
 Fibonacci-line eigenvalues.}
 \label{tab:sec4-fibonacci-eigenvalues}
\end{table}

To evaluate the action of $\mathcal L_{\mathsf q}$ on the modules in $B_3$, we first decompose an auxiliary module $\mathcal V^{B_3}_\mu$, with $\mu=[t,d,m]$, into $A_3$ representations as in~\eqref{eq:sec4-annulus-branching} for $k=3$:
\begin{align}
 \mathcal V^{B_3}_{[t,d,m]}\bigg|_{A_3}
 =\bigoplus_{\substack{r=1,\ldots,4\\t+d+r\ {\rm odd}}}
 \mathcal V^{\rm TIM}_{t,r}\otimes
 \overline{\mathcal V}^{\rm P}_{[r,m]}\,.
 \label{eq:branch_B3}
\end{align}
We further explicitly write $\mathcal L_{\mathsf q}$ in terms of its action on the chiral ``$L$'' and anti-chiral ``$R$'' pieces
\begin{align}
    \mathcal L_{\mathsf q}=\mathcal L^{L}\mathcal L^{R}\,.
\end{align}
They will act separately on the chiral and anti-chiral parts of the Ishibashi states, with eigenvalues given in the branching formula \eqref{eq:branch_B3} and data in Table~\ref{tab:sec4-fibonacci-eigenvalues}. More concretely, let $P_{\mu,r}$ be the orthogonal projector onto this summand (see Eq.~\eqref{eq:transport-chiral-maps}). Then $\mathcal L^{L/R}$ can be decomposed on each module $\mu$ as 
\begin{align}
    \mathcal L_\mu^{L/R}=\mathcal L_\mu=\sum_r\lambda_rP_{\mu,r}\,,\qquad
    {\rm with}\quad \lambda_r=\begin{cases}
        \gamma\,,\qquad\quad\ \ \, r=1,4\\
        -\gamma^{-1}\,,\qquad r=2,3
    \end{cases}
\end{align}
In addition, for each Ishibashi state, we have
\begin{align}
    |\mu;\,\Omega_{\varphi}\rangle\!\rangle\equiv \left(\Omega_\varphi\otimes\overline{\one}\right)|\mu\rangle\!\rangle\,.
\end{align}
Tracing out both the chiral and anti-chiral parts of the states, we reach 
\begin{align}
Z_{{\rm RG_\varphi},{\rm RG_\varphi}}^{A,\,\text{close}}
 =\frac1{\sqrt3}
 \sum_{\mu\in\mathscr P'}
 \operatorname{Tr}_{B_{3,\mu}}
 \left((\one+U_{\mu,\varphi}^{\dagger}\mathcal L_\mu
 U_{\mu,\varphi}\mathcal L_\mu)
 \widetilde q^{L_0-1/16}\right)\,,
 \label{eq:close_pf}
\end{align}
where $U_{\mu,\varphi}$ is the unitary representation of $\Omega_\varphi$ on each module $\mu$. It's worth noting that the quantum line action $\mathcal L_\mu$ does not commute with the inner automorphism rotation $U_{\mu,\varphi}$, and the physical spectrum can thus vary with respect to $\varphi$. A further modular transformation of \eqref{eq:close_pf} will give the open-channel partition function, from which the spectra can be read off.

At the totally reflective point, $\varphi=0$ and $U_{\mu,0}=1$, the annulus partition function can instead be written entirely in terms of product characters. Denote the TIM characters by $\chi_x^T$ and the Potts $W_3$ characters by $\chi_y^P$. The result is
\begin{align}
 Z_0(q)={}(2\chi_1^T+\chi_\epsilon^T+\chi_{\epsilon'}^T
                      +2\chi_{\epsilon''}^T)\chi_1^P
 +(\chi_1^T+3\chi_\epsilon^T+3\chi_{\epsilon'}^T
                +\chi_{\epsilon''}^T)\chi_\varepsilon^P\,.
 \label{eq:sec4-factorized-annulus}
\end{align}
In contrast, at $\varphi_*=\pi$, the RG point, the spectrum turns out to be much more involved. The identity part in Eq.~\eqref{eq:close_pf} has a known modular transform at every angle, $\chi_{110}^{B_3}(q)+\chi_{123}^{B_3}(q)$. Their first few terms are
\begin{align}
 \begin{aligned}
 \chi_{110}^{B_3}(q)
 &=q^{-1/16}(1+q+5q^2+9q^3+18q^4+\cdots)\,,\\
 \chi_{123}^{B_3}(q)
 &=q^{-1/16}(q^{1/2}+4q^{3/2}+6q^{5/2}+\cdots)\,.
 \end{aligned}
 \label{eq:sec4-known-open-sector}
\end{align}
In particular, the open spectrum contains the spin-one phantom current and five weight-two states of the auxiliary vacuum character at every modulus angle $\varphi$. The open string partition function can be in general written as
\begin{align}
    Z_{\rm RG_\varphi\,RG_{\varphi}}^{A_3,\,{\rm open}}(q)=\mathrm{Tr}_{\mathcal H_{\rm RG_\varphi\,RG_{\varphi}}}q^{L_0-\frac{1}{16}}=\chi_{110}^{B_3}(q)+\chi_{123}^{B_3}(q)+\sum_j n_j(\varphi)\chi_{h_j(\varphi)}^{\rm Vir}\,.
    \label{eq:sec4.1_open_spec}
\end{align} 

\begin{figure}[htbp]
 \centering
 \includegraphics[width=\textwidth]{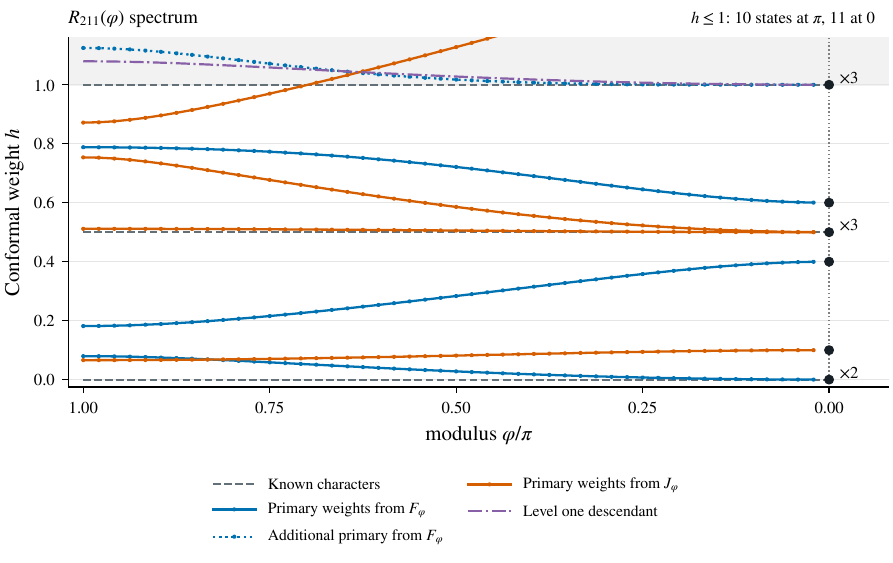}
 \caption{Self-annulus spectrum of $R_{211}(\varphi)$ between the RG angle $\varphi_*=\pi$ and the factorized point $\varphi=0$. Blue and red curves show primary weights extracted from the two remainders $F_\varphi$ and $J_\varphi$ defined in Appendix~\ref{app:open-spectrum}. Dashed lines are the known constant levels. The additional dotted primary and the descendant shown in purple approach $h=1$ from above. Black points give the independently known factorized spectrum, with multiplicities indicated when larger than one.}
 \label{fig:sec4-pf3-self-spectrum}
\end{figure}

Conformal weights and their multiplicities in the rest of the spectrum have to be extracted from the modularly transformed trace, using the numeric methods in Appendix~\ref{app:open-spectrum}. The low-lying weights at $\varphi=\pi$ are displayed in Table~\ref{tab:sec4-pf3-spectrum}. We also track the self-annulus spectrum of $R_{211}(\varphi)$ away from the RG angle, as shown in Figure~\ref{fig:sec4-pf3-self-spectrum}, while the modulus angle $\varphi$ runs from $\varphi_*=\pi$ towards $\varphi=0$, corresponding to the totally reflective brane. The constant levels at $h=0,1/2,1$ come from the known characters, while the other levels move with the modulus. At $\varphi=0$ the self-annulus again becomes Eq.~\eqref{eq:sec4-factorized-annulus}, implying its factorization: One moving level approaches zero, producing the second vacuum of the factorized boundary. Two others join the constant $h=1/2$ level, reproducing its multiplicity three. The higher levels also reorganize, and the number of states with $h\leq1$ changes from ten at the RG angle to eleven at the factorized endpoint.
\begin{table}[htbp]
 \centering
 \renewcommand{\arraystretch}{1.15}
 \begin{tabular}{c|c|c}
 boundary weight $h$&multiplicity&determination\\\hline
 $0$&$1$&vacuum\\
 $0.0660174$&$1$&numerical\\
 $0.0795566$&$1$&numerical\\
 $0.1816551$&$1$&numerical\\
 $1/2$&$1$&known character\\
 $0.5115305$&$1$&numerical\\
 $0.7537010$&$1$&numerical\\
 $0.7881030$&$1$&numerical\\
 $0.8717075$&$1$&numerical\\
 $1$&$1$&known character
 \end{tabular}
 \caption{Low-lying open-channel weights of the transported
 TIM--Potts wall with seed $211$ at $\varphi_*=\pi$. The exact entries are contributions of Eq.~\eqref{eq:sec4-known-open-sector}.}
 \label{tab:sec4-pf3-spectrum}
\end{table}

\FloatBarrier
\paragraph{UV--IR operator mixing along the modulus.}
We now evaluate bulk one-point functions along the full family to see how the UV and IR operators mix across the interface. Let $|i,\alpha\rangle$ be a normalized chiral product state of
$A_3$, with $i=(x,y)$ specifying a TIM and a Potts primary
label and $\alpha$ specifying a state within their families. As in Eq.~\eqref{eq:transport-chiral-maps}, the inclusion of the product module in an auxiliary module is denoted by $e_{\mu i}$:
\begin{align}
    &e_{\mu i}:\ \mathcal H_{A_3}\ \longrightarrow\ \mathcal H_{B_3}\,,\qquad {\rm and}\quad
    \check{e}_{i\mu}:\ \mathcal H_{B_3}\ \longrightarrow \ \mathcal H_{A_3}\,,\notag\\[6pt]
    &{\rm with}\quad \check{e}_{i\mu}e_{\mu j}=\delta_{ij}\,, \qquad{\rm and} \quad
    e_{\mu i}\check{e}_{i\nu}=\delta_{\mu\nu}P_{\nu,i}\,.
    \label{eq:e_echeck}
\end{align}
With Eqs.~\eqref{eq:e_echeck} and \eqref{eq:transport-interface-projector}, we can compute the operator mixing of two physical states in the tensor product of TIM and Potts, $|i,\alpha;\, \overline{i,\beta}\rangle$, from \eqref{eq:sec4-pf3-selected-family},
\begin{align}
 \mathcal B_{i;\alpha\beta}(\varphi)
 =\frac{{}_A\langle i,\alpha;\overline{i,\beta}|R(\varphi)\rangle_A}
        {{}_A\langle0|R(\varphi)\rangle_A}
 =e^{i\pi m_\mu/3}\sqrt{\frac{S^B_{\mathbf v\mu}S^{A}_{\mathbf v\mathbf v}}{S^B_{\mathbf v\mathbf v}S^A_{\mathbf vi}}}
 \langle i,\alpha|\check{e}_{i\mu} U_{\mu,\varphi}e_{\mu i}
                  |i,\beta\rangle\,.
 \label{eq:sec4-physical-mixing}
\end{align}
Equation~\eqref{eq:sec4-physical-mixing} applies to the six supported modules in Eq.~\eqref{eq:sec4-pf3-selected-family}, whereas all coefficients in the other three vanish.

In practice, it is more convenient to identify a tensor product state with a representative in the affine numerator of $B_3$. The coset construction further requires this state to be annihilated by the positive modes of the denominator current $H=J^3+j^3$. On such a numerator representative, from~\eqref{eq:sec4-relative-current}, the zero mode of $Q$ can be rewritten as
\begin{align}
    Q_0=J_0^3-\frac23H_0\equiv J_0^3-\frac{M}{3}\,,
\end{align}
where $M=2H_0$ is the chosen integer charge representative, with $M\equiv m_\mu\pmod6$. We then have
\begin{align}
 U_{\mu,\varphi}=e^{-i\varphi M/3}e^{i\varphi J^3_0}\,,\qquad
 M=(0,2,-2,1,3,-1)
 \label{eq:sec4-module-phase}
\end{align}
in the supported modules, in the order listed in Eq.~\eqref{eq:sec4-pf3-selected-family}. As a simple example, in module $112$ the states representing $(1,\psi)$ and $(\epsilon',\sigma_P^\dagger)$ are
\begin{align}
 |A_1\rangle=\frac{|J\rangle+|j\rangle}{\sqrt3}\,,\qquad
 |A_2\rangle=\frac{-|J\rangle+2|j\rangle}{\sqrt6}\,,\qquad
 |J\rangle=J^+_{-1}|0\rangle\,,\quad |j\rangle=j^+_{-1}|0\rangle\,.
 \label{eq:sec4-two-state-representatives}
\end{align}
Their numerator norms are $\langle J|J\rangle=2$ and
$\langle j|j\rangle=1$. The charges of $|J\rangle$ and $|j\rangle$ under $J^3_0$ are one and zero, giving
\begin{align}
 \left.e^{i\varphi J^3_0}\right|_{112}
 =\frac13\begin{pmatrix}
 1+2e^{i\varphi}&\sqrt2(1-e^{i\varphi})\\
 \sqrt2(1-e^{i\varphi})&2+e^{i\varphi}
 \end{pmatrix}\,.
 \label{eq:sec4-two-state-rotation}
\end{align}
Only its two diagonal entries contribute to the corresponding
physical primary overlaps, with the separate prefactors in
Eq.~\eqref{eq:sec4-physical-mixing}.

To collect the answer, we use the TIM primary order defined for the rows and the Potts order for the columns (see Table~\ref{tab:sec4-fibonacci-eigenvalues}). Also define the scalar functions
\begin{align}
 \begin{gathered}
 z=e^{i\varphi}\,,\qquad a(\varphi)=e^{i(\pi-\varphi)/3}\,,\qquad
 f(\varphi)=\frac{a(\varphi)^2(1+2z)}3\,,\\
 b(\varphi)=\frac{a(\varphi)^2(2+z)}3\,,\qquad
 v(\varphi)=\frac{a(\varphi)(2+z)}3\,,\qquad
 d(\varphi)=-\frac{3+4\cos\varphi}{7}\,.
 \end{gathered}
 \label{eq:sec4-primary-functions}
\end{align}
The normalized UV--IR mixing coefficients for primaries are given by
\begin{align}
 \mathcal B_{\rm primary}(\varphi)=
 \begin{pmatrix}
 1&0&f&f^*&0&0\\
 0&-\gamma^{-1}&0&0&a/\gamma&a^*/\gamma\\
 0&\gamma^{-1}&0&0&b^*/\gamma&b/\gamma\\
 d&0&v^*&v&0&0\\
 0&0&0&0&0&0\\ 0&0&0&0&0&0
 \end{pmatrix}\,,
 \label{eq:sec4-primary-mixing}
\end{align}
where the star ``*'' denotes complex conjugation. At the RG angle $\varphi_*=\pi$, $a=1$, $f=-1/3$, $b=v=1/3$ and $d=1/7$. We reproduce the UV--IR operator mixings of primaries evaluated in~\cite{Zhang:2026gqp} via a duality based analysis only available for $k=3$. Our RG brane construction yields these overlaps directly and extends them to arbitrary defect moduli $\varphi$. The construction of the states, unlike that special fermionic description, also applies to general $k$.

\begin{figure}[tbp]
 \centering
 \begin{tikzpicture}[
  x=1cm,y=1cm,
  panel edge/.style={draw=black!55,line width=0.45pt},
  top wall/.style={draw=SpinTop,line width=1.65pt,
    dash pattern=on 3pt off 1.8pt},
  rg wall/.style={draw=SpinRG,line width=1.8pt},
  motion/.style={-{Stealth[length=2mm,width=1.4mm]},
    black!75,line width=0.8pt},
  theory label/.style={font=\small,inner sep=1pt},
  field label/.style={font=\small,inner sep=2pt},
  wall label/.style={font=\small,anchor=north,inner sep=2pt},
  panel label/.style={font=\footnotesize,align=center}
 ]
  \definecolor{SpinPotts}{RGB}{226,238,249}
  \definecolor{SpinM}{RGB}{255,249,204}
  \definecolor{SpinTIM}{RGB}{247,226,226}
  \definecolor{SpinTop}{RGB}{81,143,114}
  \definecolor{SpinRG}{RGB}{181,43,114}
  \definecolor{SpinC}{RGB}{81,143,114}
  \foreach \offset in {0,5.2,10.4} {
   \begin{scope}[xshift=\offset cm]
    \path[fill=SpinPotts] (0,0) rectangle (1.5,2.25);
    \path[fill=SpinM] (1.5,0) rectangle (3,2.25);
    \path[fill=SpinTIM] (3,0) rectangle (4.5,2.25);
    \draw[panel edge] (0,0) rectangle (4.5,2.25);
    \draw[top wall] (1.5,0) -- (1.5,2.25);
    \draw[rg wall] (3,0) -- (3,2.25);
    \node[theory label] at (0.75,1.94) {Potts};
    \node[theory label] at (2.25,1.94) {$\mathcal M_{5,6}$};
    \node[theory label] at (3.75,1.94) {TIM};
    \node[wall label] at (1.5,-0.10) {$\mathcal I_{\rm top}$};
    \node[wall label] at (3,-0.10) {$\mathcal I_{\rm RG}$};
   \end{scope}
  }
  \fill (3.88,0.96) circle (2pt);
  \node[field label,anchor=south] at (3.88,1.10) {$\sigma_T$};
  \draw[motion] (3.70,0.96) -- (2.67,0.96);
  \node[panel label] at (2.25,-0.72) {(a) TIM spin insertion};
  \begin{scope}[xshift=5.2cm]
   \fill (2.38,0.96) circle (2pt);
   \node[field label,anchor=south] at (2.38,1.10) {$\Phi_{2,2}$};
   \draw[motion] (2.20,0.96) -- (1.17,0.96);
   \node[panel label] at (2.25,-0.72) {(b) A nonzero mixing channel};
  \end{scope}
  \begin{scope}[xshift=10.4cm]
   \draw[SpinC,line width=1.4pt] (0.48,0.96) -- (1.5,0.96);
   \fill (0.48,0.96) circle (2pt);
   \fill[SpinC] (1.5,0.96) circle (1.8pt);
   \node[field label,anchor=south] at (0.48,1.10) {$\mu_{\mathsf c}$};
   \node[field label,anchor=north] at (1.04,0.82)
    {$\mathcal L_{\mathsf c}$};
   \node[panel label] at (2.25,-0.72) {(c) Potts defect insertion};
  \end{scope}
 \end{tikzpicture}
 \caption{A TIM spin-mixing channel through the composed Potts--$\mathcal M_{5,6}$--TIM interface. The green dashed and magenta  lines denote $\mathcal I_{\rm top}$ and $\mathcal I_{\rm RG}$. The arrows follow a nonzero mixing channel from $\sigma_T$ to
 $\Phi_{2,2}$, not an operator identity across the RG wall. Gauging turns this $\mathsf v$-odd field into the Potts defect field $\mu_{\mathsf c}$, with $h=\bar h=1/40$, attached to the charge conjugation line. The line ends on $\mathcal I_{\rm top}$. The $\Phi_{2,4}$ channel is treated in the same way. Neither channel contributes to ordinary local Potts--TIM spin overlaps of $R_{211}$.}
 \label{fig:sec4-spin-orbifold-transport}
\end{figure}
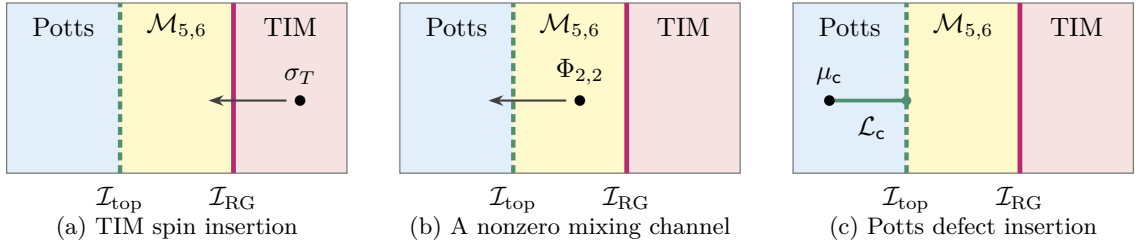

Those vanishing one-point correlators in \eqref{eq:sec4-primary-mixing} also have interesting interpretations in terms of the selection rules for invertible and non-invertible symmetries imposed on both the operators and the RG boundary: First, the Fibonacci line is transmitted through the RG brane for every value of its modulus $\varphi$. So UV and IR operators with different Fibonacci charges, $\gamma$ and $-\gamma^{-1}$, cannot mix with each other and thus have vanishing one-point correlators. We can also understand the vanishing $\sigma_T$ and $\sigma_T'$ rows, by resolving this interface into the Potts--$\mathcal M_{5,6}$ topological interface $\mathcal I_{\rm top}$ and the $\mathcal M_{5,6}$--TIM Gaiotto wall $\mathcal I_{\rm RG}$, as depicted in Figure~\ref{fig:sec4-spin-orbifold-transport}. Across the latter, $\sigma_T$ has nonzero mixing with the $\mathcal M_{5,6}$ primaries $\Phi_{2,2}$ and $\Phi_{2,4}$ of weights $h=\bar h=1/40$ and $21/40$, respectively. Both fields are odd under the gauged line $\mathcal L_\mathsf v=\mathcal L^{\mathcal M_{5,6}}_{1,5}$. They therefore correspond to endpoints $\mu_\mathsf c$ of the Potts charge conjugation line $\mathcal L_{\mathsf c}$, rather than ordinary local Potts fields. These defect fields retain nonzero spin-mixing channels, whereas the ordinary-sector projection removes them. This gives the two zero spin rows of Eq.~\eqref{eq:sec4-primary-mixing} for the $R_{211}$ family.

Descendant mixings are also obtained by the same approach. For example, for the  descendants at level one of the state $|\epsilon\varepsilon\rangle$, take the orthonormal states
\begin{align}
 \left(\sqrt5L_{-1}^{\rm IR}|\epsilon\varepsilon\rangle,
 \frac{\sqrt5}{2}L_{-1}^{\rm UV}|\epsilon\varepsilon\rangle,
 |\epsilon\phi^P_{7/5}\rangle\right)\,,
 \label{eq:sec4-descendant-basis}
\end{align}
where $\phi^P_{7/5}$ is the normalized Virasoro primary in the Potts $W^{(3)}$-module of the energy operator $\varepsilon$. The operator mixing block can be found as
\begin{align}
 \mathcal B_{\epsilon\varepsilon}(\varphi)
 =-\frac1\gamma
 \begin{pmatrix}
 \dfrac{1+4c_\varphi}{5}&\dfrac{2(1-c_\varphi)}5
                         &2i\sqrt{\dfrac3{35}}s_\varphi\\[3pt]
 \dfrac{2(1-c_\varphi)}5&\dfrac{4+c_\varphi}{5}
                         &-i\sqrt{\dfrac3{35}}s_\varphi\\[3pt]
 2i\sqrt{\dfrac3{35}}s_\varphi&-i\sqrt{\dfrac3{35}}s_\varphi
                         &\dfrac{4+3c_\varphi}{7}
 \end{pmatrix}\,,\qquad
 c_\varphi=\cos\varphi\,,\quad s_\varphi=\sin\varphi\,.
 \label{eq:sec4-descendant-mixing}
\end{align}
The imaginary entries describe mixings with the additional Potts Virasoro primary away from the RG angle. These mixings disappear at $\varphi_*=\pi$.

\subsection{RG branes for odd \texorpdfstring{$k$}{k}}
\label{sec:odd-k-branes}

So far, to construct an RG wall, we have only considered the preserved Verlinde lines transmitted through the wall. For the flow from the $\mathbb Z_k$-parafermion theory to a minimal model, the charge conjugation automorphism $\mathsf c$ of the parafermion theory, corresponding to a non-Verlinde line $\mathcal L_\mathsf c$, also plays an important role and supplies a further condition on the RG boundary when $k$ is odd.

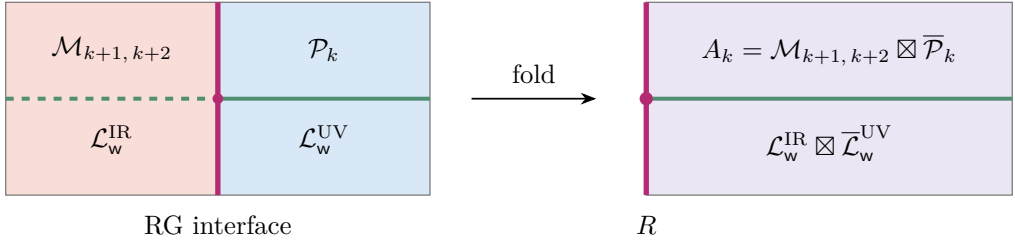
\begin{figure}[htbp]
 \centering
 \centering
 \definecolor{SecFourUV}{RGB}{217,232,247}
 \definecolor{SecFourIR}{RGB}{249,221,217}
 \definecolor{SecFourFold}{RGB}{235,229,244}
 \definecolor{SecFourWall}{RGB}{181,43,114}
 \definecolor{SecFourLine}{RGB}{81,143,114}
 \begin{tikzpicture}[x=1.1cm,y=1cm,>=Stealth,font=\small]
  \fill[SecFourIR] (0,0) rectangle (2.55,2.55);
  \fill[SecFourUV] (2.55,0) rectangle (5.1,2.55);
  \draw[black!55,line width=.45pt] (0,0) rectangle (5.1,2.55);
  \draw[SecFourWall,line width=2pt] (2.55,0)--(2.55,2.55);
  \node at (1.275,1.92) {$\mathcal M_{k+1,\,k+2}$};
  \node at (3.825,1.92) {$\mathcal P_k$};
  \draw[SecFourLine,line width=1.5pt,dashed] (0,1.275)--(2.55,1.275);
  \draw[SecFourLine,line width=1.5pt] (2.55,1.275)--(5.1,1.275);
  \node[below=6pt] at (1.275,1.275) {$\mathcal L_{\mathsf w}^{\rm IR}$};
  \node[below=6pt] at (3.825,1.275) {$\mathcal L_{\mathsf w}^{\rm UV}$};
  \fill[SecFourWall] (2.55,1.275) circle (2pt);
  \node[below=4pt] at (2.55,0) {RG interface};
  \draw[->,line width=.65pt] (5.6,1.275)--(7.1,1.275);
  \node[above] at (6.35,1.35) {fold};
  \fill[SecFourFold] (7.7,0) rectangle (12.1,2.55);
  \draw[black!55,line width=.45pt] (7.7,0) rectangle (12.1,2.55);
  \draw[SecFourWall,line width=2pt] (7.7,0)--(7.7,2.55);
  \node at (9.9,1.92) {$A_k=\mathcal M_{k+1,\,k+2}\boxtimes\overline{\mathcal P}_k$};
  \draw[SecFourLine,line width=1.5pt] (7.7,1.275)--(12.1,1.275);
  \fill[SecFourWall] (7.7,1.275) circle (2.5pt);
  \node[below=6pt] at (9.9,1.275)
       {$\mathcal L_{\mathsf w}^{\rm IR}\boxtimes
          \overline{\mathcal L}_{\mathsf w}^{\rm UV}$};
  \node[below=4pt] at (7.7,0) {$R$};
 \end{tikzpicture}
 \caption{The charge conjugation $\mathsf c$-line crosses the RG interface and, after folding, ends on the RG boundary. The dashed IR segment records its IR realization.}
 \label{fig:sec4-symmetry-endpoint}
\end{figure}

First recall that, in \eqref{eq:sec4-massless-phase}, the perturbation operators $\phi^{\rm UV}_{[1,\pm2]}$ are charged under the $\mathbb Z_k$-symmetry. Denoting its generator by $\mathsf r$, the corresponding TDL is
\begin{align}
    \mathcal L_\mathsf r\equiv \mathcal L_{[1,2]}\,,\qquad
    {\rm and}\quad \mathcal L_{\mathsf r^{-1}}=\mathcal L_\mathsf r^{-1}=\mathcal L_{[1,-2]}\,.
\end{align}
The operator $\phi^{\rm UV}_{[1,\pm2]}$ has a $\mathsf r$-charge of $\pm2$,
\begin{align}
 \mathcal L_{\mathsf r}\,\phi^{\rm UV}_{[1,\pm2]} \mathcal L_{\mathsf r^{-1}}=e^{\pm\frac{4\pi i}{k}}\phi^{\rm UV}_{[1,\pm2]}\,.
 \label{eq:sec4-perturbation-symmetry}
\end{align}
So, when $k$ is odd, one can always rotate the perturbation to 
\begin{align}
S=S_{\mathcal P_k}+\lambda\int d^2x\,
 \left(\phi_{1,2}^{\rm UV}+\phi_{1,-2}^{\rm UV}\right)\,,
 \qquad \lambda>0\,,
 \label{eq:Pk_deform}
\end{align}
which still describes a massless flow studied in~\cite{Fateev:1991bv}. Obviously, Eq.~\eqref{eq:Pk_deform} is invariant under charge conjugation $\mathsf c$:
\begin{align}
 \mathcal L_\mathsf c\,\phi^{\rm UV}_{[1,\pm2]}\,\mathcal L_\mathsf c^{-1}=\,\phi^{\rm UV}_{[1,\mp2]}\,.
\end{align}
Therefore the charge conjugation line $\mathcal L_\mathsf c$ can be moved through an RG interface into its IR realization. After folding, the pair of line segments becomes a line ending on the boundary, as shown in Fig.~\ref{fig:sec4-symmetry-endpoint}. If the IR segment is trivial, it is easy to show that the line $\mathcal L_\mathsf c$ will be mapped to the automorphism $\mathsf w$ in the auxiliary theory $B_k$, i.e.
\begin{align}
\mathcal L_\mathsf c\ \longrightarrow\ \mathcal L_{\mathsf w}\,.
\end{align}
This therefore implies that the $\mathsf w$-line can end on the seed brane, giving the condition in Eq.~\eqref{eq:w-inv_seed}:
\begin{align}
    \mathcal L_\mathsf w|R_{\rm seed}\rangle_{B_k}=|R_{\rm seed}\rangle_{B_k}
\end{align}

Now, for generic odd $k$, on an ordinary module $\mu=[t,d,m]$, we have 
\begin{align}
    e^{-2\pi iQ_0}\big|_\mu=(-1)^{d-1}e^{-i\pi m/k}\,.
    \label{eq:sec4-module-monodromy}
\end{align}
Charge conjugation reverses $Q_0$ and maps $m$ to $-m$. Combining Eq.~\eqref{eq:sec4-module-monodromy} with the modular matrix entries in Eq.~\eqref{eq:sec4-B-Smatrix}, at $\varphi=\pi$ we obtain
\begin{align}
 \mathcal L_\mathsf w|b_{[t,d,m]}(\pi)\rangle
 =|b_{[t,3-d,-m-1]}(\pi)\rangle\,.
 \label{eq:sec4-seed-reflection}
\end{align}
Choosing representatives with $d=1$, we can then use the field identification to write this action as
\begin{align}
 (t,m)\longmapsto(k+1-t,k-1-m\bmod2k)\,.
 \label{eq:sec4-row-involution}
\end{align}
An invariant seed must satisfy $t=(k+1)/2$ and $2m=k-1\pmod{2k}$. The selected one is therefore
\begin{align}
 a_*(k)=\left[\frac{k+1}{2},1,\frac{k-1}{2}\right]\,.
 \label{eq:sec4-odd-seed}
\end{align}

\paragraph{The three Potts coupling phases.}
In fact, the choice of the seed brane also depends on the non-trivial phase of the deformation couplings. For simplicity, we here only consider the case of $k=3$. For arbitrary odd $k$, the generalization is straightforward. For a generic relevant deformation considered in~\cite{Fateev:1991bv}, the action is written as
\begin{align}
 S_\theta=S_{\mathcal P_3}-\lambda\int d^2x\,
 \left(e^{i\theta/3}\phi_{[1,2]}^{\rm UV}+
       e^{-i\theta/3}\phi_{[1,-2]}^{\rm UV}\right)\,,
 \qquad \lambda>0\,,\qquad \theta\sim\theta+6\pi\,.
 \label{eq:sec4-three-phase-action}
\end{align}
Here $\theta$ is a bulk coupling phase with the identification $\theta\sim\theta+6\pi$. The RG flow is massless when $\theta=\pi,\,3\pi,$ and $5\pi$, and otherwise massive. The three massless angles can be rotated by the action of $\mathcal L_\mathsf r$ and connected to each other. However, at different phases $\theta$, the preserved symmetries are also different. The positive real perturbation~\eqref{eq:Pk_deform} at $\theta=3\pi$, which preserves charge conjugation $\mathsf c$, was discussed above. On the other hand, one can easily check that, for $\theta=\pi$ and $5\pi$, the preserved additional symmetries are respectively $\mathsf {c\, r}$ and $\mathsf {c\,r}^{-1}$, see Table~\ref{tab:sec4-three-phase-seeds}. 

\begin{table}[htbp]
 \centering
 \renewcommand{\arraystretch}{1.25}
 \begin{tabular}{c|c|c|c|c}
 $\theta$ & $e^{i\theta/3}$ & Preserved UV symmetry
 & Lines in $B_3$ & Seed at $\Omega_\pi$\\ \hline
 $\pi$ & $e^{i\pi/3}$ & $\mathsf c\,\mathsf r$
 & $\mathcal L_{\mathsf w}\mathcal L_{112}^{B_3}$ & $213$\\
 $3\pi$ & $-1$ & $\mathsf c$
 & $\mathcal L_\mathsf w$ & $211$\\
 $5\pi$ & $e^{-i\pi/3}$ & $\mathsf c\,\mathsf r^{-1}$
 & $\mathcal L_{\mathsf w}\mathcal L_{114}^{B_3}$ & $215$
 \end{tabular}
 \caption{Massless Potts perturbations with different phases determine distinguished seed branes.}
 \label{tab:sec4-three-phase-seeds}
\end{table}

The $\mathsf r$-line lifts through the extension interface $\mathcal I_{A_3\to B_3}$ to 
\begin{align}    \mathcal L_\mathsf r\ \longrightarrow\ \mathcal L_{112}^{B_3}\,,\qquad 
    \mathcal L_{\mathsf r^{-1}}\ \longrightarrow\ \mathcal L_{114}^{B_3}\,.
\end{align}
One then can find that the line $\mathcal L_{\mathsf w}\mathcal L_{112}^{B_3}$ fixes the brane $213$ and $\mathcal L_{\mathsf w}\mathcal L_{114}^{B_3}$ fixes $215$ (see Table~\ref{tab:sec4-three-phase-seeds}). The $\mathsf r$-line also transports the three boundaries as
\begin{align}
 |R_{211}\rangle\xrightarrow{\ \mathsf r\ }|R_{213}\rangle
 \xrightarrow{\ \mathsf r\ }|R_{215}\rangle
 \xrightarrow{\ \mathsf r\ }|R_{211}\rangle\,.
 \label{eq:sec4-three-seed-orbit}
\end{align}
In the ordered basis $(211,213,215)$, they furnish a regular NIM-representation of the $\mathbb Z_3$-symmetry category
\begin{align}
 n_\mathsf r=\begin{pmatrix}0&1&0\\0&0&1\\1&0&0\end{pmatrix}\,,
 \qquad n_{\mathsf r^2}=n_\mathsf r^2\,,\qquad n_\mathsf r^3=\one\,,
 \qquad \mathcal L_\mathsf r|R_{a_i}\rangle=\sum_j(n_\mathsf r)_{ij}|R_{a_j}\rangle\,,
 \label{eq:sec4-three-seed-nimrep}
\end{align}
where $a_j=[2,1,1+2j]$, $j=0,1,2$. Clearly, the rotation relates the three coupling phases. For each distinguished phase, a different RG brane is chosen.

The UV--IR operator mixing coefficients for three different RG branes can be also related directly through the seed brane relations: For every supported module $\mu=[u,e,m_\mu]$, the coefficients of the Cardy brane $R_{a_j}$ differ from those of $R_{211}$ by $\omega^{j m_\mu}$, and all others are unchanged. Defining $\mathcal B^{(a_j)}_{xy}$ by the normalized one-point correlator as in Eq.~\eqref{eq:sec4-physical-mixing}, we can evaluate the three mixing coefficient matrices for primaries:
\begin{align}
 \mathcal B^{(a_j)}=
 \begin{pmatrix}
 1&0&-\omega^{2j}/3&-\omega^j/3&0&0\\
 0&-\gamma^{-1}&0&0&\omega^j/\gamma&\omega^{2j}/\gamma\\
 0&\gamma^{-1}&0&0&\omega^j/(3\gamma)&\omega^{2j}/(3\gamma)\\
 1/7&0&\omega^{2j}/3&\omega^j/3&0&0\\
 0&0&0&0&0&0\\
 0&0&0&0&0&0
 \end{pmatrix}\,,\qquad j=0,1,2\,,
 \label{eq:sec4-three-seed-mixing}
\end{align}
where $\omega=e^{2\pi i/3}$ is a cubic root of unity. Defining the charge matrix
$D_\mathsf r=\operatorname{diag}(1,1,\allowbreak\omega^2,\omega,\omega,\omega^2)$, we have
\begin{align}
 \mathcal B^{(213)}=\mathcal B^{(211)}D_\mathsf r\,,\qquad
 \mathcal B^{(215)}=\mathcal B^{(211)}D_\mathsf r^2
 =\bigl(\mathcal B^{(213)}\bigr)^*\,.
 \label{eq:sec4-three-seed-mixing-relations}
\end{align}
In addition, all three boundaries have the same entropy, transmission and self-annulus spectrum, but differ in the phases of one-point correlators of charged operators.

\paragraph{The infrared realization of charge conjugation}
In previous discussions, we assumed that the charge conjugation line transmitted through the RG domain wall becomes a trivial line acting on the IR theory. Actually preservation of the UV reflection does not by itself determine its IR realization. Another candidate is that the $\mathsf c$-line becomes the $\mathbb Z_2$-Verlinde line $\mathsf v$, $\mathcal L_\mathsf v=\mathcal L_{1,k+1}^{\mathcal M_{k+1,k+2}}$, in the minimal model. After folding, the corresponding line is $\mathsf v\boxtimes\overline{\mathsf c}$. In the auxiliary theory $B_k$, the $\mathsf c$-line is identified with the $\mathsf w$-line, while the line $\mathcal L_\mathsf v$ can be checked to map to $\mathcal L_{[1,2,k]}^{B_k}$ for odd $k$.

This possibility will impose different endpoint constraints on a compatible seed and thus on the RG boundary as follows:
\begin{align}
 \begin{aligned}
 &\mathcal L_\mathsf c\longrightarrow\one:
 \qquad \ \,\mathcal L_\mathsf w|b\rangle=|b\rangle\,,\\[4pt]
 &\mathcal L_\mathsf c\longrightarrow \mathcal L_\mathsf v:
 \qquad \mathcal L^{B_k}_{[1,2,k]}\,\mathcal L_\mathsf w|b\rangle=|b\rangle\,.
 \end{aligned}
 \label{eq:sec4-reflection-ir-matching}
\end{align}
In the case of $k=3$, among the inner seeds at $\varphi=\pi$, the first condition selects the brane $211$, whereas the second also admits $114$ and $121$, besides $211$. Both $\mathcal L_\mathsf w$ and $\mathcal L^{B_3}_{123}$ exchange $114$ with $121$, so their combination fixes both. Meanwhile the seed $211$ is fixed by the two operations separately. Its additional invariance forces the one-point correlators involving $\sigma_T$ and $\sigma_T'$ to vanish, as seen in Eq.~\eqref{eq:sec4-primary-mixing}. In contrast, the combined symmetry alone allows a Potts $\mathsf c$-odd operator to mix with a TIM $\mathsf v$-odd one, as suggested in~\cite{Klassen:1991dz}. This scenario suggests that the chosen seed brane should be $114$ or $121$, enabling non-zero mixings between $\mathsf c$- and $\mathsf v$-odd fields. However, this is not consistent with the orbifolding picture and $g$-factor computation (see Eqs.~\eqref{eq:sec4-pf3-g-table}, \eqref{eq:g-factor_orb} and Figure~\ref{fig:sec4-spin-orbifold-transport}).

We retain $R_{211}$ and its odd-$k$ generalization as the $\mathcal L_\mathsf c$-invariant family studied here. Identifying the microscopic RG wall requires additional physical handles, including IR symmetry matching and the associated junction data, or independently obtained data for some UV--IR operator mixings. Nevertheless, the relations between the three candidate branes are clear: If the IR realization of the $\mathsf c$-line in Potts is the $\mathsf v$-line in TIM, orbifolding the $\mathsf c$-line requires simultaneously orbifolding the $\mathsf v$-line. Notice that TIM is self-dual under gauging the $\mathsf v$-line, but its gauging interface is the non-invertible duality line $\mathcal N=\mathcal L^{\rm TIM}_{2,1}$, with fusion rule $\mathcal N^{\,2}=\one+\mathcal L_\mathsf v$. It then modifies the previous orbifolding picture, see Figure~\ref{fig:sec4-spin-orbifold-transport}, by fusing a further $\mathcal N$-line on the brane $R_{211}$, and gives
\begin{align}
 \mathcal N|R_{211}(\pi)\rangle
 =|R_{114}(\pi)\rangle+|R_{121}(\pi)\rangle\,.
 \label{eq:sec4-ir-gauging-resolution}
\end{align}
As a consistency check, one can find that the $g$-factors match well from Eq.~\eqref{eq:sec4-pf3-g-table}, as the $\mathcal N$-line has quantum dimension $g_\mathcal N=\sqrt 2$.

\FloatBarrier

\subsection{Monodromy of the boundary families}
\label{sec:boundary-monodromy}

In previous subsections, we discussed in detail  how to pick out the RG brane, and we discussed its physical properties. In fact, the approach we have developed can be applied to construct generic non-factorizable conformal interfaces between two distinct CFTs. In this subsection, we will disclose an interesting monodromy property that connects these conformal interfaces with the same $g$-factors, when varying the defect modulus $\varphi$.

In fact the gluing automorphism does not retain all the information in a boundary state. Although $\Omega_\varphi$ is $2\pi$-periodic on the vacuum module, its implementation on a non-vacuum module can return with a phase. This phase results in a non-trivial monodromy of a boundary state when the defect conformal modulus $\varphi$ completes one revolution. We demonstrate this phenomenon for the conformal interfaces between Potts and TIM, where there are 9 inner-twisted and 3 outer-twisted Cardy branes with $g$-factors summarized in~\eqref{eq:sec4-pf3-g-table}. Generalization to arbitrary $k$ is straightforward.

\begin{figure}[htbp]
 \centering
\begin{tikzpicture}[x=1cm,y=1cm,
  font=\fontsize{10}{12}\selectfont,
  >={Stealth[length=2.2mm,width=1.6mm]},
  every node/.style={inner sep=2pt},
  brane arc/.style={line width=1.6pt,line cap=round},
  direction/.style={line width=1.6pt,-Stealth},
  marked point/.style={circle,fill=black,draw=white,line width=.6pt,
    minimum size=5.2pt,inner sep=0pt}]
 \definecolor{MonoWall}{RGB}{181,43,114}
 \definecolor{MonoBlue}{RGB}{93,137,175}
 \definecolor{MonoGreen}{RGB}{81,143,114}
 \path[use as bounding box] (0,-2.7) rectangle (13.4,3.5);

 \coordinate (B) at (3.0,.15);
 \coordinate (G) at (10.6,.15);
 \node at (3.0,3.05) {Brane family};
 \node at (10.6,3.05) {Defect moduli};

 \draw[brane arc,MonoWall] ($(B)+(90:1.7)$)
   arc[start angle=90,end angle=210,radius=1.7];
 \draw[brane arc,MonoBlue] ($(B)+(210:1.7)$)
   arc[start angle=210,end angle=330,radius=1.7];
 \draw[brane arc,MonoGreen] ($(B)+(330:1.7)$)
   arc[start angle=330,end angle=450,radius=1.7];
 \draw[direction,MonoWall] ($(B)+(137:1.7)$)
   arc[start angle=137,end angle=160,radius=1.7];
 \draw[direction,MonoBlue] ($(B)+(257:1.7)$)
   arc[start angle=257,end angle=280,radius=1.7];
 \draw[direction,MonoGreen] ($(B)+(377:1.7)$)
   arc[start angle=377,end angle=400,radius=1.7];
 \node[align=center] at (B) {$\varphi\bmod 6\pi$};

 \node[marked point] at ($(B)+(90:1.7)$) {};
 \node[marked point] at ($(B)+(210:1.7)$) {};
 \node[marked point] at ($(B)+(330:1.7)$) {};
 \node[align=center,anchor=south] at (3.0,1.98)
   {$R_{211}(\pi)$};
 \node[align=center,anchor=north] at (1.10,-.92)
   {$R_{215}(\pi)$};
 \node[align=center,anchor=north] at (4.90,-.92)
   {$R_{213}(\pi)$};

 \draw[->,black!75,line width=.8pt] (5.35,.42)--(8.43,.42);
 \node[align=center] at (6.89,.91)
   {threefold cover};

 \draw[black!60,line width=1.2pt] (G) circle[radius=1.35];
 \draw[black!60,line width=1.2pt,->] ($(G)+(142:1.35)$)
   arc[start angle=142,end angle=168,radius=1.35];
 \node at (G) {$\varphi\bmod 2\pi$};
 \node[marked point] at ($(G)+(90:1.35)$) {};
 \node[align=center,anchor=south] at (10.6,1.7)
   {$\varphi_*=\pi$};
%
\end{tikzpicture}%
 \caption{The connected family $R_{211}(\varphi)$ is a threefold cover of the defect conformal moduli space $\rm U(1)$ parametrized by $\varphi$. The three marked boundaries lie over the same $\Omega_\pi$: at $\varphi=\pi,3\pi,5\pi$ they are  $R_{211}(\pi)$, $R_{215}(\pi)$ and $R_{213}(\pi)$, respectively. Each coloured arc maps onto one full circuit of the gluing circle. The full boundary state returns after a period of $6\pi$.}
 \label{fig:sec4-brane-cover}
\end{figure}
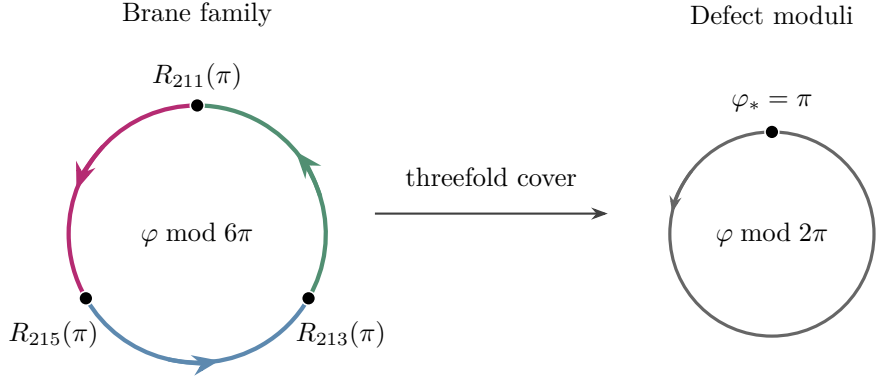

\paragraph{The two inner families at $k=3$.}
Recall that, for an ordinary module $\mu=[u,e,n]$, with $u=1,2,3$, $e=1,2$ and
$n\in\mathbb Z_6$ subject to Eq.~\eqref{eq:sec4-B-labels}, the relation
$Q_0=H_0/3-j^3_0$ gives
\begin{align}
 \left.e^{2\pi iQ_0}\right|_\mu=(-1)^{e-1}e^{i\pi n/3}\,.
 \label{eq:sec4-positive-monodromy-phase}
\end{align}
Multiplying all the  S-matrix entries by
this phase, using Eq.~\eqref{eq:sec4-B-Smatrix}, gives
\begin{align}
 \begin{split}
 \left|b_{[t,d,m]}(\varphi+2\pi)\right\rangle
 =\left|b_{[t,3-d,m+1]}(\varphi)\right\rangle=\left|b_{[4-t,d,m+4]}(\varphi)\right\rangle \,,
 \end{split}
 \label{eq:sec4-potts-period-label}
\end{align}
where the second equality uses field identification. The topological interface $\check{\mathcal I}_{B_3\to A_3}$ preserves this identity when applied to the seeds. For the physical boundaries in the tensor product of Potts and TIM, we find two orbits of periods $12\pi$ and $6\pi$, starting from the RG angle $\varphi_*=\pi$:
 \begin{align}
 &R_{211}\longrightarrow R_{215}\longrightarrow R_{213}\longrightarrow R_{211}\,,\\[5pt] 
 &R_{110}\longrightarrow R_{121}\longrightarrow R_{112}\longrightarrow R_{123}
       \longrightarrow R_{114}\longrightarrow R_{125}\longrightarrow R_{110} \,,
 \label{eq:sec4-potts-monodromy}
 \end{align}
where every arrow denotes an increment $\Delta\varphi=2\pi$, with the resulting brane expressed at the original RG angle $\varphi_*=\pi$. The threefold cover for the $211$ family is illustrated in Figure~\ref{fig:sec4-brane-cover}. One circuit of the gluing circle connects two different boundaries. The other six boundaries give the analogous sixfold cover.

\paragraph{The outer families at $k=3$.}
For the outer gluing $\Omega_\varphi\mathsf w$, we follow the notation~\eqref{eq:w-twist-cardy} and~\eqref{eq:sec4-outer-three-seeds},
\begin{align}
    \left|R_{110_\mathsf w}(\varphi)\right\rangle=\check{\mathcal I}_{B_3\to A_3}\left|b_1^{\mathsf w}(\varphi)\right\rangle\,,
    \left|R_{123_\mathsf w}(\varphi)\right\rangle=\check{\mathcal I}_{B_3\to A_3}\left|b_2^{\mathsf w}(\varphi)\right\rangle\,,
    \left|R_{213_\mathsf w}(\varphi)\right\rangle=\check{\mathcal I}_{B_3\to A_3}\left|b_3^{\mathsf w}(\varphi)\right\rangle
\end{align}

Write $W_\mu$ for the fixed intertwiner implementing $\mathsf w$ on a supported module. The Ishibashi state is $(U_{\mu,\varphi}W_\mu\otimes\overline{\one})|\mu\rangle\!\rangle$. After one turn, the phases in Eq.~\eqref{eq:sec4-positive-monodromy-phase} multiply the three columns of the coefficient matrix $\Psi$ by $(1,1,-1)$. The coefficient matrix $\Psi$ satisfies
\begin{align}
 \Psi\cdot\begin{pmatrix}1&0&0\\0&1&0\\0&0&-1\end{pmatrix}=
 \begin{pmatrix}0&1&0\\1&0&0\\0&0&1\end{pmatrix}\cdot\Psi\,.
 \label{eq:sec4-outer-monodromy-matrix}
\end{align}
Consequently, we have
\begin{align}
 \begin{aligned}
 \left|R_{110_\mathsf w}(\varphi+2\pi)\right\rangle&=\left|R_{123_\mathsf w}(\varphi)\right\rangle\,,\\
 \left|R_{123_\mathsf w}(\varphi+2\pi)\right\rangle&=\left|R_{110_\mathsf w}(\varphi)\right\rangle\,,\\
 \left|R_{213_\mathsf w}(\varphi+2\pi)\right\rangle&=\left|R_{213_\mathsf w}(\varphi)\right\rangle\,.
 \end{aligned}
 \label{eq:sec4-outer-monodromy}
\end{align}

Overall, we summarize the four connected families of boundary states
in Table~\ref{tab:sec4-all-pf3-covers}.

\begin{table}[htbp]
 \centering
 \renewcommand{\arraystretch}{1.2}
 \begin{tabular}{c|l|c|c|c}
 Gluing & Labels & $g$-factor
 & Cover degree & Period of $\varphi$\\ \hline
 $\Omega_\varphi$ & $110,112,114,121,123,125$ & $g_0$ & $6$ & $12\pi$\\
 $\Omega_\varphi$ & $211,213,215$ & $\sqrt2g_0$ & $3$ & $6\pi$\\
 $\Omega_\varphi\mathsf w$ & $110_\mathsf w,123_\mathsf w$ & $\sqrt3g_0$ & $2$ & $4\pi$\\
 $\Omega_\varphi\mathsf w$ & $213_\mathsf w$ & $\sqrt6 g_0$ & $1$ & $2\pi$
 \end{tabular}
 \caption{Monodromy of all inner- and outer-twisted $B_3$ Cardy seeds and their transported boundaries. The cover degree counts the number of distinct boundaries at the RG angle.}
 \label{tab:sec4-all-pf3-covers}
\end{table}

\FloatBarrier

\section{Orbifold RG flows and transport of RG branes}
\label{sec:orbifold-transport}

We now relate RG branes for flows whose UV theories are connected by a finite orbifold. We require that the orbifold symmetries are preserved under the UV perturbation and leave the IR fixed point unchanged. The two RG branes of the two flows can then be connected by the topological interface associated with the orbifolding. When the orbifolding projects out the spin-one phantom currents, it will lift the associated defect moduli, while finite automorphisms can survive in the smaller auxiliary algebra. This transport can therefore connect the outer and inner constructions of the preceding sections. It also provides a practical way to identify the RG brane in a continuous family, where both the gluing RG angle and the Cardy label must be determined. Since the topological interface transmits the stress tensor, the two transported branes must have the same energy transmission coefficient. Matching their transmission can locate the candidate RG angles, which in the examples below are precisely the points of maximal transmission. When the RG brane on one side is known, this brane and the topological interface determine the Cardy label on the other side, as we have seen in Section~\ref{sec:odd-k-branes}.

Let $G$ be a gaugeable finite symmetry of $\mathcal T_{\rm UV}$, and let $\delta S$ be the perturbation action, which is $G$-invariant. The orbifold perturbation $\delta S'$ is its invariant local image in \mbox{$\mathcal T_{\rm UV'}=\mathcal T_{\rm UV}/G$}. We consider
\emph{orbifold RG flows} of the form
\begin{align}
 \mathcal T_{\rm UV}\xrightarrow{\ \ \ \delta S\ \ \ }\mathcal T_{\rm IR}\,,
 \qquad\qquad
 \mathcal T_{\rm UV'}\xrightarrow{\ \ \ \delta S'\ \ \ }\mathcal T_{\rm IR}\,,
 \label{eq:sec5-orbifold-flows}
\end{align}
with the same infrared theory. We assume the orbifold interface to act trivially on this IR part. If the preserved symmetry instead has a nontrivial IR realization, it should also be included in the transport. The IR realization matching discussed at the end of Section~\ref{sec:odd-k-branes} is an example of this distinction. 

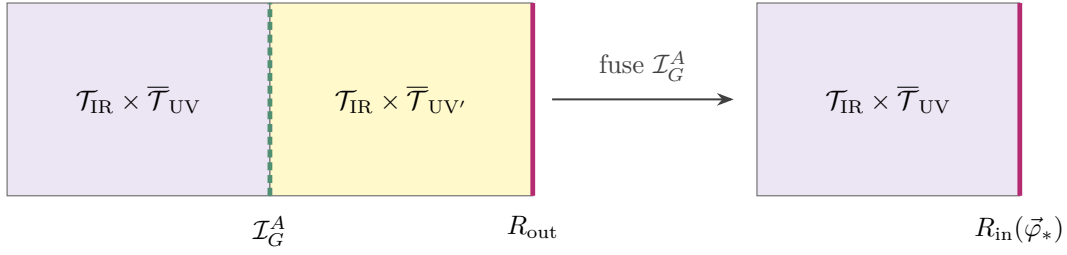
\begin{figure}[tbp]
\centering
\definecolor{PCOrbifoldFill}{RGB}{255,249,204}
\definecolor{PCFoldFill}{RGB}{235,229,244}
\definecolor{PCWall}{RGB}{181,43,114}
\definecolor{PCTopWall}{RGB}{81,143,114}
\begin{tikzpicture}[
 x=1.022296cm,y=1.213775cm,>=Stealth,
 panel edge/.style={draw=black!55,line width=0.45pt},
 rg wall/.style={PCWall,line width=1.8pt},
 top wall/.style={PCTopWall,line width=1.8pt,
   dash pattern=on 3pt off 1.8pt},
 process arrow/.style={->,black!75,line width=0.75pt},
 theory label/.style={font=\small},
 wall label/.style={font=\small,anchor=north,inner sep=2pt}
]
 \path[panel edge,fill=PCFoldFill] (0,0) rectangle (3.4,2.1);
 \path[panel edge,fill=PCOrbifoldFill] (3.4,0) rectangle (6.8,2.1);
 \draw[top wall] (3.4,0) -- (3.4,2.1);
 \draw[rg wall] (6.8,0) -- (6.8,2.1);
 \node[theory label] at (1.7,1.05)
  {$\mathcal T_{\rm IR}\times\overline{\mathcal T}_{\rm UV}$};
 \node[theory label] at (5.1,1.05)
  {$\mathcal T_{\rm IR}\times\overline{\mathcal T}_{\rm UV'}$};
 \node[wall label] at (3.4,-0.16) {$\mathcal I_G^A$};
 \node[wall label] at (6.8,-0.16) {$R_{\rm out}$};
 \draw[process arrow] (7.05,1.05) --
  node[above=3pt,font=\small] {fuse $\mathcal I_G^A$} (9.4,1.05);
 \path[panel edge,fill=PCFoldFill] (9.7,0) rectangle (13.1,2.1);
 \draw[rg wall] (13.1,0) -- (13.1,2.1);
 \node[theory label] at (11.4,1.05)
  {$\mathcal T_{\rm IR}\times\overline{\mathcal T}_{\rm UV}$};
 \node[wall label,anchor=north east] at (13.75,-0.16)
  {$R_{\rm in}(\vec\varphi_*)$};
\end{tikzpicture}
\caption{Transport between RG branes for orbifold RG flows. The green dashed line is the topological interface $\mathcal I_G^A$ determined by the original and orbifold UV theories, with the common IR theory unchanged. Fusing it into the outer RG brane (magenta) removes the intervening orbifold strip and produces the inner RG brane at the distinguished moduli point $\vec\varphi_*$, the RG angle.}
\label{fig:sec5-orbifold-transport}
\end{figure}

After folding, write $A=\mathcal T_{\rm IR}\boxtimes\overline{\mathcal T}_{\rm UV}$ and $A'=\mathcal T_{\rm IR}\boxtimes\overline{\mathcal T}_{\rm UV'}$. The UV orbifold defines a topological interface $\mathcal I_G^A:A'\to A$, acting trivially on the IR factor. Suppose that the phantom sectors and their junctions admit a $G$-compatible condensation. Condensing them gives auxiliary algebras $B$ and $B'$, and induces an interface $\mathcal I_G^B:B'\to B$. We then have the following commutative diagram for boundary transport:
\begin{align}
\begin{tikzpicture}[baseline=(current bounding box.center),>=Stealth,
 line width=0.6pt,
 theory/.style={inner sep=4pt,minimum height=0.65cm},
 transport/.style={->,shorten <=3pt,shorten >=3pt}]
 \node[theory] (Bp) at (0,1.8) {$B'$};
 \node[theory] (B) at (5,1.8) {$B$};
 \node[theory] (Ap) at (0,0) {$A'$};
 \node[theory] (A) at (5,0) {$A$};
 \draw[transport] (Bp)--node[above=3pt]{$\mathcal I_G^B$}(B);
 \draw[transport] (Ap)--node[below=3pt]{$\mathcal I_G^A$}(A);
 \draw[transport] (Bp)--node[left=4pt]
 {$\check{\mathcal I}_{B'\to A'}$}(Ap);
 \draw[transport] (B)--node[right=4pt]
 {$\check{\mathcal I}_{B\to A}$}(A);
\end{tikzpicture}
\label{diag:brane_transport}
\end{align}
The vertical interfaces are defined by the condensation of phantom currents specified by $\mathfrak a'$ and $\mathfrak a$, respectively. The commutative diagram gives
\begin{align}
 \mathcal I_G^A\circ\check{\mathcal I}_{B'\to A'}
 =\check{\mathcal I}_{B\to A}\circ\mathcal I_G^B\,.
 \label{eq:sec5-square}
\end{align}
Therefore, we can first work out how the seed branes are transported in the auxiliary theories to identify the corresponding RG angles and brane labels. In Figure~\ref{fig:sec5-orbifold-transport}, we show the transport of RG branes from $A'$ to $A$ via the topological interface $\mathcal I_G^A$.

Since the perturbation $\delta S$ is required to be $G$-invariant, the preserved fusion category $\mathscr C_{\rm pre.}$ is transmitted to the topological interface $\mathcal I_G$, and is thus a sub-fusion category of $\mathscr C'_{\rm pre.}$ of the RG flow between $\mathcal T_{\rm UV}'$ and $\mathcal T_{IR}$. Nevertheless, the allowed endpoint operators of these lines could have different spins after the orbifold, as the original endpoint operators will be projected out if they are not invariant under the $G$-action. Consequently the orbifold can remove spin-one phantom fields while keeping the corresponding non-invertible symmetries. In the auxiliary theory this is the projection $B\to B'=B^G$. If $B'$ contains no spin-one current, the continuous gluing automorphisms are absent. The remaining finite  symmetries may nevertheless descend to nontrivial outer automorphisms of $B'$. This is how a continuous family of inner automorphisms can be replaced by discrete outer ones. For an outer seed $|b_{\rm out}\rangle_{B'}$ with elementary image $|b_{a_*};\Omega_*\rangle_B$,
the square gives
\begin{align}
 \mathcal I_G^B|b_{\rm out}\rangle_{B'}
 =|b_{a_*};\Omega_*\rangle_B\,,\qquad
 \mathcal I_G^A|R_{\rm out}\rangle_{A'}
 =|R_{a_*,\Omega_*}\rangle_A\,.
 \label{eq:sec5-transport-seeds}
\end{align}
Both the automorphism $\Omega_*$ and the Cardy label $a_*$ are determined by this transport. When one flow has an independently identified RG wall, it will provide a physical selection of its image. Otherwise it establishes a relation between candidate walls and locates their gluing automorphisms within the continuous family.

\FloatBarrier

\subsection{Parafermion flows at odd level}
\label{sec:orbifold-odd-pf}

For odd $k\geq3$, with the coupling phase chosen as in \eqref{eq:Pk_deform}, the deformation will preserve the charge conjugation $\mathsf c$. We compare the boundary constructions for the following two orbifold RG flows,
\begin{align}
 \mathcal P_k\longrightarrow\mathcal M_{k+1,\,k+2}\,,
 \qquad\qquad
 \mathcal P_k^{\mathsf c}\longrightarrow\mathcal M_{k+1,\,k+2}\,.
 \label{eq:sec5-odd-pair}
\end{align}
Here $\mathcal P_k^{\mathsf c}$ denotes the charge conjugation orbifold CFT, whose chiral algebra is the $\mathsf c$-invariant subalgebra of $\mathcal P_k$. The corresponding folded algebra is $A'_k=\mathcal M_{k+1,\,k+2}\boxtimes \overline{\mathcal P_k^{\mathsf c}}$. At $k=3$ the second flow is the familiar $\mathcal M_{5,\,6}\to\mathcal M_{4,\,5}$ one. This construction is the one giving $R_{211}$ in Section~\ref{sec:inner-walls}.

\paragraph{Phantom sectors after projection.}
For $a=0,\ldots,(k-1)/2$, let $F_a$ denote the UV parafermion module $\mathcal V^{\rm UV}_{[2a+1,0]}$ and $V_a$ the IR module $\mathcal V^{\rm IR}_{1,2a+1}$. These are the same matched representations introduced in Eq.~\eqref{eq:sec4-matched-sector}. Their primary conformal weights are
\begin{align}
 h(F_a)=\frac{a(a+1)}{k+2}\,,\qquad
 h(V_a)=a^2-\frac{a(a+1)}{k+2}\,.
 \label{eq:sec5-parent-phantom}
\end{align}
The original matched endpoint phantom current has weight $a^2$. Charge conjugation $\mathsf c$ sends $[r,m]$ to $[r,-m]$ and preserves each neutral module $F_a$. In the orbifolded theory, the module $F_a$ is a fixed point under the $\mathsf c$-action and resolves into two distinct modules,
\begin{align}
 F_a\big|_{\mathcal P_k^{\mathsf c}}
 =F_a^+\oplus F_a^-\,,
\end{align}
where the signs ``$\pm$'' refer to the charges of $\mathsf c$. The conformal weights of $F_a^{\pm}$ are given as follows~\cite{jiang2017representationsmathbbz2orbifoldparafermionvertex}:
\begin{align}
\begin{array}{c|c|c}
 a & \text{weight of } F_a^{+} & \text{weight of } F_a^{-}\\ \hline
 0 & 0 & 3\\
 \text{even},\ a>0 & h(F_a) & h(F_a)+1\\
 \text{odd} & h(F_a)+1 & h(F_a)
\end{array}\,.
\end{align}
The new phantom sectors contain the $F_a^+$ branch, in place of the $F_a$ appearing in Eq.~\eqref{eq:sec4-matched-sector}, 
\begin{align}
    \mathfrak a'=\bigoplus_{a=0}^{\lfloor k/2\rfloor}V_a\,\boxtimes \overline{F}_a^{+}\,,
\end{align}
where, as throughout, the bar denotes the reversal of the UV factor under folding. The occurrence of $F_a^{+}$ in the phantom sectors is more clearly seen in the auxiliary theories $B_k$ and $B_k'$. Recall that, in $B_k$, $\mathsf c$ is lifted to $\mathsf w$. We thus have $B_k'=B_k^{\mathsf w}$. Thus, after gauging $\mathsf w$, the phantom sectors retain only the modules with positive $\mathsf w$ eigenvalue. Therefore we have
\begin{align}
 &B'_k=B_k^{\mathsf w}\,,\qquad
 B'_k\big|_{A'_k}
 =\mathfrak a'=\bigoplus_{a=0}^{(k-1)/2}V_a\boxtimes\overline{F}^+_a\,,\notag\\[5pt]
 &h(F^{+}_a)=\frac{a(a+1)}{k+2}+(a\bmod2)\,,\qquad
 h_a'=a^2+(a\bmod2)\,.
 \label{eq:sec5-orbifold-phantom}
\end{align}
The nonvacuum phantom sectors now have weights $2,4,10,\ldots$, truncated at $a=(k-1)/2$. In particular, the first phantom sector starts with a spin-two rather than a spin-one field. The resolved modules close under fusion:
\begin{align}
 F^+_a\otimes F^+_b
 =\bigoplus_{d=|a-b|}^{\min(a+b,k-a-b)}F^+_d\,,
 \label{eq:sec5-orbifold-fusion}
\end{align}
satisfying the fusion rules of $\mathrm{SO}(3)_k$, as before the orbifold \cite{Jiang:2019ghv}.

\paragraph{Outer automorphisms and seed branes.}
The spin-one current $Q$ in $B_k$ is $\mathsf w$-odd and is therefore absent from $B'_k=B_k^{\mathsf w}$. The rotation $\Omega_\pi$ nevertheless commutes with $\mathsf w$ on the chiral algebra $\mathfrak B_k$ and restricts to an involution of $\mathfrak B_k'$
\begin{align}
 \omega_k=\Omega_\pi|_{\mathfrak B'_k}\,,\qquad
 \operatorname{Aut}(\mathfrak B'_k)=\{1,\omega_k\}\simeq\mathbb Z_2\,.
 \label{eq:sec5-odd-auto}
\end{align}
It is outer because $B'_k$ has no spin-one zero modes. One can also explicitly verify that $\omega_k$ is the only non-trivial action on the vacuum module of $B_k'$ from the OPE covariance equations~\eqref{eq:graded-OPE-covariance} level by level\,\footnote{For an all level proof, we refer readers to~\cite{shimakura2006liftsautomorphismsvertexoperator}.}. Its lift within the continuous inner family thus selects the RG angle $\varphi_*=\pi$ in $B_k$. At $k=3$, the orbifold flow is $\mathcal M_{5,6}\to\mathcal M_{4,5}$, with the auxiliary theory
\begin{align}
 B'_3=
 \frac{\widehat{\mathfrak{su}}(2)_2\times
 \widehat{\mathfrak{su}}(2)_1\times\widehat{\mathfrak{su}}(2)_1}
 {\widehat{\mathfrak{su}}(2)_4}\,,
 \label{eq:sec5-k3-outer-coset}
\end{align}
and $\omega_3$ exchanges its two level-one numerator factors. Gaiotto's construction uses the seed brane labelled by the identity, as discussed in Section~\ref{sec:outer-walls}. For general odd $k$, an analogous identification of the RG seed is not known independently in $B_k'$. We can nevertheless establish its relation to the $\mathsf w$-invariant seed branes proposed in Section~\ref{sec:odd-k-branes}.

Let $\mathcal L_{\mathsf q}$ denote the quantum $\mathbb Z_2$ symmetry line of the auxiliary orbifold theory $B_k'$. It is a Verlinde line of $B'_k$, labelled by the $\mathsf w$-odd module $\mathfrak B_k^-$ appearing in the decomposition 
\begin{align}
    \mathfrak B_k|_{\mathfrak B_k'}=\mathfrak B^+_k\oplus \mathfrak B_k^-\,,\qquad{\rm with}\ \ \mathfrak B_k'\equiv \mathfrak B_k^+\,.
\end{align}
Gauging this quantum symmetry condenses the module $\mathfrak B_k^-$ into $\mathfrak B_k'$ and recovers the theory $B_k$. It defines the topological interface
\begin{align}
 \mathcal I_{\mathbb Z_2}:B'_k\longrightarrow B_k\,,
 \label{eq:sec5-quantum-interface}
\end{align}
the auxiliary interface in the commutative diagram~\eqref{diag:brane_transport}. Let $\mathcal L_{\mathsf w}$ denote the reflection symmetry line of $B_k$, which realizes the preserved UV reflection. We thus have
\begin{align}
 \check{\mathcal I}_{\mathbb Z_2}\circ\mathcal I_{\mathbb Z_2}
 &=\mathbf1_{B'_k}+\mathcal L_{\mathsf q}\,,&
 \mathcal I_{\mathbb Z_2}\circ\check{\mathcal I}_{\mathbb Z_2}
 &=\mathbf1_{B_k}+\mathcal L_{\mathsf w}\,.
 \label{eq:sec5-quantum-interface-fusion}
\end{align}
At $\varphi_*=\pi$, notice that, on the $B_k$ side, the $\mathsf w$-invariant seed brane is given in~\eqref{eq:sec4-odd-seed},
\begin{align}
    \left|b_{a_*};\,\Omega_\pi\right\rangle\ {\rm with}\,\ a_*=\left[\frac{k+1}{2},1,\frac{k-1}{2}\right]\,,\quad {\rm satisfying}
    \ \ \mathcal L_\mathsf w\left|b_{a_*};\,\Omega_\pi\right\rangle=\left|b_{a_*};\,\Omega_\pi\right\rangle\,.
\end{align}
Orbifolding this fixed boundary gives two elementary $\omega_k$-glued seeds, denoted
$|\Gamma_\pm;\,\omega_k\rangle$, which are exchanged by the quantum symmetry:
\begin{align}
 \mathcal L_{\mathsf q}|\Gamma_\pm;\,\omega_k\rangle
 &=|\Gamma_\mp;\,\omega_k\rangle\,.
 \label{eq:sec5-quantum-seed-action}
\end{align}
Conversely, both $|\Gamma_\pm;\,\omega_k\rangle$ are mapped to $\left|b_{a_*};\,\Omega_\pi\right\rangle$ via $\mathcal I_{\mathbb Z_2}$, i.e.
\begin{align}
 \mathcal I_{\mathbb Z_2}|\Gamma_\pm;\omega_k\rangle
 =|b_{a_*};\Omega_\pi\rangle\,, \qquad\qquad
 \check{\mathcal I}_{\mathbb Z_2}|b_{a_*};\Omega_\pi\rangle
 =|\Gamma_+;\omega_k\rangle+|\Gamma_-;\omega_k\rangle\,,
 \label{eq:sec5-gamma-transport}
\end{align}
in accordance with the standard transitive/fixed boundary relations under $\mathbb Z_2$ orbifolding~\cite{Fukusumi:2021zme}. At $k=3$, $\Gamma_+$ is Gaiotto's seed labelled by the identity, $[1,1,1;1]_{\omega_3}$, and $\Gamma_-$ is $[1,1,1;5]_{\omega_3}$. Both seeds map to $b_{211}(\pi)$ in $B_3$ and hence to $R_{211}$ via the transport diagram~\eqref{diag:brane_transport}. For higher odd $k$, $\Gamma_\pm$ are identified by this transport relation, rather than an assumed identity label. We summarize a few examples of the $\Gamma_\pm$ twisted modules, together with the lowest twisted module weight of $\Gamma_+$, in Table~\ref{tab:sec5-odd-sectors}.

\begin{table}[tbp]
\centering
\renewcommand{\arraystretch}{1.2}
\begin{tabular}{c|c|c|c}
 $k$ & Non-vacuum $h'_a$ & $h^{\rm tw}(\Gamma_+)$ & Parent seed $b_*$\\ \hline
 $3$ & $2$ & $1/16$ & $[2,1,1]$\\
 $5$ & $2,4$ & $1/12$ & $[3,1,2]$\\
 $7$ & $2,4,10$ & $3/32$ & $[4,1,3]$\\
 odd $k$ & $a^2+(a\bmod2)$ & $\dfrac{k-1}{8(k+1)}$
 & $\left[\frac{k+1}{2},1,\frac{k-1}{2}\right]$
\end{tabular}
\caption{Odd-level orbifold transport. The second column gives the lowest weights of matched phantom sectors. The third is the ground weight of the resolved twisted module labeling $\Gamma_+$. Both resolutions $\Gamma_\pm$ have the parent image in the
last column.}
\label{tab:sec5-odd-sectors}
\end{table}

Finally, we comment on the $g$-factors of the three candidate RG branes $R_{a_*}$ and $R'_{\Gamma_{\pm}}$ after pulling them back to the tensor product theories $A_k$ and $A_k'$. Noting that the topological interface $\mathcal I_{\mathbb Z_2}$ has a $g$-factor of $\sqrt 2$, we conclude, from Eq.~\eqref{eq:sec5-gamma-transport}, that
\begin{align}
    g\left(R_{a_*}\right)=\sqrt 2\,g\left(R'_{\Gamma_{\pm}}\right)\,,
\end{align}
a straightforward generalization of Eq.~\eqref{eq:g-factor_orb}.

\FloatBarrier

\subsection{A generalized parafermion flow}
\label{sec:orbifold-generalized}

We will end  this section with a generalized parafermion RG flow proposed in~\cite{Ahn:2022pia, Ahn:2025xvb}: 
\begin{align}
 U=\frac{\widehat{\mathfrak{su}}(3)_2}
 {\widehat{\mathfrak u}(1)^2}
 \simeq\frac{\widehat{\mathfrak{su}}(2)_1^3}
 {\widehat{\mathfrak{su}}(2)_3}
 \quad\longrightarrow\quad
 P=\frac{\widehat{\mathfrak{su}}(2)_3}
 {\widehat{\mathfrak u}(1)_3}=W_3(4,5)\,,
 \qquad c_U=\frac65\,,\quad c_P=\frac45\,.
 \label{eq:sec5-generalized-flow}
\end{align}
This flow has two spin-one phantom currents and is known to have an orbifold counterpart from the second to the first minimal $W_3$ model, $W_3(5,6)\to W_3(4,5)$~\cite{Poghosyan:2022mfw, Poghosyan:2023brb}. A more systematic study of this and higher rank generalized parafermion flows and their domain walls will appear elsewhere~\cite{Chen:generalizedPFfuture}.

\begin{figure}[tbp]
\centering
\definecolor{PCWall}{RGB}{181,43,114}
\begin{tikzpicture}[
 >=Stealth, line width=0.8pt,
 theory/.style={inner sep=5pt,align=center},
 topological/.style={->,draw=green!45!black},
 rg/.style={->,draw=PCWall,line width=1pt},
 flow label/.style={font=\small,fill=white,inner sep=2pt}
]
 \node[theory] (U) at (-3.6,0)
  {$U=\dfrac{\widehat{\mathfrak{su}}(3)_2}
                    {\widehat{\mathfrak u}(1)^2}$};
 \node[theory] (C) at (3.5,0) {$U'=W_3(5,6)$};
 \node[theory] (P) at (0,-4.05) {$P=W_3(4,5)$};
 \draw[topological] (-2.05,0.23)--
  node[above=4pt,font=\small]
  {$\mathbb Z_3^{\rm outer}$-orbifolding} (2.05,0.23);
 \draw[topological] (2.05,-0.23)--
  node[below=4pt,font=\small]
  {$\mathbb Z_3^{\rm Verlinde}$-orbifolding} (-2.05,-0.23);
 \draw[rg] (U.south)
  .. controls (-3.6,-1.2) and (-2.45,-1.05) .. (-2.45,-1.65)
  .. controls (-2.45,-2.18) and (-3.2,-2.13) .. (-3.2,-2.72)
  .. controls (-3.2,-3.5) and (-2.0,-4.05) .. (P.west);
 \draw[rg] (C.south)
  .. controls (3.5,-0.95) and (2.5,-1.05) .. (2.5,-1.55)
  .. controls (2.5,-2.0) and (3.25,-2.0) .. (3.25,-2.55)
  .. controls (3.25,-3.45) and (1.8,-4.05) .. (P.east);
 \node[flow label] at (-3.65,-1.75) {RG flow};
 \node[flow label] at (3.7,-1.55) {RG flow};
\end{tikzpicture}
\caption{The generalized parafermion theory $U$ and its cyclic orbifold $U'$ flow to the same three-state Potts model $P$. The upper arrows gauge the outer $\mathbb Z_3$ symmetry of $U$ and the quantum Verlinde $\mathbb Z_3$ symmetry of $U'$, respectively. The curved arrows denote the two massless RG flows.}
\label{fig:sec5-generalized-orbifold}
\end{figure}
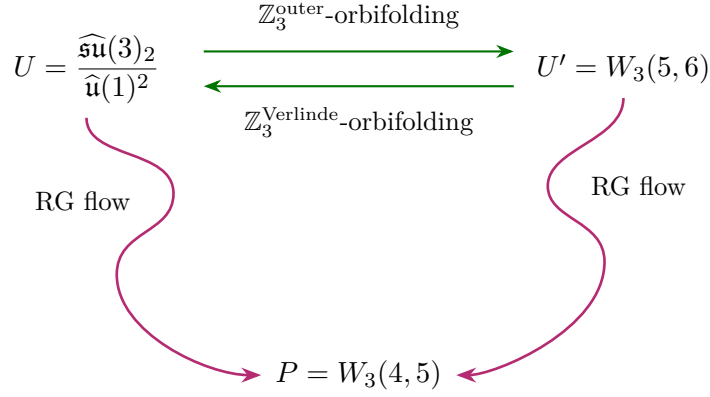

The $U\to P$ flow is triggered by the following perturbation
\begin{align}
 \delta S=\lambda\int d^2x\,
 \frac{\Phi_{12}+\Phi_{23}+\Phi_{31}}{\sqrt3}\,,
 \qquad \lambda>0\,,
 \label{eq:sec5-root-perturbation}
\end{align}
where the three real scalar fields $\Phi_{ij}$ have weights $(1/2,1/2)$ in $U$. It is interesting to note that $U=\frac{\widehat{su}(3)_2}{\widehat{u}(1)^2}$ is actually level-rank dual to the auxiliary theory $B_2=\frac{\widehat{\mathfrak{su}}(2)_1^3}  {\widehat{\mathfrak{su}}(2)_3}$ in~\eqref{eq:sec3-gaiotto-extension} of the flow between TIM and Ising~\cite{Jiang:2014tdi}. In the dual description, the three scalar operators have labels $[2,1,1;4],\,[1,2,1;4],\,[1,1,2;4]$ and are permuted under the cyclic permutation $\rho$ of the three $\widehat{\mathfrak {su}}(2)_1$ factors in the numerator. Gauging the outer automorphism $\rho$ gives
\begin{align}
 U'=U^{\langle\rho\rangle}=W_3(5,6)
 =\frac{\widehat{\mathfrak{su}}(3)_2\times
 \widehat{\mathfrak{su}}(3)_1}{\widehat{\mathfrak{su}}(3)_3}\,.
 \label{eq:sec5-W3-UV}
\end{align}
In $U'$, a scalar primary labeled by $(0,0;\mathrm{adj})$ with $h=\bar h=1/2$ is obtained from the $\mathbb Z_3^{\rho}$ invariant sum of the fields $\Phi_{ij}$ in $U$. It triggers a massless RG flow to $P=W_3(4,5)$, and its RG wall was constructed in~\cite{Poghosyan:2022mfw,Poghosyan:2023brb}. Conversely, gauging the quantum $\mathbb Z_3$ symmetry in $U'$, which is made of three Verlinde lines, will recover $U$. Figure~\ref{fig:sec5-generalized-orbifold} summarizes the two orbifold related flows.

\paragraph{Phantom sectors and the auxiliary coset.}
The theory $U$ has $8$ primaries and thus $8$ corresponding Verlinde lines. They furnish a fusion category with fusion rules of $\textbf{Fib}\,\boxtimes\mathbb Z_2^2$, where $\textbf{Fib}$ is the Fibonacci-like category. The deformation~\eqref{eq:sec5-root-perturbation} preserves the $\textbf{Fib}$ part, with endpoint operators of weight $3/5$, whereas its IR realization in $P$ has another endpoint of weight $2/5$, i.e.
\begin{align}
    \chi_{\frac35}^U=2q^{\frac35-\frac{c_U}{24}}+\cdots\,,\qquad\qquad
    \chi_{\frac25}^P=q^{\frac25-\frac{c_P}{24}}+\cdots\,.
\end{align}
As the former ground state space is two-dimensional, the matched phantom sector supplies two independent spin-one phantom currents. The Frobenius algebra is then given by
\begin{align}
    \mathfrak a=\mathcal X_0\oplus \mathcal X_1\,,\qquad 
    \mathcal X_0=\mathcal V_\textbf{v}^{P}\boxtimes\overline{\mathcal V}_\textbf{v}^{U}\,,\,\quad
    \mathcal X_1=\mathcal V_{\frac25}^{P}\boxtimes\overline{\mathcal V}_{\frac35}^{U}\,.
    \label{eq:generalized_parafermion_interface}
\end{align}
Condensation of $\mathfrak a$ defines the auxiliary theory $B$,
\begin{align}
 A=P\,\boxtimes\,\overline{U}\subset B\,,\quad
 B=\frac{\widehat{\mathfrak {su}}(2)^3_1}
      {\widehat{\mathfrak u}(1)_3}\simeq V_{\sqrt 2A_2}\,,
 \quad \mathcal V_{\textbf{v}}^B=\mathfrak a=\mathcal X_0\oplus\mathcal X_1\,,
 \quad
 c_B=c_U+c_P=2
 \label{eq:sec5-lattice-auxiliary}
\end{align}
where $V_L$ is the chiral algebra of two compact bosons with momentum lattice $L=\sqrt 2A_2$. It is also convenient to use the $A_2$-lattice to formulate the modular data of $B$ and perform OPE computations later: we choose simple lattice vectors $\beta_1,\beta_2$ with Gram matrix
\begin{align}
    G_{A_2}=\begin{pmatrix}4&-2\\-2&4\end{pmatrix}\,,
\end{align}
and their dual vectors $\beta^i$ satisfying $\beta^i\cdot\beta_j=\delta^i_j$. The irreducible modules are labeled by the twelve momentum classes in $\mathcal D=L^*/L$ where $L^*$ is the dual lattice. On the other hand, in the coset notation, we label the $12$ primaries by 
\begin{align}
 [\boldsymbol r;m]\equiv[s,t,d;m]:\quad &s,t,d=1,2\,,\ m\in\mathbb Z_6\,;\quad
 s+t+d+m\in2\mathbb Z+1\,;\notag\\[5pt]
 &[s,t,d;m]\sim[3-s,3-t,3-d;m+3]\,.
\end{align}
A set of representatives can be chosen as
\begin{align}
 \boldsymbol r\in\{111,221,212,122\}\,,\qquad m=0,2,4\,,\quad \text{with vacuum label }
 \textbf{v}=[111;0]\,.
 \label{eq:sec5-E-module-labels}
\end{align}
We summarize more details on the modular data of $B$ in Appendix~\ref{app:lattice-E-data}.

We now determine how the phantom currents act on this algebra. In terms of the lattice bosons $X^i$, the two weight-one bosonic currents are given by
\begin{align}
    H^i\equiv i\partial X^i\quad {\rm for}\ \ i=1,2\,,\quad
    {\rm satisfying}\ \ H^{i}(z)H^j(0)\sim\frac{\delta^{ij}}{z^2}\,.
\end{align}
For convenience, we choose the spin-one phantom currents
\begin{align}
 Q_i(z)=\beta^i\cdot H(z)
 =\sum_{a=1}^{2}(\beta^i)_a H^a(z)\,,
 \qquad i=1,2\,,
\end{align}
where $a$ labels the two orthonormal boson directions, $H=(H^1,H^2)$, and $\beta^i$ are the dual lattice vectors defined before. These currents span the weight-one quasiprimary subspace of $B$,
\begin{align}
 \mathcal Q_1(B)=\operatorname{span}\{Q_1,Q_2\}\,.
\end{align}
Their zero modes $Q_{i,0}$ thus generate the following action on a vertex operator $V_{n_1\beta_1+n_2\beta_2}$:
\begin{align}
\Omega_{\vec\varphi}\equiv\operatorname{Ad}\left(e^{i\varphi_1Q_{1,0}+i\varphi_2Q_{2,0}}\right)\,,\qquad
\Omega_{\vec\varphi}\cdot V_{n_1\beta_1+n_2\beta_2}=e^{i(n_1\varphi_1+n_2\varphi_2)} V_{n_1\beta_1+n_2\beta_2}\,,
\label{eq:two_current_zero_modes}
\end{align}
with $\varphi_i\sim\varphi_i+2\pi$. To see how OPE covariance recovers both the continuous and discrete transformations, we set $\beta_3=\beta_1+\beta_2$. The six shortest momenta $\pm\beta_a$, $a=1,2,3$, have norm squared four, so their vertex fields first appear at weight two. Together with the three
normal-ordered current quadratics, they form the nine-dimensional quasi-primary space, with a basis
\begin{align}
 \mathcal Q_2(B)=\left\{
 :(H^1)^2:\,,\,:H^1H^2:\,,\, :(H^2)^2:\,,\,
 V_{\pm\beta_1}\,,\,V_{\pm\beta_2}\,,\,V_{\pm\beta_3}\right\}\,.
 \label{eq:sec5-E-Q2-basis}
\end{align}
Solving Eq.~\eqref{eq:graded-OPE-covariance} jointly at weights one and two, we find $12$ possible matrices $M_1$ in the orthonormal frame $\beta_1=(2,0)$ and $\beta_2=(-1,\sqrt3)$,
\begin{align}
 M_1&\in\left\{\mathsf R_n,\;
 \mathsf R_n\begin{pmatrix}1&0\\0&-1\end{pmatrix}
 \right\}_{n=0}^{5}\,,\qquad
 {\rm for}\quad
 \mathsf R_n=
 \begin{pmatrix}
 \cos\frac{n\pi}{3}&-\sin\frac{n\pi}{3}\\
 \sin\frac{n\pi}{3}&\cos\frac{n\pi}{3}
 \end{pmatrix}\,,
 \label{eq:sec5-E-M1}
\end{align}
They precisely form the isometry group of the lattice $L=\sqrt{2}A_2$
\begin{align}
    \operatorname{Iso}(L)=\{g\in O(2)\,|\,g\cdot L=L\}\simeq S_3\times \mathbb Z_2^\theta\,,
\end{align}
where $S_3$ is the Weyl group of $A_2$ and $\theta$ is the lattice inversion. On the identity component $M_1=\mathbf 1_2$, $M_2$ can be solved as
\begin{align}
 M_2=\operatorname{diag}
 \bigl(1,1,1,z_1,z_1^{-1},z_2,z_2^{-1},
 z_1z_2,(z_1z_2)^{-1}\bigr)\,,
 \qquad z_i=e^{i\varphi_i}\,.
 \label{eq:sec5-E-M2}
\end{align}
In particular, the two tangent matrices $A_i=\partial_{\varphi_i}M_2|_{\boldsymbol\varphi=0}$ precisely represent the action of the two phantom current zero modes in~\eqref{eq:two_current_zero_modes}. Overall, we obtain twelve continuous components, giving the full automorphism group of $\mathfrak B$~\cite{dong1998automorphismgroupstwistedmodules},
\begin{align}
 \operatorname{Aut}(\mathfrak B)
 =\mathrm U(1)^2\rtimes
 (S_3\times\mathbb Z_2^{\theta})\,,\qquad
 {\rm with}\quad
 \theta(H^i)=-H^i\,,\quad\theta(V_\alpha)=V_{-\alpha}\,.
 \label{eq:sec5-lattice-auto}
\end{align}
For given $\vec\varphi=(\varphi_1,\,\varphi_2)$ and $g\in \operatorname{Iso}(L)$, one notices that the UV and IR stress tensors are invariant under $g$,
\begin{align}
    g(T_U)=T_U\,,\qquad g(T_P)=T_P\,.
\end{align}
Therefore, for $\Omega_{\vec\varphi,\,g}\equiv \Omega_{\vec\varphi}\circ g$, we can compute the energy transmission rate in~\eqref{eq:stress-overlap-definition},
\begin{align}
    \mathcal T\left(\Omega_{\vec\varphi,\,g}\right)
 =\frac4{25}\left(3-\cos\varphi_1-\cos\varphi_2
 -\cos(\varphi_1+\varphi_2)\right)\,.
 \label{eq:sec5-lattice-transmission}
\end{align}
On each branch labelled by $g$, the transmission reaches its maximum at
\begin{align}
    \mathcal T_{\max}=\frac{18}{25}\,,\qquad \vec\varphi=\pm\left(\frac{2\pi}{3},\frac{2\pi}{3}\right)\,.
\end{align}

\paragraph{The $W_3$ RG wall and its transport.}
For the flow $U'\to P$, the same Fibonacci condensation extends
$A'=P\boxtimes\overline {U'}$ to
\begin{align}
 B'=B^{\langle\hat\rho\rangle}
 =\frac{\widehat{\mathfrak{su}}(3)_1^{(1)}\times
 \widehat{\mathfrak{su}}(3)_1^{(2)}\times
 \widehat{\mathfrak{su}}(3)_1^{(3)}}
 {\widehat{\mathfrak{su}}(3)_3}\,,\qquad {\rm with}\ \ \mathcal Q_1(B')=0\,.
 \label{eq:sec5-W3-auxiliary}
\end{align}
The cyclic automorphism $\rho\in\operatorname{Aut}(\mathfrak U)$ lifts to $\hat\rho\in\operatorname{Aut}(\mathfrak B)$, which rotates the two spin-one currents, so neither survives the projection to $B'$. The diagonal endpoints instead start at weight two, with UV and IR weights $8/5$ and $2/5$. Thus the Fibonacci symmetry survives while the two continuous gluing directions disappear. The known RG wall uses $\omega=(23)$, exchanging the last two level-one factors in the numerator of $B'$, and the identity-labelled permutation seed~\cite{Gaiotto:2012np,Poghosyan:2022mfw,Poghosyan:2023brb}.

We label the ordinary $B'$ modules by $[a_1,a_2,a_3;\mu]$, where $a_i=0,1,2$ label the vacuum, fundamental and antifundamental modules of the three level-one factors, and
$\mu=(\mu_1,\mu_2)$ is an integrable weight of $\widehat{\mathfrak{su}}(3)_3$.
After imposing the coset selection rule and applying the field identification, precisely ten ordinary modules are fixed by $\omega$. A set of representatives is
\begin{align}
 \lambda_\mu=[t(\mu),0,0;\mu]\,,\qquad
 t(\mu)=\mu_1+2\mu_2\pmod3\,,\qquad
 \mu_1,\mu_2\geq0\,,\quad\mu_1+\mu_2\leq3\,.
 \label{eq:sec5-W3-fixed-modules}
\end{align}
Each supplies one $\omega$-twisted Ishibashi state $|\lambda_\mu;\,\omega\rangle\!\rangle_{B'}$. Using them, we can find ten $\omega$-twisted Cardy branes,
\begin{align}
 |\nu;\,\omega\rangle_{B'}
 &=\sum_\mu
 \frac{\Psi^{\omega}_{\nu,\lambda_\mu}}
 {\sqrt{S^{B'}_{\textbf{v}\lambda_\mu}}}
 |\lambda_\mu;\,\omega\rangle\!\rangle_{B'}\,,\qquad
 \Psi^{\omega}_{\nu,\lambda_\mu}=\left(S^{(3)}_{\nu\mu}\right)^*\,,
 \quad S^{B'}_{\textbf{v}\lambda_\mu}=\frac{S^{(3)}_{\textbf{v}\mu}}{\sqrt 3}\,,
 \label{eq:sec5-W3-twisted-Cardy}
\end{align}
where $S^{(3)}$ is the affine $\widehat{\mathfrak{su}}(3)_3$ S-matrix. In \cite{Poghosyan:2023brb}, the seed brane was chosen to be
\begin{align}
    |b';\,\omega\rangle_{B'}
 =|\textbf{v};\,\omega\rangle_{B'}
 =3^{1/4}\sum_\mu\sqrt{S^{(3)}_{0\mu}}\,
 |\lambda_\mu;\omega\rangle\!\rangle_{B'}\,.
 \label{eq:sec5-W3-seed}
\end{align}
In particular, $S^{(3)}_{00}=1/6$ and $S^{B'}_{00}=1/(6\sqrt3)$, so the vacuum overlap of this seed is
\begin{align}
 g(b')={}_{B'}\langle0|b';\,\omega\rangle_{B'}
 =\frac{1/6}{\sqrt{1/(6\sqrt3)}}=12^{-1/4}\,.
 \label{eq:sec5-W3-seed-g}
\end{align}
Now, for the two orbifold related RG flows,
\begin{align}
    U'=U^{\mathbb Z_3^\rho}\,,\quad\rho\in\operatorname{Aut(\mathfrak U)}\,;\qquad
    U=(U')^{\mathbb Z_3^\mathsf v}\,,\quad\mathcal L_\mathsf v\ \ \text{Verlinde lines in }U' \,;\qquad
\end{align}
the $\rho$-lines lift to the symmetry lines implementing $\hat\rho$ in $B$, while the $\mathsf v$-lines lift to the Verlinde lines $\mathcal L_{\hat{\mathsf v}}$ in $B'$. Following the brane transport diagram~\eqref{diag:brane_transport}, one can establish the topological interface between $B$ and $B'$, whose fusion relations are
\begin{align}
 \check{\mathcal I}_{\mathbb Z_3}\circ\mathcal I_{\mathbb Z_3}
 &=\one_{B'}+\mathcal L_{\hat{\mathsf v}}+\mathcal L_{\hat{\mathsf v}}^{\,2}\,,\qquad
 \mathcal I_{\mathbb Z_3}\circ\check{\mathcal I}_{\mathbb Z_3}
 =\one_B+\mathcal L_{\hat\rho}+\mathcal L_{\hat\rho}^{\,2}\,.
 \label{eq:sec5-Z3-interface-fusion}
\end{align}
Using $\mathcal I_{\mathbb Z_3}$, one can identify the seed brane in $B$,
\begin{align}
    \left|b_*;\,\Omega_{\vec\varphi_*,\theta}\right\rangle=\mathcal I_{\mathbb Z_3}|b';\,\omega\rangle_{B'}\,.
\end{align}
Here the RG angle is identified with 
\begin{align}
    \vec\varphi_*=\left(\frac{2\pi}{3},\frac{2\pi}{3}\right)\,,
\end{align}
precisely at the maximal transmission rate, accompanied by the outer automorphism $\theta$. It's easy to check that there are four $\Omega_{\vec\varphi_*,\theta}$-twisted modules from the labels $[\boldsymbol r;0]$, denoted by $[\boldsymbol r;0]_\theta$. The label of the seed brane can be verified as $b_*=[111;0]_\theta$. In addition, it is worth  noting that the seed brane $|b';\,\omega\rangle_{B'}$ belongs to an orbit in $B'$ on which $\mathbb Z_3^{\hat{\mathsf v}}$ acts transitively. Therefore, the seed brane in $B$, $\left|b_*;\,\Omega_{\vec\varphi_*,\theta}\right\rangle$, is invariant under $\hat\rho$, providing a $\mathbb Z_3$ analogue of the parafermion construction in Section~\ref{sec:orbifold-odd-pf}.  

Pull the seed back through the interface defined by $\mathfrak a$ in~\eqref{eq:generalized_parafermion_interface} to
the two physical theories $U$ and $P$. We have the RG brane in $A=P\boxtimes \overline U$, 
\begin{align}
 |R\rangle_{\!A}
 =\check{\mathcal I}_{B\to A}\left|b_*;\,\Omega_{\vec\varphi_*,\theta}\right\rangle_B
 =\check{\mathcal I}_{B\to A}\circ\mathcal I_{\mathbb Z_3}|b';\,\omega\rangle_{B'}
 \label{eq:sec5-physical-walls}
\end{align}
The computation of its $g$-factor is also clear: 
\begin{align}
    g(R)=g\left(\check{\mathcal I}_{B\to A}\right)\cdot g\left({\mathcal I}_{\mathbb Z_3}\right)\cdot g(b')=\sqrt{1+\gamma^2}\cdot\sqrt{3}\cdot 12^{-1/4}=\sqrt{\frac{\sqrt 3}{2}(1+\gamma^2)}\,,
\end{align}
with $\gamma=\frac{\sqrt 5+1}{2}$.

\paragraph{All brane families and their monodromy.}
Finally, it is also interesting to classify the monodromy of all branes in the $12$ branches defined by the $\operatorname{Aut(\mathfrak B)}$ in~\eqref{eq:sec5-lattice-auto}. Since $B$ can be realized as a free boson lattice, the computation of ($\Omega_{\vec\varphi_*,g}$-twisted) Cardy branes is standard. For brevity, we omit the details of the computation. There are $12+3\times6+4+2\times3+3\times2+2=48$ branes in total. In particular, the selected $\Omega_{\vec\varphi_*,\theta}$ seed belongs to the four-sheeted $\mathbb Z_2\times\mathbb Z_2$ cover. These are boundary monodromies over fixed bulk theories, analogous to the one-angle covers in Section~\ref{sec:boundary-monodromy}. The topological interface pullback keeps the same monodromy of these branes in the physical product theory. We summarize all branes in the $12$ branches  in Table~\ref{tab:sec5-lattice-covers}.

\begin{table}[tbp]
\centering
\renewcommand{\arraystretch}{1.2}
\begin{tabular}{c|c|c|c|c|c}
 Isometry $g$ & Components & $N_g$ & Monodromy group & $\operatorname{ord}(g)$
 & Torus brane\\ \hline
 $1$ & $1$ & $12$ & $\mathbb Z_2\times\mathbb Z_6$ & $1$ & D0\\
 $s_1,s_2,s_3$ & $3$ & $6$ & $\mathbb Z_6$ & $2$ & D1\\
 $\theta$ & $1$ & $4$ & $\mathbb Z_2\times\mathbb Z_2$ & $2$ & D2\\
 $\hat\rho,\hat\rho^{-1}$ & $2$ & $3$ & $\mathbb Z_3$ & $3$ & D2/D0\\
 $\theta s_1,\theta s_2,\theta s_3$ & $3$ & $2$ & $\mathbb Z_2$ & $2$ & D1\\
 $\theta\hat\rho,\theta\hat\rho^{-1}$ & $2$ & $1$ & trivial & $6$ & D2/D0
\end{tabular}
\caption{The 48 elementary seed branes at the RG angle, organized into 12 continuous branches. The $s_i$ are the three Weyl reflections and $\hat\rho$ is the order-three Weyl rotation. $N_g$ is the degree of the cover associated with the component labelled by $g$ over the angle torus. ``D2/D0'' denotes D2/D0 bound states.}
\label{tab:sec5-lattice-covers}
\end{table}

The auxiliary theory $B=V_{\sqrt 2A_2}$ with $c_B=2$ is at a special locus of the torus CFT. These branes, before pulling back, admit a further interpretation connecting to the geometric constructions in~\cite{Gukov:2002nw}. Our current gluing $gH_n+\bar H_{-n}=0$ corresponds to their oscillator matrix $R_{\rm GV}=-g^{-1}$. Thus $g=1$ gives D0, $g=\theta$ gives D2, and the six reflections give D1 branes. Their involutive D0/D1/D2 enumeration, evaluated at this lattice point, accounts for $12+4+3\times6+3\times2=40$ branes. The eight additional branes arise from the order-three and order-six isometries and can be interpreted as D2/D0 bound states~\cite{Bachas:2007td}, denoted by D2/D0 in the table.

\FloatBarrier

\section{Discussion}
\label{sec:discussion}

We have constructed the RG domain wall between two rational CFTs, connected by an RG flow, by using the phantom currents associated with transmitted topological defects. Condensation of these phantom currents into the vacuum of the product UV--IR theory defines an auxiliary chiral algebra. An automorphism of this algebra implements the mixing of UV and IR operators at the interface. Therefore, an automorphism-twisted Cardy brane in the auxiliary theory, after pullback to the original theory, will define the wanted RG brane. This construction gives two mechanisms for RG walls: Outer automorphisms exchanging identical current algebra factors reproduce Gaiotto-type walls. We review this construction and apply it to construct RG domain walls for a family of non-unitary minimal model flows. On the other hand, spin-one phantom currents generate continuous inner automorphisms which can be used to define twisted Cardy branes with continuous moduli. For flows between parafermions and minimal models, we use the second mechanism to construct candidate RG walls for odd $k$ with continuous defect moduli. In addition, we compute their various physical observables non-perturbatively, including boundary entropies and low-lying spectra, and UV--IR operator mixings. Finite orbifolds relate these two types of constructions. When a symmetry preserves the perturbation of a UV theory, its orbifolding may project out the spin-one phantom currents without removing the matched non-invertible symmetries in the UV and IR theories. Therefore the corresponding topological interface will transport the RG branes of the two flows. We demonstrate this phenomenon via examples of flows from parafermions to minimal models, and also a flow between generalized parafermions. We also find an interesting monodromy property of the RG branes constructed by the second mechanism: after a closed cycle in the defect conformal moduli space, the RG brane does not return to itself but is instead permuted into another nonfactorizable conformal interface. A common feature of the RG wall candidates constructed here is that their selected automorphisms are involutions, $\Omega_{\varphi_*}^2=1$. For RG branes with a continuous moduli space, the RG angle $\vec\varphi_*$ also lies at a maximum of the energy transmission. It would be interesting to understand whether these observations have a further physical origin or whether counterexamples exist.  

The constructions considered here can be applied to a wider set of massless flows. In particular, the three known families reviewed in~\cite{Benedetti:2026drn} admit the auxiliary algebras and automorphisms summarized in Table~\ref{tab:coset-flow-survey}. For this table, we write
\begin{align}
 S_\ell=\widehat{\mathfrak{su}}(2)_\ell\,,\qquad
 D(a,b)=\frac{S_a\times S_b}{S_{a+b}}\,,\qquad
 \mathcal P_k=\frac{S_k}{\widehat{\mathfrak u}(1)_k}\,,
 \label{eq:discussion-coset-notation}
\end{align}
where the levels are positive integers. Condensation cancels the intermediate algebra in these coset presentations. The remaining numerator either contains identical factors that can be exchanged or contains currents whose zero modes generate inner automorphisms. Our construction therefore extends to all three families.

\begin{table}[htbp]
 \centering
 \small
 \setlength{\tabcolsep}{5pt}
 \renewcommand{\arraystretch}{1.6}
 \begin{tabular}{@{}p{0.35\textwidth}p{0.29\textwidth}p{0.28\textwidth}@{}}
 \hline
 Massless flow & Auxiliary algebra & Gluing automorphism\\
 \hline
 $D(a,b)\longrightarrow D(a,b-a)$\newline
 $b>a\geq1$
 & $\displaystyle\frac{S_{b-a}\times S_a\times S_a}{S_{a+b}}$
 & \raggedright Outer automorphism exchanging the two $S_a$ factors\tabularnewline[5pt]
 $\mathcal P_k\longrightarrow D(k-1,1)$\,,
 $k\geq3$
 & $\quad\ \displaystyle\frac{S_{k-1}\times S_1}{\widehat{\mathfrak u}(1)_k}$
 & Inner ${\rm U}(1)$ rotations\\[5pt]
 $\mathcal P_k\longrightarrow \widehat{\mathfrak u}(1)_k$\,,
 $k\geq5$
 & $\qquad\quad S_k$
 & Inner ${\rm SO}(3)$ rotations\\[2pt]
 \hline
 \end{tabular}
 \caption{Auxiliary algebras and gluing automorphisms for the three known
 families of massless flows reviewed in Ref.~\cite{Benedetti:2026drn}. The first row includes the unitary minimal model flows of Section~\ref{sec:outer-walls}, while the second is the family studied in Section~\ref{sec:parafermion-walls}. In the last row, the affine currents of $S_k$ generate the inner automorphism.}
 \label{tab:coset-flow-survey}
\end{table}

Our framework opens numerous new avenues: The construction of non-unitary minimal model walls allows us to compute energy transport, operator mixing and boundary spectra. Their negative reflection coefficients and the complex weights call for a physical interpretation. In a companion paper, we consider conformal interfaces with defect conformal manifolds between two RCFTs which contain at least one non-unitary RCFT. We show that two operators of complex weights can collide into a logarithmic pair at a certain exceptional point in the moduli space of the interface~\cite{Chen:logarithmicPoint}. This observation indicates very rich structures of the defect conformal manifolds. On the other hand, flows between generalized parafermion models offer a second direction. Their several spin-one phantom currents produce higher-dimensional conformal defect moduli spaces. We will present a systematic study of their massless flows and domain walls separately~\cite{Chen:generalizedPFfuture}. Furthermore, our construction is not restricted to CFTs connected by an RG flow. Given two CFTs with a common fusion category of topological lines, conformal interfaces transparent to these TDLs can also be constructed using our method. We are developing this application for classifying rational interfaces between Ising and TIM CFTs~\cite{AntunesChenRong:interfaces}.

Finally, it is natural to ask whether non-invertible quantum symmetries of the auxiliary theory can be used to define boundary states beyond automorphism gluing. Recent work on symmetry actions on conformal boundaries~\cite{Choi:2024tri} and boundary families generated by continuous non-invertible symmetries~\cite{Delmastro:2025ksn} provides useful starting points. It would be interesting to understand whether such symmetries can define gluing conditions that mix phantom currents with the UV and IR stress tensors. If so, such constructions could then enlarge the class of conformal interfaces accessible by our method.

\FloatBarrier

\section*{Acknowledgments}
I am grateful to Ant\'onio Antunes, Junchen Rong and Yunqin Zheng for useful discussions. Some of the computations were done and checked with the help of ChatGPT/Codex. I am supported by the Fujian Provincial Natural Science Foundation of China (No. 2025J01004), and the National Natural Science Foundation of China (Grant No. 12247103).

\appendix
\section{Numerical boundary spectra for the TIM--Potts family}
\label{app:open-spectrum}
\label{app:pf3-spectrum}

In this appendix, we provide detailed computations of the spectrum of the physical TIM--Potts interface from its self-annulus partition function. The closed string partition function is given by~\eqref{eq:close_pf}. Since the non-invertible quantum line $\mathcal L$ does not commute with the inner automorphism rotation, the modular transform of the closed string partition function is not a finite sum of the ordinary auxiliary characters. We therefore have to evaluate it numerically and extract the exponential contributions of individual Virasoro modules.

\subsection{Open string partition functions}

Let $Z_a(\beta;\varphi)$ denote the physical self-annulus partition function obtained from the auxiliary Cardy label $a=110$ or $211$. Define
\begin{align}
 q=e^{-\beta}\,,\qquad \widetilde q=e^{-\ell}\,,\qquad
 \ell=\frac{4\pi^2}{\beta}\,,\qquad \beta>0\,,
 \label{eq:app-annulus-moduli}
\end{align}
where $q$ and $\widetilde q$ are the open and closed-channel parameters,
respectively, and the total central charge is $c_A=3/2$.
For label $211$, the closed-channel expression is Eq.~\eqref{eq:close_pf}. We evaluate its trace on each module of the auxiliary theory $B_3$. 

It turns out to be useful to first establish the relations between the branes $R_{211}$ and $R_{110}$. In the TIM and Potts product theory, it is easy to show that acting with the line $\mathcal N\boxtimes \overline{\mathcal L}_{\mathsf r^{-1}}$ on $R_{110}$ gives
\begin{align}
    |R_{211}\rangle=\mathcal N\boxtimes \overline{\mathcal L}_{\mathsf r^{-1}}|R_{110}\rangle\,,
\end{align}
where, as discussed in Section~\ref{sec:odd-k-branes}, $\mathcal N$ and $\mathcal L_\mathsf r$ are the Kramers-Wannier line in TIM and $\mathbb Z_3$-line in Potts, respectively. Furthermore, the $\mathbb Z_2$-line $\mathcal L_{\mathsf v}$ in TIM maps $R_{110}$ to $R_{123}$, i.e.,
\begin{align}
    \mathcal L_{\mathsf v}|R_{110}\rangle=|R_{123}\rangle\,.
\end{align}
Therefore the open string partition function of $R_{211}$ can be rewritten as
\begin{align}
    Z_{211}(\beta;\,\varphi)=Z_{110}(\beta;\,\varphi)+Z_{110,\,123}(\beta;\,\varphi)\,,
\end{align}
where $Z_{110,\,123}$ is the open string partition function between the branes $R_{110}$ and $R_{123}$, and we have used the property that the self-annuli of $R_{110}$ and $R_{123}$ are equal, i.e. $Z_{110}=Z_{123}$. 

The known characters $\chi_{110}^{B_3}$ and $\chi_{123}^{B_3}$ in \eqref{eq:sec4.1_open_spec} appear in $Z_{110}$ and $Z_{110,\,123}$, respectively. As they are independent of $\varphi$, we subtract them from the partition functions and define
\begin{align}
 \begin{aligned}
 &P(\beta)=e^{-\beta/16}
              \prod_{n=1}^{\infty}(1-e^{-n\beta})\,,\\
 &F_\varphi(\beta)
  =\left(Z_{110}(\beta;\varphi)
       -\chi_{110}^{B_3}(e^{-\beta})\right)P(\beta)\,,\\[7pt]
 &J_\varphi(\beta)
  =\left(Z_{110,\,123}(\beta;\varphi)
       -\chi_{123}^{B_3}(e^{-\beta})\right)P(\beta)\,.
 \end{aligned}
 \label{eq:app-pf3-remainders}
\end{align}
Here we introduce the function $P(\beta)$ to remove the descendant contributions in the partition function $Z_{211}$. This is because, for a Virasoro module in $B_3$ with central charge $c_B=\frac{3}{2}$, the character is given by
\begin{align}
    \chi_{h}^{\rm Vir}(q)\equiv\chi_{h}^{\rm Vir}(e^{-\beta})=\frac{e^{-\beta(h-1/16)}}{\prod_{n=1}^{\infty}(1-e^{-n\beta})}\,\quad\Longrightarrow\quad
    e^{-\beta h}=P(\beta)\,\chi_{h}^{\rm Vir}(e^{-\beta})\,.
\end{align}
So overall we write 
\begin{align}
Z_{211}=\chi_{110}^{B_3}+\chi_{123}^{B_3}+P^{-1}\left(F_{\varphi}+J_{\varphi}\right)\,,
\end{align}
and our aim is to extract the spectrum from $F_\varphi$ and $J_{\varphi}$.

Our input data are the closed string partition functions, expressed in terms of the variable $\widetilde q=e^{-\ell}$,
\begin{align}
 Z_a^{(H_{\max})}(\beta;\varphi)
 =\sum_{H\leq H_{\max}}A_{a,H}(\varphi)
       e^{-\ell(H-1/16)}\,.
 \label{eq:app-annulus-cutoff}
\end{align}
Here we introduce $H_{\max}$ as the cutoff of the highest chiral conformal weight for each module. The data used for Fig.~\ref{fig:sec4-pf3-self-spectrum} extend to $H_{\max}=20$. One can further expand the coefficients $A_{a,H}(\varphi)$ as
\begin{align}
    A_{a,H}(\varphi)=\sum_{n=-n_H}^{n_H}\widetilde{A}_{a,H,n}e^{in\varphi},
\end{align}
where the frequency ``$n_H$" depends on the $Q_0$-charge on the specific module. The Fourier coefficients $\widetilde{A}_{a,H,n}$ can be computed exactly from Eq.~\eqref{eq:close_pf}.

\subsection{Extract the numerical spectrum}

At fixed $\varphi$, we approximate either $F_\varphi(\beta)$ or $J_\varphi(\beta)$ in Eq.~\eqref{eq:app-pf3-remainders} by a sum of $r$ exponentials representing its lowest spectral contributions, and sample it at equally spaced moduli, e.g.
\begin{align}
 \begin{aligned}
 &F_\varphi(\beta)\simeq\sum_{j=1}^{r}d_j e^{-\beta h_j}\,,\qquad m_n\equiv F_\varphi(\beta_0+n\delta)\simeq\sum_{j=1}^{r}A_jx_j^n\,,\\[5pt]
 \ {\rm with}\quad 
 &x_j=e^{-\delta h_j}\,,\  A_j=d_j e^{-\beta_0h_j}\,,
 \end{aligned}
 \label{eq:app-annulus-fit}
\end{align}
where $\beta_0>0$ is the starting modulus, $\delta>0$ is the
sampling step, $h_j$ is a fitted primary weight, and $d_j$ is its
fitted multiplicity. We introduce the following $r\times r$ matrices
\begin{align}
 H_{ab}=m_{a+b}\,,\qquad K_{ab}=m_{a+b+1}\,,
 \qquad 0\leq a,b<r\,,
 \label{eq:app-annulus-pencil}
\end{align}
which have the factorizations
$H=V\operatorname{diag}(A_j)V^{\mathsf T}$ and
$K=V\operatorname{diag}(A_jx_j)V^{\mathsf T}$, with $V_{aj}=x_j^a$.
For a finite sum of distinct exponentials, the generalized
eigenvalues of the matrix pair $(K,H)$ are therefore the $x_j$~\cite{HuaSarkar:1990matrixpencil}.
Solving for the amplitudes then gives
\begin{align}
 h_j=-\frac{\log x_j}{\delta}\,,\qquad
 d_j=A_j e^{\beta_0h_j}\,.
 \label{eq:app-annulus-log}
\end{align}
We assess numerical stability by varying the closed-channel cutoff, the sampling window and the number of fitted exponentials. Finally we test the reconstructed annulus at samples not used in the fit. For this unitary example, we also check that the extracted weights are real and nonnegative and that their multiplicities are consistent with positive integers. The multiplicities are determined by the fit, rather than fixed in advance. At $\varphi=\pi$, the first few extracted weights are
\begin{align}
 \begin{array}{c|l}
 \text{remainder}&\text{lowest primary weights}\\ \hline
 F_\pi&0.0795566,\quad 0.1816551,\quad 0.7881030\\
 J_\pi&0.0660174,\quad 0.5115305,\quad 0.7537010,\quad 0.8717075
 \end{array}\,.
 \label{eq:app-pf3-fit}
\end{align}

\FloatBarrier

\section{Modular data of \texorpdfstring{$V_L$}{VL} in the generalized parafermion flow}
\label{app:lattice-E-data}

The auxiliary theory $B=\frac{\widehat{\mathfrak{su}}(2)_1^3}{\widehat{\mathfrak u}(1)_3} =V_L$, with $L=\sqrt2 A_2$, has twelve irreducible modules. We give their coset labels and modular data here to fix the conventions used for the seed brane in Section~\ref{sec:orbifold-generalized}. We label each irreducible module of $V_L$ by $\mu=(\boldsymbol r;m)$, where $\boldsymbol r=(r_1,r_2,r_3)$, each $r_i\in\{1,2\}$ labels an integrable module of $\widehat{\mathfrak{su}}(2)_1$, and $m\in\mathbb Z_6$ is the denominator $\widehat{\mathfrak u}(1)$ charge. Define $\ell_i=r_i-1\in\{0,1\}$. The selection rule and field identification in $V_L$ are
\begin{align}
 m&\equiv\ell_1+\ell_2+\ell_3\pmod2\,,
 \nonumber\\
 (r_1,r_2,r_3;m)
 &\sim(3-r_1,3-r_2,3-r_3;m+3)\,.
 \label{eq:app-E-identification}
\end{align}
There are no fixed labels under this identification. We choose
$\boldsymbol r\in\{111,221,212,122\}$ and $m=0,2,4$ as representatives.

We obtain the lattice description by representing each numerator factor
by a compact boson $X_i$, normalized by $X_i(z)X_j(0)\sim-\delta_{ij}\log z$, with $i,j=1,2,3$. The denominator is the diagonal combination $X_{\mathrm{diag}}=(X_1+X_2+X_3)/\sqrt3$, so the coset retains the two orthogonal bosons. To see how this acts on the module labels, choose orthonormal vectors $e_1,e_2,e_3$ along the numerator momentum directions. The allowed momenta in the numerator module $\boldsymbol r$ are
\begin{align}
 p=\frac1{\sqrt2}\sum_{i=1}^3q_i e_i\,,\qquad
 q_i\in\mathbb Z\,,\qquad q_i\equiv\ell_i\pmod2\,,
 \notag
\end{align}
where $q_i$ is twice the $J_0^{3}$-charge of the $i$th factor. For an integer representative of the denominator charge $m$, we take $\sum_iq_i=m$. The diagonal momentum is then $m/\sqrt6$, and its contribution to the conformal weight is $m^2/12$. Separating this boson leaves the transverse momentum
\begin{align}
 p_\perp=p-\frac{m}{3\sqrt2}(e_1+e_2+e_3)
 =\frac1{\sqrt2}\sum_{i=1}^3\left(q_i-\frac m3\right)e_i\,.
 \notag
\end{align}

The vacuum momenta of the remaining two bosons form the lattice $L=\mathbb Z\beta_1+\mathbb Z\beta_2$, with $\beta_1=\sqrt2(e_1-e_2)$ and $\beta_2=\sqrt2(e_2-e_3)$. At fixed $m$, two allowed numerator momenta differ by $\sqrt2\sum_i n_i e_i$ with $n_i\in\mathbb Z$ and $\sum_i n_i=0$, so their transverse momenta differ by a vector in $L$. Moreover, $p_\perp\cdot\beta_1=q_1-q_2$ and $p_\perp\cdot\beta_2=q_2-q_3$ are integers. Therefore $p_\perp$ belongs to the dual lattice
\begin{align}
 L^*=\{\lambda\in\operatorname{span}_{\mathbb R}L\mid
                  \lambda\cdot\alpha\in\mathbb Z\text{ for all }\alpha\in L\}\,.
 \notag
\end{align}
Each irreducible module of $V_L$ contains the momenta in $L+\lambda$, together with their bosonic oscillator excitations~\cite{nguyen2019fusionruleslatticevertex}. The module, denoted by $V_{L+\lambda}$, therefore depends only on the equivalence class $[\lambda]\in\mathcal D=L^*/L$.

For the coset label $\mu$, we choose
$(q_1,q_2,q_3)=(m-\ell_2-\ell_3,\ell_2,\ell_3)$, which obeys
the parity condition by Eq.~\eqref{eq:app-E-identification}.
Substituting into $p_\perp$ gives the representative
\begin{align}
 \lambda_\mu
 =\left(\frac m3-\frac{\ell_2+\ell_3}{2}\right)\beta_1
  +\left(\frac m6-\frac{\ell_3}{2}\right)\beta_2\,.
 \label{eq:app-E-lattice-dictionary}
\end{align}
The field identification in Eq.~\eqref{eq:app-E-identification} can be
implemented by $q_i\mapsto q_i+1$ for all three factors, which changes
only the diagonal momentum. With the representative chosen above, it
changes $\lambda_\mu$ by a vector in $L$, and hence leaves the module
$V_{L+\lambda_\mu}$ unchanged. The lowest conformal weight is obtained by
minimizing the transverse momentum energy:
\begin{align}
 h_\mu
 &=\min_{\alpha\in L+\lambda_\mu}\frac{\alpha^2}{2}\,,
 &
 h_\mu&\equiv\frac{\ell_1+\ell_2+\ell_3}{4}
                 -\frac{m^2}{12}\pmod1\,.
 \label{eq:app-E-weights}
\end{align}
For example, $\mu=(111;2)$ allows $(q_1,q_2,q_3)=(2,0,0)$ and gives
\begin{align}
 \lambda_{(111;2)}=\frac23\beta_1+\frac13\beta_2\,,\qquad
 \frac{\lambda_{(111;2)}^2}{2}=\frac23\,.
 \notag
\end{align}
Applying the same minimization to all twelve classes gives Table~\ref{tab:app-E-weights}.

\begin{table}[htbp]
\centering
\renewcommand{\arraystretch}{1.25}
\begin{tabular}{c|ccc}
 $\boldsymbol r$ & $h_{(\boldsymbol r;0)}$
 & $h_{(\boldsymbol r;2)}$ & $h_{(\boldsymbol r;4)}$\\ \hline
 $111$ & $0$ & $2/3$ & $2/3$\\
 $221$ & $1/2$ & $1/6$ & $1/6$\\
 $212$ & $1/2$ & $1/6$ & $1/6$\\
 $122$ & $1/2$ & $1/6$ & $1/6$
\end{tabular}
\caption{Lowest conformal weights of the twelve irreducible $B$ modules.
The vacuum is $(111;0)$.}
\label{tab:app-E-weights}
\end{table}

Let $q=e^{2\pi i\tau}$, where $\tau$ is the torus modular parameter, and let $\eta(\tau)$ be the Dedekind $\eta$-function. The characters are
\begin{align}
 \chi^B_\mu(\tau)
 =\frac1{\eta(\tau)^2}
   \sum_{\alpha\in L+\lambda_\mu}q^{\alpha^2/2}\,.
 \label{eq:app-E-characters}
\end{align}
Poisson resummation gives the modular matrix
$S^B_{\mu\nu}=12^{-1/2}e^{-2\pi i\lambda_\mu\cdot\lambda_\nu}$.
For $\nu=(\boldsymbol s;n)$ and $\ell'_i=s_i-1$, this becomes
\begin{align}
 S^B_{(\boldsymbol r;m),(\boldsymbol s;n)}
 &=\frac1{\sqrt{12}}\,
 (-1)^{\sum_{i=1}^3\ell_i\ell'_i}
 e^{i\pi mn/3}\,,
 &
 T^B_{\mu\nu}
 &=\delta_{\mu\nu}e^{2\pi i(h_\mu-1/12)}\,.
 \label{eq:app-E-modular-data}
\end{align}
Here $T^B$ includes the central-charge contribution $c_B/24=1/12$.
\FloatBarrier

\newpage
\bibliographystyle{JHEP}
\bibliography{references}
\end{document}